%% file: main.tex
\documentclass[sigconf]{acmart}

\setcopyright{none}
\renewcommand\footnotetextcopyrightpermission[1]{}

\usepackage[absolute,showboxes]{textpos}

\usepackage[utf8]{inputenc}
\usepackage{listings}
\usepackage{balance}

\usepackage{tikz}
\usepackage{xcolor}
\usepackage{color}
\usepackage{pgfplots}

\graphicspath{{../figs/}}

\usetikzlibrary{
	arrows.meta,
	calc,
	colorbrewer,
	external,  
	fit,
	matrix,
	positioning,
	shapes,
	shapes.misc,
	backgrounds
}

\begin{document}

\setlength{\TPHorizModule}{\paperwidth}
\setlength{\TPVertModule}{\paperheight}
\TPMargin{5pt}
\begin{textblock}{0.8}(0.1,0.02)
    \noindent
    \footnotesize
    If you cite this paper, please use the AGERE!@SPLASH reference:
    Raphael Hiesgen, Dominik Charousset and Thomas C. Schmidt.
    A Configurable Transport Layer for CAF. In \emph{Proc. of ACM SIGPLAN SPLASH}, ACM, 2018.
\end{textblock}

\title{A Configurable Transport Layer for CAF}

\author{Raphael Hiesgen}
\affiliation{
  \department{Dept. Computer Science}
  \institution{HAW Hamburg}
  \country{Germany}
}
\email{r.hiesgen@haw-hamburg.de}

\author{Dominik Charousset}
\affiliation{
  \department{Dept. Computer Science}
  \institution{HAW Hamburg}
  \country{Germany}
}
\email{dcharousset@acm.org}

\author{Thomas C. Schmidt}
\affiliation{
  \department{Dept. Computer Science}
  \institution{HAW Hamburg}
  \country{Germany}
}
\email{t.schmidt@haw-hamburg.de}

\renewcommand{\shortauthors}{R. Hiesgen, D. Charousset, and T. C. Schmidt}

\input{abstract.tex}



\lstset{
  language=C++,
  morekeywords={constexpr,nullptr,size\_t,this,receive,\_\_global,\_\_kernel},%
  frame=top,frame=bottom,
  basicstyle=\small\normalfont\ttfamily,  
  stepnumber=1,                           
  numbersep=10pt,                         
  tabsize=2,                              
  extendedchars=true,                     %
  breaklines=true,                        
  captionpos=t,                           
  numbers=left,
  numberstyle=\small\color{midgrey},
  showspaces=false,                       
  showtabs=false,                         
  xleftmargin=17pt,
  framexleftmargin=17pt,
  framexbottommargin=0pt,
  framextopmargin=0pt,
  showstringspaces=false                  
}

\maketitle

\input{introduction}
\input{related}
\input{caf}
\input{background}
\input{design}
\input{evaluation}
\input{conclusion}

\subsection*{A Note on Reproducibility}

We explicitly support reproducible research~\cite{acmrep,swgsc-terrc-17}.
Our experiments have been conducted in a transparent standard environment.
The source code of our implementations (including scripts to setup the experiments, CAF measurement apps etc.) are available on GitHub at \url{https://github.com/inetrg/agere-2018}.

\begin{acks} 
We are grateful for many lively discussions and the inspiring environment of the INET team in Hamburg. In particular, we want to thank Jakob Otto for his helping hands in experimentation.

This work was supported in parts by the German Federal Ministry of Education and Research within the projects Scalecast and X-Check.
\end{acks}

\balance

\bibliographystyle{ACM-Reference-Format}
\bibliography{main}



\end{document}

%% file: abstract.tex
\begin{abstract}

The message-driven nature of actors lays a foundation for developing scalable and distributed software. While the actor itself has been thoroughly modeled, the message passing layer lacks a common definition. Properties and guarantees of message exchange often shift with implementations and contexts. This adds complexity to the development process, limits portability, and  removes transparency from distributed actor systems.

In this work, we examine actor communication, focusing on the implementation and runtime costs of reliable and ordered delivery. Both guarantees are often based on TCP for remote messaging, which mixes network transport with the semantics of messaging. However, the choice of transport may follow different constraints and is often governed by deployment.
As a first step towards re-architecting actor-to-actor communication, we decouple the messaging guarantees from the transport protocol. We validate our approach by redesigning the network stack of the C++ Actor Framework (CAF) so that it allows to combine an arbitrary transport protocol with additional functions for remote messaging. An evaluation quantifies the cost of composability and the impact of individual layers on the entire stack.

\end{abstract}

%% file: introduction.tex
\section{Introduction}
\label{sec:intro}

Concurrency and distribution are an inherent part of modern systems and prevalent in most areas from personal computing and data centers to mobile platforms and the IoT. One challenge apparent in those areas is the dynamic adaption to the environment of deployment. Personal devices---often mobile notebooks, tablets, or phones---change locations, rely on cloud services, and regularly communicate through NATs and firewalls. For cloud scenarios and Mobile Edge Computing (MEC), this  leads to service mobility and unpredictable location of nodes, which change application deployment according to  user behavior. The IoT is still an emerging field with applications in home, infrastructure, and industrial automation targeted at a heterogeneous variety of deployments that include gateways.

The actor model of computation~\cite{hbs-umafa-73} seamlessly integrates concurrency and distribution, and gains popularity for designing and developing applications that meet the demands of flexible adaptivity and high scalability. Actors solely communicate via network-transparent message passing while applying a strong failure model. In reaction to a message, an actor can send messages, create new actors, or configure its future behavior. Actors offer a high level of abstraction that allows developers to focus on their application while the underlying framework takes responsibility for error prone tasks such as synchronization and networking---the implementations of which require domain-specific knowledge and experience.

\paragraph*{Problem Statement}

Although the behavior of actors has been carefully modeled, their message passing layer lacks a clear definition.  Communication guarantees regarding message delivery or ordering often diverge between implementations and contexts. For example, Armstrong~\cite{a-mrdsp-03} assumes message passing in Erlang ``[$\dots$] to be unreliable with no guarantee of delivery''. The Erlang software documentation closely couples reliability to the reliability of TCP transport. Similarly, according to its documentation\footnote{\url{https://doc.akka.io/docs/akka/current/general/message-delivery-reliability.html}, accessed Aug'18} Akka delivers messages with an ``at-most-once'' semantics, even though authors acknowledge that the guarantees are much stronger in local deployment. The same discrepancies can be found for ordering guarantees. Here non-local messages between a pair of actors often follow \textit{FIFO} ordering while local messages are usually ordered \textit{causally}---a result of synchronous calls to enqueue messages into a local mailbox. 

There are many reasons for these discrepancies in the implementation of local and remote contexts \cite{hcs-rrdas-16}. First, local guarantees are much easier achieved than by protocols involving network communication, where uncertainty and unreliability need explicit treatment. Often an analysis of the alternatives and a reasoning for the offered guarantees is missing.

In practice, guarantees are often enforced by a tight coupling with the transport protocol of the desired characteristics. This approach may be acceptable for a large number of applications. However, it must be considered a severe limitation when scaling from small embedded devices over mobile and desktop to cloud services. While TCP is the dominant protocol throughout the Internet, HTTP tunnels and WebRTC can enable communication between nodes hidden behind firewalls and NATs. Scaling to high performance environments, technologies such as DCTCP or InfiniBand are optimized for closely coupled clusters. On the low end of the scale, constrained environments depend on specialized standards such as 6LoWPAN or CoAP over UDP to address a loose coupling in lossy networks. 

Choosing a transport protocol is a trade-off between the scope of services a protocol offers and the environment of its operation. Simply deploying the protocol that offers the best guarantees is not a viable solution. While constrained environments might not be able to handle the messages sizes or network load, other applications may require low-latency and would rather loose messages than wait for retransmits. This trade-off further motivates the need for decoupling messaging guarantees from transport. Instead, an exchangeable transport layer augmented by configurable services can address scalability, adaptivity, as well as dynamic deployment at the same time. 

Developers should be able to rely on a set of guarantees offered by a framework instead of rewriting applications to handle these tasks or tying communication to a specific protocol. These guarantees should be enforced across protocol choices and layers and allow transparent deployment of data transport based on the use case or environment.

\paragraph*{Contributions}

In this work, we re-examine the duties and workings of actor communication with the goal to identify a set of reasonable guarantees for message passing between actors. The C++ Actor Framework \cite{cshw-nassp-13,chs-rapc-16} is used as a reference. 
Specifically, we contribute:
\begin{enumerate}
  \item A survey and discussion of communication aspects relevant to actor frameworks, focusing on reliable delivery and ordering.
  \item A redesign of the CAF network layer to address our observations and allow a composable transport implementation.
  \item A first evaluation of our design focusing on the cost of composability.
\end{enumerate}

\paragraph*{Overview}

Section \S~\ref{sec:related_work} discusses related work while Section \S~\ref{sec:caf} introduces CAF, the framework hosting our subsequent work. \S~\ref{sec:background} reflects the main aspects considered in our design: reliable delivery and ordering. We present the  redesign of the CAF network layer in \S~\ref{sec:design}, and evaluate  our  implementation in \S~\ref{sec:evaluation}. Finally, \S~\ref{sec:conclusion} concludes with an outlook.

%% file: related.tex
\section{Related Work}
\label{sec:related_work}

\paragraph*{Reliable Delivery}

This aspect signifies how likely it is for a sent message to reach the destination and whether feedback is available in case of failure. Akka delivers messages unreliably with ``at-most-once'' semantics per default\footnote{\url{https://doc.akka.io/docs/akka/current/general/message-delivery-reliability.html}, accessed Aug'18}. This means a message is delivered either once or not at all to the destination mailbox. Included in the framework is a solution for ``at-least-once'' delivery in form of a persistence module which additionally allows actors to recover their state after a crash. Erlang is named as an inspiration for defaulting to weak delivery guarantees as it successfully uses a similar approach.

Armstrong defined message passing in Erlang ``[$\dots$] to be unreliable with no guarantee of delivery'' in his thesis~\cite{a-mrdsp-03}. The additional effort required to write applications that can handle unreliable message passing furthers scalability and increases robustness against errors. A later publication~\cite{a-he-07} goes into more detail on the topic and couples the reliability of messages passing to the reliability of TCP\@. However, TCP itself is not enough to guarantee delivery to an actor. Errors in the runtime environment (RE) can occur after a message was accepted at the application endpoint, but before it was passed on. An example for this type of failure in a simple distributed Erlang setup is provided by Svensson et al.~\cite{sf-pdeap-07}.

Microsoft released Orleans~\cite{bbgkt-odvap-14}, an implementation of the actor model that targets clusters. It hides most of the distribution and error handling from developers. Failed actors are detected by the runtime environment and redeployed transparently before delivering a message. The RE favors availability over consistency when redeploying actors and accepts temporary inconsistencies such as actors performing redundant calculations. Per default, messages are exchanged with a ``maybe'' delivery guarantee to avoid the associated costs in every message exchange. However, ``at-least-once'' delivery can be enabled, which retransmits messages until reception is acknowledged\footnote{\url{http://dotnet.github.io/orleans/Documentation/clusters_and_clients/configuration_guide/messaging_delivery_guarantees.html}, acc. Aug'18}. Since the RE does not detect duplicates, implementing ``at-least-once'' delivery burdens developers with deduplication in their implementation.

The blog post ``Nobody Needs Reliable Messaging''\footnote{\url{http://www.infoq.com/articles/no-reliable-messaging}, accessed Aug'18} analyzes reliability in the context of SOA, Web Services and REST\@. It argues that reliability requires conformation on the application layer which makes an implementation on a lower layer redundant. A similar conclusion is drawn for duplicate message detection, e.g., a duplicate order in an online market would lead to the same messages with different sequence numbers on the transport layer. Related to this discussion, Saltzer et al.~\cite{src-eeasd-84} explore the implications of end-to-end communication. Without knowledge of higher layers it might be tempting to provide more functionality than needed. While functionality can be implemented on top of communication systems, in some cases it may be beneficial to implement partial functionality on lower levels to enhance the overall performance and reduce the complexity and overhead. As a result, the assumption that avoiding redundancy improves performance should be viewed with care.

In his dissertation, Agha argues the guarantee of communications delivery should be modeled as it eases the reasoning about the system in regard to its correctness or termination properties~\cite{a-amccd-86}. However, he notes that the buffers required for the communication are limited by nature which makes it impossible to ensure delivery in all cases.

\paragraph*{Reliable Ordering}

Ordering describes relationships among messages exchanged between two or more actors, i.e., whether messages arrive in the same order they were sent. This is usually limited to the order of arrival in mailboxes. Actors are free to process messages out-of-order or deploy mailboxes that sort incoming messages by priority. There are four orderings with increasingly strong assurances that we consider here: \textit{non-deterministic}, \textit{first in - first out}, \textit{causal} and \textit{total}. There are several opportunities to establish ordering. While some guarantees could be implemented by a suitable transport or application layer protocol, other require more complex synchronization between nodes.

First-in, first-out ordering (\textit{FIFO}) means that messages sent first arrive first. This guarantee only creates a relation between messages from a single sender and is not transitive. Transitivity would maintain order even if a message is received and forwarded by an intermediate node.

The ``happened before'' relation~\cite{l-tcoed-78, m-vtgsd-89} describes the logic of \textit{causal} message ordering. Unrelated messages are determined to be ``concurrent'' or ``independent''. Hence, \textit{causal} ordering is not restricted to messages exchanged by a pair of actors, but can establish a relationship between messages throughout the whole system.

A \textit{total} order extends \textit{causal} order and gives order to all messages in the system not only to causally related ones. Hence, all messages arrive in the same order at all receivers. Introducing a \textit{total} order requires the synchronization of all participants. To achieve this, the totem protocol~\cite{ammac-fmomu-93} passes a token around in a logical ring, which allows the owner to broadcast messages. Until the token is acquired, messages are buffered locally. An alternative approach could be a central sequencer that provides sequence numbers for all messages and advances the time.

The actor system Orleans~\cite{bbgkt-odvap-14} is an example of a framework that does not enforce ordering at all. It wants to avoid the related impact on scalability as well as the overhead in processing power and state that is required to restore the order of received messages. CAF follows a similar approach and currently does not maintain the order of messages actively and instead relies on the ordering implicitly inherited from TCP. This leads to \textit{causal} ordering in a local context and transport-dependent ordering for remote messaging.

Erlang and Akka both enforce \textit{FIFO} ordering. Although Erlang defines this ordering as part of their basic rules of message passing~\cite{a-mrdsp-03}, the decision is not further explained besides stating that it eases application development. Akka stresses that this is only true for the order in which messages are enqueued into the mailbox~\cite{l-asd-18}. 
In particular, system messages such as errors use special mailboxes and may be delivered out-of-order. Akka implements ordering on top of TCP, but utilizes additional per-connection queues to sort messages and handle errors such as TCP reconnects and full buffers. 

Long et al.~\cite{lblur-opmps-16} explore reasons for ordering problems in message passing systems. The three main criteria they identify are 1) \textit{synchronization}, i.e., either asynchronous or synchronous messaging, 2) \textit{processing}, comparing non-deterministic against in-order delivery and processing, as well as 3) the \textit{sharing} aspects data sharing and data isolation. For example, code that looks sequential but depends on asynchronous, unordered messages may lead to undefined behavior. They build a message passing model by combining different aspects of these semantics. Their base model uses asynchronous message passing, with non-deterministic message delivery and processing as well as data sharing semantics. The other models are built by exchanging different aspects as well as adding transitive in-order delivery. A static analysis is used to evaluate how programs are affected by ordering problems when exchanging messages with these models. Their evaluation shows that synchronous, in-order, and data isolation have the biggest effect on ordering problems for applications. In contrast, transitivity only helps in for very few cases. For framework designers, they see in-order delivery and data isolation as the most critical semantics. This analysis can help to weigh guarantees against their costs when choosing what to provide as a default.

Blessing et al.~\cite{bcd-ttcmd-17} propose implementing \textit{causal} ordering by arranging participating nodes in a tree topology. While the approach further relies on \textit{FIFO} ordering between each pair of nodes, it does not require additional meta data. This work related to the Pony actor language which aims to implement transparent distribution, i.e., hide the characteristics of distribution from the programmer.

%% file: caf.tex
\section{The C++ Actor Framework}
\label{sec:caf}

The C++ Actor framework (CAF)~\cite{chs-ccafs-14, chs-rapc-16} combines the benefits of native program execution with a high level of abstraction. The best known implementations of the actor model, Erlang and Akka, are both implemented in languages that rely on virtual machines. In contrast, CAF is implemented in C++, thus compilies to native code and has shown significant performance benefits. C++ is used across the industry from high performance computing installations running on thousands of computing nodes all the way down to systems on a chip. CAF fits into the gap between the high level of abstraction offered by the actor model and an efficient, native runtime environment.

Following the tradition of the actor model, actors are created using \lstinline^spawn^. The function takes a C++ functor or class and returns a handle to the created actor. Hence, functions are first-class citizens and developers can choose whether they prefer an object-oriented or a functional software design. Per default, actors are sub-thread entities scheduled cooperatively using a work-stealing algorithm~\cite{bl-smcws-99}. This results in a lightweight and scalable actor implementation that does not rely on system-level calls as required when mapping actors to threads. Uncooperative actors that require access to blocking function calls can still be bound to separate threads by the programmer to avoid starvation. Recent optimization work by Torquati et al.~\cite{tmmsm-rmlcu-18} pushed CAF into the direction of low latency communication by reducing messaging latency by up to two orders of magnitude for low and moderate data rates.

The network stack in CAF consists of several components that manage network communication. The \textit{middleman} provides the user-facing API of CAF in a distributed context. When communicating with actors on remote nodes, a local \textit{proxy} is created for each directly known actor. \textit{Brokers} are actors that abstract over a network interface and provide an actor interface for sending and receiving data. CAF deploys a system broker to parse and handle the application layer protocol BASP (Binary Actor System Protocol) for the management and communication between CAF nodes. Finally, a \textit{multiplexer} uses a system-dependent multiplexing implementation to bridge the gap between socket operations and the broker interface.

\begin{figure}[htb]
  \centering
  \includegraphics[width=1.0\columnwidth]{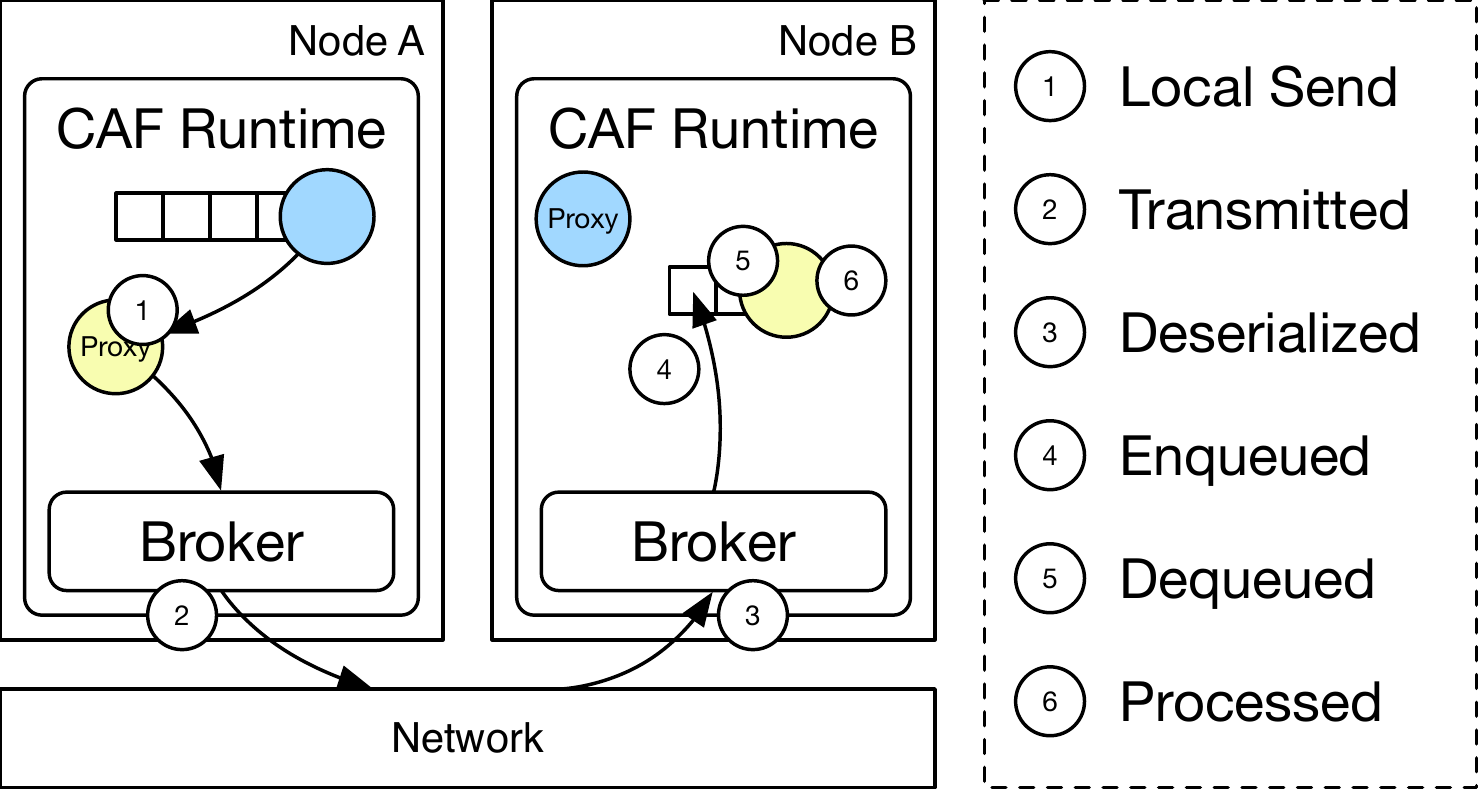}
	\caption{Message path through CAF.}
  \label{fig:caf:message_path}
\end{figure}

Figure \ref{fig:caf:message_path} shows the path that a message takes in CAF. The first step is a synchronous local operation (1). For messages to remote actors, the local proxy transparently forwards messages to the local system broker which  serializes the message (2). Then, the Broker resolves the address of the receiver and transmits the message. After reception on the remote node a broker deserializes the message (3). Then, it is enqueued into the mailbox of the receiver (4). When the receiver is scheduled and its mailbox contains no messages that arrived previously or have a higher priority, it dequeues the message (5) and processes it (6).

%% file: background.tex
\section{Communication Guarantees for Actors}
\label{sec:background}

Message passing is the central communication primitive of the actor model for exchanging data and driving the application logic. Hence, the characteristics of the messaging layer dictate failure models and API decisions.

In this work, we focus on reliability and ordering as central aspects of any message exchange. Both concepts are well understood in packet-switching networking and implemented in transport protocols such as TCP or QUIC~\cite{draft-ietf-quic-transport} on UDP. However, most implementations of the actor model simply rely on guarantees made by the transport protocol rather than thoroughly defining actor messaging. This is convenient for implementers of the actor model, but tightly couples fundamental system properties to deployment technologies. We leave failure detection, error propagation, reachability and security to future work. Incorporating these aspects is a natural extensions to our results presented here.

There are many choices when deciding on communication guarantees. We argue that a good design reduces complexity to a minimum for both users and implementers. Users of the system must be able to quickly form a consistent mental model without being overwhelmed by having to consider diverse edge and corner cases. Implementers likewise must be able to understand and---most crucially---debug many layers of interdependent software modules.

Reducing complexity is especially important when modeling a distributed system. Many sources of errors combined with unpredictable control flow timing pose a different challenge than designing a software stack for a single-node or even a single-thread program. Operational overhead requires careful consideration as well. Finally, incorporating a choice of desired guarantees enables developers to rely on the default implementation without adding additional hand-crafted layers on top.

This sections examines reliable delivery and ordering in regard to actor systems and discusses which guarantees impl

\subsection{Reliable Message Delivery}

At the lowest level, physically available memory is always limited. Messages cannot reach their destination if an actor mailbox or network buffer fails to allocate sufficient storage. However, to model delivery guarantees is still valuable for reasoning about program correctness~\cite{a-amccd-86}.

Message delivery is reliable if each message either reaches its destination eventually or gets discarded with an error report to the sender. In other words, the system must never drop messages silently. However, the actor model is based on asynchronous message passing and does not specify errors for dropped messages. Likewise, limiting mailbox sizes is typically neither addressed nor implemented. Dedicated communication channels that signal status and demand go beyond the scope of this work but are addressed by the forthcoming streaming API in CAF.

Messages to remote actors travel through several software layers until they reach the destined actor. Figure~\ref{fig:caf:message_path} depicts this path specifically for CAF, but each implementation of the actor model will have similar steps. Hence, using CAF for examining individual steps and discussing algorithm choices translates well to other systems.

\textbf{Local Sending} is a synchronized operation that only fails when running out of memory. Remote actors are represented by proxies that transparently forward messages---along with meta information for reaching the remote actor---to the system broker. Mailbox state of remote actors remains opaque, as proxies act only as a message relay. Tightly synchronizing proxies with remote actors could potentially catch overloads early, but ultimately would only shift stress between nodes and impair performance by inducing very high communication overhead.

\textbf{Transmitting} packets to remote nodes requires peer management and serialization of messages to a portable format. The BASP broker in CAF acts as the central network hub and provides all required functionality. In particular, the broker 1) maps node IDs to sockets, 2) forwards EXIT and DOWN messages between local and remote actors, and 3) generates EXIT and DOWN messages for monitored / linked remote actors on node failures. The latter requires liveliness detection of remote nodes. Most frameworks simply rely on TCP by interpreting connection aborts as node failure. Trying to re-establish communication requires extensive buffering and sychronization when trying to maintain exactly once delivery between nodes. Alternatively, raising errors early can at least reduce the amount of buffered and potentially lost messages by giving actors immediate feedback about potentially unreachable remotes.

\textbf{Deserializing} at the remote node follows successful transmission. Network communication is inherently unreliable and bares additional sources of errors such as packet loss, packet duplication, or link failure. Moreover the exact failure is often hard to detect. As an example, nodes cannot distinguish between loss and delay until data arrives. Timeouts, retransmissions, and deduplication offset or solve some issues at the cost of increased communication, slow buffering, and additional complexity. Transport protocols such as TCP offer increased reliability by implementing guarantees for communication between two endpoints. Failures on the transport layer still give vague feedback to determine the liveliness of remote nodes. Initiating and managing reconnects after communication errors is not part of transport protocols. Instead, applications need to deploy necessary state and connection tracking manually. Hence, simply relying on TCP neither prevents message loss nor failures~\cite{sf-pdeap-07}. Overall, improving reliability of the network transport improves usability of the communication primitives as it relieves developers from the complexity to implement their own protocols. Deserialization fails when running out of memory. Dropping messages under temporary heavy load can become an option in the presence of an application layer protocol that handles retransmissions. Observing repeated retransmission requests from remotes also allows nodes to detect likely overloaded peers and to raise related error or status messages.

\textbf{Enqueueing} messages into the mailbox of the receiving actor concludes the processing steps involving brokers. Again, this operation can only fail when lacking sufficient memory. An error at this stage usually indicates an imbalance between the message arrival rate and processing capacity. Detecting and managing such issues requires some form of flow control between actors.

\textbf{Dequeuing} messages from the mailbox is the final step under control of the framework before handing control to user code. Estimating wait time of messages is bound to be very imprecise because it depends on processing time, fairness of the scheduling, prioritization of messages by the actor, etc. Unlike network packets, the framework could track individual messages to reproduce a global view of all messages in the system. However, considerable performance impacts due to the high synchronization overhead make global tracking undesirable in practice. The framework could still check for per-message timeouts at the point of dequeuing and drop timed-out messages. Such user-defined timeouts could force errors, but require very precise estimates by developers to add any value, in particular, trigger neither too aggressively nor too generous during ordinary program flow. In the worst case, a timeout is triggered while the response message is already traveling through the system back to the sender.

\textbf{Processing} messages can fail due to exceptions in user code. Such errors automatically terminate the actor and the framework propagates this failure through DOWN and EXIT messages. Estimating processing times again is very imprecise at best, unless developers have provided information for predicting runtime from message content. Actors yield control back to the framework after completing a message, optionally producing a response message.

\paragraph*{Discussion}

Three messaging steps stand out among the six that were discussed: 1) local send, 2) enqueueing into the mailbox, and 3) receiving a processing confirmation.

The first one, a ``fire and forget'' send, stands out because it is the simplest, most bare-bone step. Its messaging model remains asynchronous with little complexity, overhead, and state. Combined with messages that propagate liveliness of actors and nodes, complex systems can be built on top. While this approach burdens developers with error handling for basic messaging, the resulting applications are robust to a variety of failures. Most notably, this leaves the implementations with a discrepancy between local and remote message passing, thus breaking transparent distribution.

Reliable delivery to the destination mailbox extends the local guarantees to the remote messaging. The assurances of this step go beyond network transport and additionally address deserialization as well as buffering issues. The actor model does not include means to propagate these failures. Both failures categories are not easy to address generically. If deserialization fails there is no solution to fix it at runtime. Adding a specific message to propagate the error is  possible although well defined message passing interfaces seem to be a better way to address the problem. When running out of memory a system has limited options to address the failure. Simply dropping messages that could not be processed for such a reason might allow an application layer protocol to retransmit messages until the receiver acknowledges receipt.

Acknowledging message processing provides the most insightful information about end-to-end communication~\cite{src-eeasd-84}. At the same time, addressing a generalized use-case is a very complex task heavily dependent on the application logic. Processing time per message, average delay in the mailbox, current load and the messaging interface of the receiver all influence whether a message is processed and how long it takes. As a result, a reasonable failure case cannot be defined across all scenarios. Propagating related information requires messages with custom handlers at the sender side since a generic reaction cannot be assigned.

From a model perspective reliable delivery that raises remote to local guarantees is valuable for modeling and makes it easier for developers to argue about their code. In practice, the step from delivery over the network to enqueuing messages into mailboxes does not provide enough benefit to merit an additional application layer protocol. Cases that merit overhead to ensure delivery are often interested in the processing results and not only the delivery, thus falling into the category of the end-to-end argument.

\subsection{Reliable Message Ordering}

Reading code and understanding side effects is easier when messages sent by sequential statements are delivered in the same order~\cite{lblur-opmps-16}. Relying on the same ordering for local and remote messages prevents deployment specific bugs and eases porting local applications to distributed systems. In the same way, reproducing failures is easier to achieve if communication is predictable. While priority messages naturally break ordering, users expect that effect.

\textbf{Non-deterministic} ordering is easy to implement. Although dependent on the implementation details of local message passing, this often leaves developers with different guarantees for local and remote communication~\cite{hcs-rrdas-16}.

\textbf{First in, first out (FIFO)} ordering can be implemented for actor-to-actor or node-to-node communication. It requires sequence numbers to determine order and buffering to restore it in both cases, but with differing granularity. Implementing ordering per actor distributes the problem and avoids ordering unrelated messages between different actors. In practice, this not only introduces an additional step between the application layer protocol and actor messages, but requires state that scales with the number of actors in the system. Moving ordering to a protocol between each pair of nodes offers much better scalability as the state to track sequence numbers and buffers only scales with the number of peers. On the downside, a delayed message also impacts unrelated messages.

\textbf{Causal} ordering can be established in various ways. Restricting communication to synchronous message passing is often easy to implement, but heavily impacts the application behavior. The asynchronous nature of actor messages does not map well to such a restriction. Annotating each message with metadata is another option. The additional information that needs to be exchanged are time vectors with a size equal to the number of processes $n$~\cite{b-cslcd-91}. Moreover, message sizes increase further to determine \textit{causal} dependencies for transitive message passing with more than one intermediate node~\cite{as-rcmom-95}. A third alternative is a fixed routing topology such as a ring or a tree as discussed in the context of the Pony language~\cite{bcd-ttcmd-17}. This overloads routing and leads to a worst cases where messages are routed from one leaf through the root to another leaf, thus introducing latency. Maintaining the topology with nodes joining, leaving, or failing is a complex task that becomes more tedious in mobile environments.

\textbf{Total} ordering requires a straight forward but very expensive implementation. One node in the system is chosen as a sequencer that determines the message order. Such an approach introduces a strong coupling in the system as even local messages would have to be subject to this process.

\paragraph*{Discussion}

Local delivery in CAF leads to a \textit{causal} ordering of messages enqueued into a mailbox. This is a result of implementing mailboxes as lock-free \textit{FIFO} queues which are accessed by actors in a single non-blocking but synchronous call when sending messages.

\textit{Total} order is not a desirable property for messages exchanged between actors. By definition, actors are concurrent and isolated entities. Adding a strong coupling in the form of a central sequencer to all communications impacts scalability and performance without significant benefit. While some use cases may justify the overhead to maintain a \textit{total} order, the majority of cases does not. As such, it is not a good candidate for default ordering.

Implementing \textit{causal} ordering also comes at significant cost. Relying on synchronous communication introduces a strong coupling between actors and nodes. Although synchronization on a local machine is cheap, extending it to a remote context significantly impacts performance and scalability. The cost of synchronization over the network is several orders of magnitude higher and introduces undesirable delay. Developers would have a strong incentive to avoid remote communication breaking transparency on another axis. Alternatively, adding vector timestamps to messages largely increases the amount of data exchanged in the system. In addition, hosts schedule high amounts of actors and frequently spawn new ones that only run for a limited time or task. A changing amount of participants is generally not handled well by vector clocks. Neither approach comes without tradeoffs that significantly impact performance and scalability of an actor system.

While ordering eases software development, strong ordering guarantees are costly and introduce the need for synchronization. \textit{FIFO} ordering has a comparably low overhead and provides part of the ordering characteristics of local messaging to remote messaging. For each pair of actors reasoning about exchanged messages is straight forward. As such it is a tradeoff between desirable properties and overhead.

%% file: design.tex
\section{A Composable Network Stack}
\label{sec:design}

Maintaining a consistent set of communication guarantees across exchangeable transport protocols requires design changes to the CAF network stack. Although support for UDP was added recently, developers who want to integrate new transport protocols are still required to adjust various components throughout the I/O library. Extending the guarantees of transport protocol requires implementation on top of a broker and is not easily reusable.

The redesign addresses these issues and leads to a composable network stack that can be extended with new transport protocols and augmented with reusable protocol layers to add to its functionality. With the goal to enable use of arbitrary transport protocols, this concept uses TCP and UDP as examples for the design. These two protocols do not cover all functionality that transport protocols can offer, but differ greatly in their included guarantees. While UDP is a bare-bones protocol that provides connectionless transmission of datagrams with few guarantees, TCP streams bytes with strong reliability and ordering guarantees among others. Thus, this protocol selection provides the opportunity to examine how our concept could integrate them.

\begin{figure}[htb]
  \centering
  \includegraphics[width=\columnwidth]{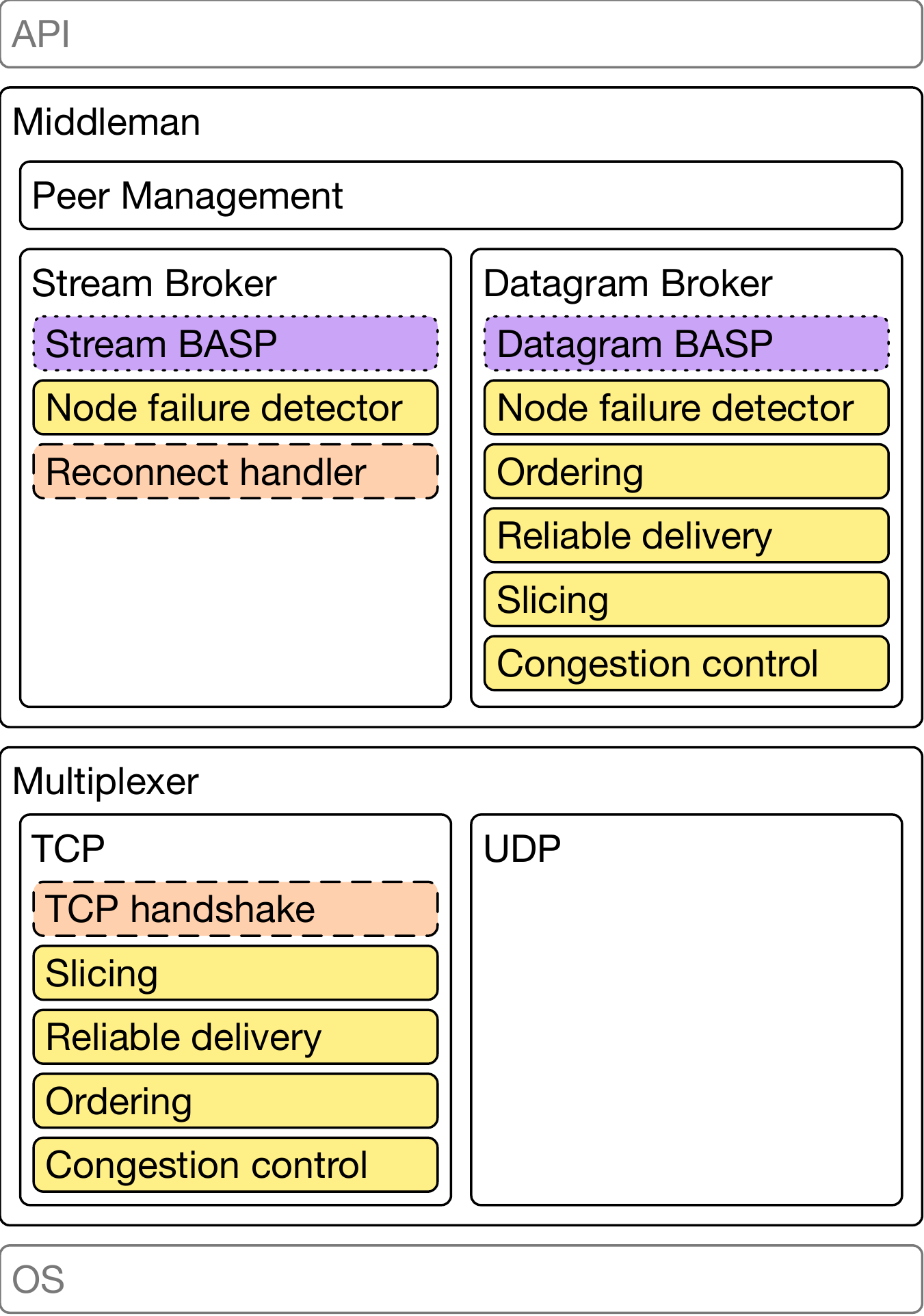}
  \caption{Composition of the CAF network stack deploying TCP and UDP.}
  \label{fig:design:opt_final}
  \vspace{-15pt}
\end{figure}

The design of the network stack is shown in Figure \ref{fig:design:opt_final}. Yellow boxes (normal border) signify general functionality that can be provided by and for various protocols. Orange boxes (dashed border) are TCP specific and purple boxes (dotted border) are specific to CAF. Management of sockets is handled by the \textit{multiplexer} which interfaces with the OS to provide asynchronous socket access. Located above the multiplexer in the stack are \textit{brokers} which are managed by the \textit{middleman}. A broker bundles a transport protocol with additional layers and wraps it in a message passing interface.

In the example case, a TCP-based stream broker adds layers to detect remote node failures and attempt reconnects in addition to a component for reading and writing BASP messages from and to a byte stream. The second broker handles datagrams characteristic for UDP. Similar to the stream broker it deploys a failure detector and a layer to translate between datagrams and BASP messages. Additionally, it is extended to slice messages into datagrams of suitable size to avoid IP fragmentation, order incoming datagrams and ensure their delivery. A reconnect handler is not needed due the connectionless nature of UDP. While some layers might be valuable in both protocols, they could benefit from a protocol specific implementation. As an example, a TCP failure detector could monitor the connection state to detect failures.

Brokers are suitable components for this functionality. They sit in-between the actor abstraction and low-level socket API. As such, they already require protocol dependent code to translate between the incoming bytes and application data. Placing such functionality higher up would make it part of the application logic. While developers are free to do so, the default approach should cleanly separate the networking logic from the application logic. In contrast moving lower down the stack hinders access as the functionality would be colocated with low level code. Brokers are a fitting abstraction for this task.

\paragraph*{Implementation}

In CAF, a broker is a component that abstracts over an endpoint for a specific transport protocol. Instead of running in the system scheduler, it is scheduled in the event loop of the multiplexer. The multiplexer executes it when an I/O event occurs on its socket or when it receives a message.

There are two types of brokers. The first one handles regular events on a socket. Similar to other actors it handles messages according to its behavior which has to include a handler for the message type it receives for new data on its sockt. It is configured by two policies: a \textit{transport policy} and a \textit{protocol policy}. Policies are a way to implement configurable components in C++.

A \textit{transport policy} wraps a transport protocol by implementing functionality to read from and write to a socket and manage the related buffers.

The layers that augment guarantees or functionality are thereby define the overall protocol are bundled in a \textit{protocol policy}. Each layer accepts the type of the next layer as a template argument and instantiates it as a member. An exception is the upmost layer which does not have another layer as a member. In addition, it dictates the message type passed to the broker for new data.
  
Before sending data, the protocol policy gives each layer the opportunity to write headers, set timeouts, and augment the send buffer. Similarly, upon receipt each layer can read its header, set timeouts, and sent messages. The order of layers is meaningful and is reverse for sending and receiving.

The second broker type is responsible for accepting new endpoints or multiplexing over a single socket, if desired. It creates a new broker of the first type to handle new endpoints. It is configured by an \textit{accept policy} that determines how to react to incoming data. For TCP, an accept policy could simply accept new connections and pass the sockets to its broker, which in turn spawns brokers to handle regular communication.

%% file: evaluation.tex
\section{Evaluation}
\label{sec:evaluation}

Network performance is critical when building a framework that enables horizontal and vertical scalability. Using C++ for such a task further raises the expectation that the implementation performs well and its abstraction comes at little cost. Our initial evaluations focus on the cost of layers in a composable network stack.

\subsection{Experimental Setup}

Measurements were performed on a 2017 MacBook Pro with a 2.9\,GHz Intel Core i7 and 16\,GB RAM running macOS 10.14. The benchmark \S~\ref{sub:eval:network-performance} uses Mininet\,\cite{hhjlm-rneuc-12} to simulate a network link with loss. Mininet offers a VM image\footnote{http://mininet.org/download/} with a configured environment. We used the image running in Virtual Box Version 5.2.18 to perform all benchmarks.

Our benchmarks can be found online on GitHub\footnote{https://github.com/inetrg/agere-2018} and are based on the CAF branch linked in the repository. For \S~\ref{sub:eval:layers} we used Google Benchmark\footnote{https://github.com/google/benchmark} in version 1.4.1 to perform the measurements.

\subsection{The Costs of Layers in CAF}
\label{sub:eval:layers}

Passing data through the layers of a protocol policy happens on every send and receive call. Quantifying the time spent in the new broker class when sending and receiving data is valuable to evaluate the implementation in general and find performance problems. The measurements in this section do not include calls to the socket API. Instead, a mock transport policy offers buffers to read from and write to. Benchmarks that send data write their header and payload into a buffer of the policy, thus introducing a dependency between runtime and payload size. Since the mock data that is received can mostly be prepared in advance, benchmarks that receive data only have to copy data that they require to parse headers.

All benchmarks were performed for payload sizes from 128 to 8,192 bytes in increasing powers of two. All graphs show the mean real time in microseconds over ten runs as a function of the payload size and plot the standard deviation as error bars.

\paragraph*{Sending}

The first benchmark examines the cost to prepare a message for sending. It compares the policies used for the TCP and UDP-related implementations. For both protocols we measure the operation cost to handle a raw protocol that does nothing but write to the send buffer and a simplified BASP protocol that prefixes data with a header consisting of a source and destination actor as well as a payload size. The UDP measurements additionally include an ordering layer that adds a sequence number.

\begin{figure}[htb]
  \centering
  \vspace{-5pt}
  \input{send_combined.tikz}
  \vspace{-15pt}
  \caption{Cost to prepare a message for sending with different protocol layers. (left: TCP, right: UDP)}
  \label{fig:eval:send_combined}
  \vspace{-5pt}
\end{figure}
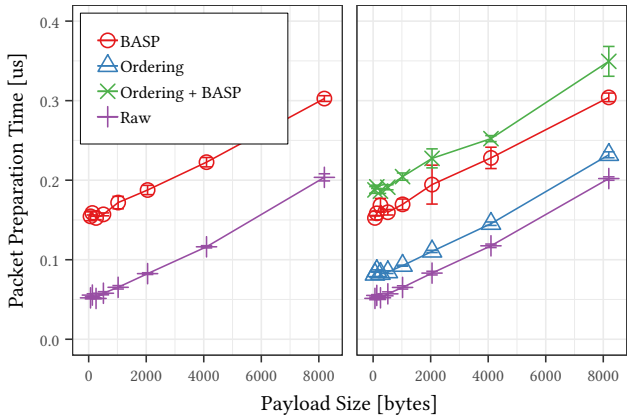

Figure~\ref{fig:eval:send_combined} shows the results for TCP on the left and UDP on the right. The time to send data rises linearly with the payload size due to the copy operations. As expected, using the raw protocol induces the least overhead in both cases with similar time requirements for both protocols.

Adding a layer introduces additional overhead depending on the layer implementation. The BASP layer comes at the same cost for both protocol. The additional time requirements stem from the serialization of its three fields: actor ids (64 bit) of the sender and receiver as well as a size parameter (32 bit).

UDP additionally includes measurements for ordering. The ordering header is smaller than the BASP header, only including a single sequence number (16 bit). The cost for adding ordering seems constant, whether it is deployed only with the raw protocol or in addition to BASP.

The error bars are small overall with the exception of a few measurement points. BASP for UDP with a payload size of 2000\,bytes and 4000\,bytes shows small error bars as does ordering with BASP for UDP with a payload of 8000\,bytes. This could be a result of measurements on such a small time scale.

\paragraph*{Receiving}

In general, message receipt promises to show a greater impact on performance. Depending on the protocol, it requires not only deserialization and parsing of the protocol headers but may include checks such as the validation of sequence numbers for ordering. In this benchmark the packet to receive is prepared in advance but adjusted during each receive call to include the expected sequence number and payload size. This means that no message is received out of order.

\begin{figure}[htb]
  \centering
  \vspace{-5pt}
  \input{receive_combined.tikz}
  \vspace{-15pt}
  \caption{Cost to prepare a received packet for processing with different protocol layers. (left: TCP, right: UDP)}
  \label{fig:eval:receive_combined}
  \vspace{-5pt}
\end{figure}
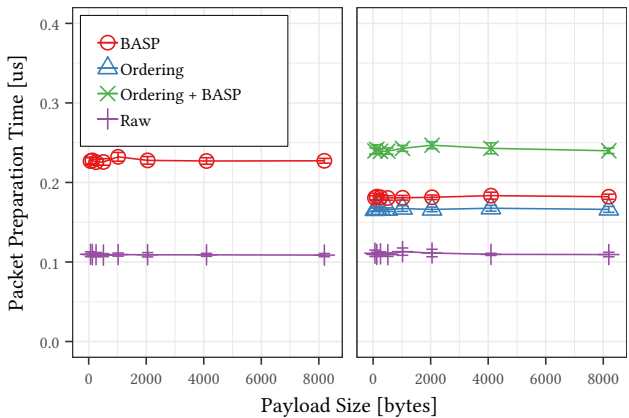

Figure~\ref{fig:eval:receive_combined} depicts the time required to prepare a single received message for processing by the broker, showing TCP on the left and UDP on the right. The measurements show constant performance across all payload sizes due to the lack of a copy operation.

Once again, the raw protocol has the least overhead as it only passes a pointer to the data through the stack. The difference in performance of the raw protocol compared to the send operation (Figure~\ref{fig:eval:send_combined}) is likely the overhead to activate the receiving broker to handle the message with the new data.

The BASP layer has varying costs depending on the underlying transport semantics. On top of a stream protocol (left graph) it parses the stream and reads twice to parse a complete message, a first read to get the header and deserialize the payload size and a second read to get a number of matching bytes. In contrast, BASP for datagrams (right graph) expects the message to arrive in one datagram and only requires a single read as a result.

Adding the ordering layer to the datagram broker is slightly cheaper than BASP. Note that all messages arrive in order. As a results, the layer only has to check the sequence number but never perform buffering to reorder messages. The cost of ordering is approximately constant whether it is deployed only with the raw protocol or in combination with BASP.

The error bars are negligibly small for all measurements.

\paragraph*{Receiving UDP Sequences}

An ordering layer that never has to reorder is very cheap. Costs only arise once packets arrive out of order or not at all. Since ordering can be deployed without reliability, missing message are dropped eventually to avoid or the message flow just stops. There are two triggers to drop a missing message: a timeout triggers or the buffer of pending messages runs full. In both cases the runtime delivers buffered messages starting with the smallest buffered sequence number. This benchmark evaluates the cost for our ordering layer to process a sequence of ten messages in three scenarios:

\begin{enumerate}
  \item \textit{Ordered}: All messages arrive in the expected order.
  \item \textit{Late}: One message arrives late by one.
  \item \textit{Dropped}: One message is dropped during transport.
\end{enumerate}

The maximum length of the pending message buffer is configured to five messages. Timeouts are complicated to benchmark as they rely on time and are generally long compared to execution times for the operations measured here. Thus timeouts are not represented in the benchmark. In general, triggering a timeout can be expected to be more expensive than delivering messages due to a full buffer as it requires interaction with the clock in CAF.

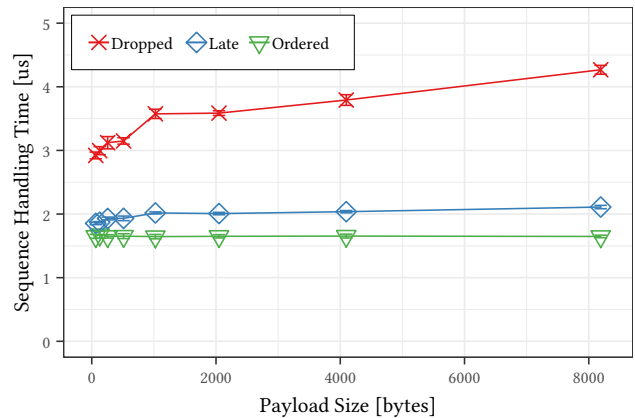
\begin{figure}[htb]
  \centering
  \vspace{-5pt}
  \input{udp_receive_sequence.tikz}
  \vspace{-15pt}
  \caption{Cost to handle a sequence of ten packets in the presence of message loss and out-of-order delivery.}
  \label{fig:eval:udp_receive_sequence}
  \vspace{-5pt}
\end{figure}

Figure~\ref{fig:eval:udp_receive_sequence} depicts the time required to handle the message sequence. Delivering all messages in order naturally performs best and shows a constant runtime. As soon as a message arrives late the handling time increases and no longer remains constant. Subsequent messages are buffered until the missing message is received. Since the copy operation depends on the size of the received payload we can see an increase in handling time. This behavior is more prominent when more messages need buffering. When a single message is dropped, others are buffered until the pending message buffer runs full. This behavior is hard to avoid as the bytes have to be copied from the receive buffer for later delivery. The error bars are negligible for all measurements.

\subsection{Network Performance}
\label{sub:eval:network-performance}

Having implemented a composable network stack for CAF, we took the opportunity to implement a reliability layer for UDP. A virtual network built with Mininet\,\cite{hhjlm-rneuc-12} allows testing its behavior over links with configurable loss and delay. In contrast to the previous benchmarks, these measurements now include network operations.

Two brokers bounce a message back and forth 4000 times over a lossy link until each broker sent and received the message 2000 times. The Mininet topology for the benchmark consists of two hosts connected directly via a link with no delay. Our retransmit timeout for UDP is configured to be 40\,ms and the minimum retransmit timeout for TCP on the routes is configured to the same value. The measurements are performed for different transport and layer combinations: TCP, reliable UDP, and reliable, ordered UDP.

\begin{figure}[htb]
  \centering
  \vspace{-5pt}
  \input{pingpong.tikz}
  \vspace{-15pt}
  \caption{Two actors sequentially exchange messages over a lossy link without delay.}
  \label{fig:eval:pingpong}
  \vspace{-5pt}
\end{figure}
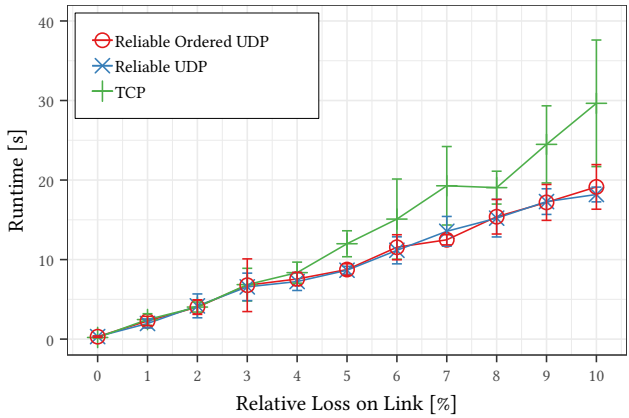

Figure~\ref{fig:eval:pingpong} displays the total runtime as a function of the configured packet loss percentage with error bars for the 5 and 95 percentile. The linear increase in runtime for both UDP implementations is expected. It shows that our reliability layer performs retransmits and the program works despite the loss. Since only one message is sent at a time, every lost packet adds the retransmit timeout to the runtime. Adding the ordering layer does not impact performance as messages should not arrive out of order.

Below 3\% loss TCP shows similar performance to our simple reliability layer. Thereafter, TCP is increasingly slower than UDP. A key difference here is that TCP adjusts its congestion windows and retransmit timeouts continuously in reaction to individual losses of the specific run. This is also reflected by the large error bars for TCP.

\begin{figure}[htb]
  \centering
  \vspace{-5pt}
  \input{pingpong-10.tikz}
  \vspace{-15pt}
  \caption{Two actors sequentially exchange messages over a lossy link with 10ms delay.}
  \label{fig:eval:pingpong:delay}
  \vspace{-5pt}
\end{figure}
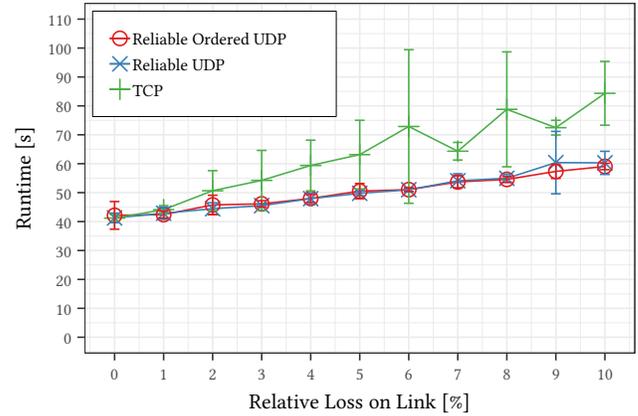

We repeated the benchmark in the same setup with a link delay of 10ms. Figure~\ref{fig:eval:pingpong:delay} shows the results. Note that the y-axis has a different scale. The results for UDP look similar with an offset of about 40\,s. This matches the expected increase, a total of 4000 messages with 10\,ms for each transmission. In contrast, the delay impacts the performance of TCP to a greater extent. Here, the mean runtime increase is larger than for UDP.

Both protocols show larger error bars. While this can be seen for UDP especially for 0\% and 9\% loss, TCP shows much more variation overall and has largely increased error bars. Once again, the individual loss pattern in each run has a greater impact on TCP due to its adaptability. TCP interprets loss as network overload and re-adjusts its congestion control algorithm.

The benchmarks validate that a working retransmit protocol can be implemented as a layer in our network stack. In addition to optimizing the reliability layer for a more general use including adaptive retransmit timeouts, we want to ship a slicing layer to make it easy for users to configure the network layer for their needs.

%% file: send_combined.tikz
\begin{tikzpicture}[x=1pt,y=1pt]
\definecolor{fillColor}{RGB}{255,255,255}
\path[use as bounding box,fill=fillColor,fill opacity=0.00] (0,0) rectangle (245.72,166.22);
\begin{scope}
\path[clip] (  0.00,  0.00) rectangle (245.72,166.22);
\definecolor{drawColor}{RGB}{255,255,255}
\definecolor{fillColor}{RGB}{255,255,255}

\path[draw=drawColor,line width= 0.6pt,line join=round,line cap=round,fill=fillColor] (  0.00,  0.00) rectangle (245.72,166.22);
\end{scope}
\begin{scope}
\path[clip] ( 30.54, 28.25) rectangle (132.63,160.72);
\definecolor{fillColor}{RGB}{255,255,255}

\path[fill=fillColor] ( 30.54, 28.25) rectangle (132.63,160.72);
\definecolor{drawColor}{gray}{0.92}

\path[draw=drawColor,line width= 0.3pt,line join=round] ( 30.54, 49.32) --
	(132.63, 49.32);

\path[draw=drawColor,line width= 0.3pt,line join=round] ( 30.54, 79.43) --
	(132.63, 79.43);

\path[draw=drawColor,line width= 0.3pt,line join=round] ( 30.54,109.54) --
	(132.63,109.54);

\path[draw=drawColor,line width= 0.3pt,line join=round] ( 30.54,139.65) --
	(132.63,139.65);

\path[draw=drawColor,line width= 0.3pt,line join=round] ( 47.54, 28.25) --
	( 47.54,160.72);

\path[draw=drawColor,line width= 0.3pt,line join=round] ( 69.31, 28.25) --
	( 69.31,160.72);

\path[draw=drawColor,line width= 0.3pt,line join=round] ( 91.07, 28.25) --
	( 91.07,160.72);

\path[draw=drawColor,line width= 0.3pt,line join=round] (112.84, 28.25) --
	(112.84,160.72);

\path[draw=drawColor,line width= 0.6pt,line join=round] ( 30.54, 34.27) --
	(132.63, 34.27);

\path[draw=drawColor,line width= 0.6pt,line join=round] ( 30.54, 64.37) --
	(132.63, 64.37);

\path[draw=drawColor,line width= 0.6pt,line join=round] ( 30.54, 94.48) --
	(132.63, 94.48);

\path[draw=drawColor,line width= 0.6pt,line join=round] ( 30.54,124.59) --
	(132.63,124.59);

\path[draw=drawColor,line width= 0.6pt,line join=round] ( 30.54,154.70) --
	(132.63,154.70);

\path[draw=drawColor,line width= 0.6pt,line join=round] ( 36.66, 28.25) --
	( 36.66,160.72);

\path[draw=drawColor,line width= 0.6pt,line join=round] ( 58.43, 28.25) --
	( 58.43,160.72);

\path[draw=drawColor,line width= 0.6pt,line join=round] ( 80.19, 28.25) --
	( 80.19,160.72);

\path[draw=drawColor,line width= 0.6pt,line join=round] (101.96, 28.25) --
	(101.96,160.72);

\path[draw=drawColor,line width= 0.6pt,line join=round] (123.72, 28.25) --
	(123.72,160.72);
\definecolor{drawColor}{RGB}{228,26,28}

\path[draw=drawColor,line width= 0.6pt,line join=round] ( 37.36, 80.80) --
	( 38.05, 82.06) --
	( 39.45, 80.18) --
	( 42.23, 81.54) --
	( 47.80, 86.01) --
	( 58.95, 90.76) --
	( 81.24,101.30) --
	(125.81,125.38);
\definecolor{drawColor}{RGB}{152,78,163}

\path[draw=drawColor,line width= 0.6pt,line join=round] ( 37.36, 49.94) --
	( 38.05, 50.98) --
	( 39.45, 49.72) --
	( 42.23, 51.68) --
	( 47.80, 53.93) --
	( 58.95, 59.11) --
	( 81.24, 69.27) --
	(125.81, 95.55);

\path[draw=drawColor,line width= 0.6pt,line join=round,line cap=round] ( 33.52, 49.94) -- ( 41.19, 49.94);

\path[draw=drawColor,line width= 0.6pt,line join=round,line cap=round] ( 37.36, 46.11) -- ( 37.36, 53.77);

\path[draw=drawColor,line width= 0.6pt,line join=round,line cap=round] ( 34.22, 50.98) -- ( 41.89, 50.98);

\path[draw=drawColor,line width= 0.6pt,line join=round,line cap=round] ( 38.05, 47.15) -- ( 38.05, 54.82);

\path[draw=drawColor,line width= 0.6pt,line join=round,line cap=round] ( 35.61, 49.72) -- ( 43.28, 49.72);

\path[draw=drawColor,line width= 0.6pt,line join=round,line cap=round] ( 39.45, 45.88) -- ( 39.45, 53.55);

\path[draw=drawColor,line width= 0.6pt,line join=round,line cap=round] ( 38.40, 51.68) -- ( 46.07, 51.68);

\path[draw=drawColor,line width= 0.6pt,line join=round,line cap=round] ( 42.23, 47.85) -- ( 42.23, 55.52);

\path[draw=drawColor,line width= 0.6pt,line join=round,line cap=round] ( 43.97, 53.93) -- ( 51.64, 53.93);

\path[draw=drawColor,line width= 0.6pt,line join=round,line cap=round] ( 47.80, 50.09) -- ( 47.80, 57.76);

\path[draw=drawColor,line width= 0.6pt,line join=round,line cap=round] ( 55.11, 59.11) -- ( 62.78, 59.11);

\path[draw=drawColor,line width= 0.6pt,line join=round,line cap=round] ( 58.95, 55.27) -- ( 58.95, 62.94);

\path[draw=drawColor,line width= 0.6pt,line join=round,line cap=round] ( 77.40, 69.27) -- ( 85.07, 69.27);

\path[draw=drawColor,line width= 0.6pt,line join=round,line cap=round] ( 81.24, 65.43) -- ( 81.24, 73.10);

\path[draw=drawColor,line width= 0.6pt,line join=round,line cap=round] (121.98, 95.55) -- (129.65, 95.55);

\path[draw=drawColor,line width= 0.6pt,line join=round,line cap=round] (125.81, 91.72) -- (125.81, 99.39);
\definecolor{drawColor}{RGB}{228,26,28}

\path[draw=drawColor,line width= 0.6pt,line join=round,line cap=round] ( 37.36, 80.80) circle (  2.71);

\path[draw=drawColor,line width= 0.6pt,line join=round,line cap=round] ( 38.05, 82.06) circle (  2.71);

\path[draw=drawColor,line width= 0.6pt,line join=round,line cap=round] ( 39.45, 80.18) circle (  2.71);

\path[draw=drawColor,line width= 0.6pt,line join=round,line cap=round] ( 42.23, 81.54) circle (  2.71);

\path[draw=drawColor,line width= 0.6pt,line join=round,line cap=round] ( 47.80, 86.01) circle (  2.71);

\path[draw=drawColor,line width= 0.6pt,line join=round,line cap=round] ( 58.95, 90.76) circle (  2.71);

\path[draw=drawColor,line width= 0.6pt,line join=round,line cap=round] ( 81.24,101.30) circle (  2.71);

\path[draw=drawColor,line width= 0.6pt,line join=round,line cap=round] (125.81,125.38) circle (  2.71);
\definecolor{drawColor}{RGB}{152,78,163}

\path[draw=drawColor,line width= 0.6pt,line join=round] ( 35.18, 50.32) --
	( 39.53, 50.32);

\path[draw=drawColor,line width= 0.6pt,line join=round] ( 37.36, 50.32) --
	( 37.36, 49.56);

\path[draw=drawColor,line width= 0.6pt,line join=round] ( 35.18, 49.56) --
	( 39.53, 49.56);

\path[draw=drawColor,line width= 0.6pt,line join=round] ( 35.88, 51.48) --
	( 40.23, 51.48);

\path[draw=drawColor,line width= 0.6pt,line join=round] ( 38.05, 51.48) --
	( 38.05, 50.49);

\path[draw=drawColor,line width= 0.6pt,line join=round] ( 35.88, 50.49) --
	( 40.23, 50.49);

\path[draw=drawColor,line width= 0.6pt,line join=round] ( 37.27, 49.86) --
	( 41.62, 49.86);

\path[draw=drawColor,line width= 0.6pt,line join=round] ( 39.45, 49.86) --
	( 39.45, 49.57);

\path[draw=drawColor,line width= 0.6pt,line join=round] ( 37.27, 49.57) --
	( 41.62, 49.57);

\path[draw=drawColor,line width= 0.6pt,line join=round] ( 40.06, 52.28) --
	( 44.41, 52.28);

\path[draw=drawColor,line width= 0.6pt,line join=round] ( 42.23, 52.28) --
	( 42.23, 51.09);

\path[draw=drawColor,line width= 0.6pt,line join=round] ( 40.06, 51.09) --
	( 44.41, 51.09);

\path[draw=drawColor,line width= 0.6pt,line join=round] ( 45.63, 54.58) --
	( 49.98, 54.58);

\path[draw=drawColor,line width= 0.6pt,line join=round] ( 47.80, 54.58) --
	( 47.80, 53.28);

\path[draw=drawColor,line width= 0.6pt,line join=round] ( 45.63, 53.28) --
	( 49.98, 53.28);

\path[draw=drawColor,line width= 0.6pt,line join=round] ( 56.77, 59.44) --
	( 61.13, 59.44);

\path[draw=drawColor,line width= 0.6pt,line join=round] ( 58.95, 59.44) --
	( 58.95, 58.78);

\path[draw=drawColor,line width= 0.6pt,line join=round] ( 56.77, 58.78) --
	( 61.13, 58.78);

\path[draw=drawColor,line width= 0.6pt,line join=round] ( 79.06, 69.69) --
	( 83.41, 69.69);

\path[draw=drawColor,line width= 0.6pt,line join=round] ( 81.24, 69.69) --
	( 81.24, 68.84);

\path[draw=drawColor,line width= 0.6pt,line join=round] ( 79.06, 68.84) --
	( 83.41, 68.84);

\path[draw=drawColor,line width= 0.6pt,line join=round] (123.64, 96.91) --
	(127.99, 96.91);

\path[draw=drawColor,line width= 0.6pt,line join=round] (125.81, 96.91) --
	(125.81, 94.20);

\path[draw=drawColor,line width= 0.6pt,line join=round] (123.64, 94.20) --
	(127.99, 94.20);
\definecolor{drawColor}{RGB}{228,26,28}

\path[draw=drawColor,line width= 0.6pt,line join=round] ( 35.18, 82.48) --
	( 39.53, 82.48);

\path[draw=drawColor,line width= 0.6pt,line join=round] ( 37.36, 82.48) --
	( 37.36, 79.12);

\path[draw=drawColor,line width= 0.6pt,line join=round] ( 35.18, 79.12) --
	( 39.53, 79.12);

\path[draw=drawColor,line width= 0.6pt,line join=round] ( 35.88, 82.80) --
	( 40.23, 82.80);

\path[draw=drawColor,line width= 0.6pt,line join=round] ( 38.05, 82.80) --
	( 38.05, 81.31);

\path[draw=drawColor,line width= 0.6pt,line join=round] ( 35.88, 81.31) --
	( 40.23, 81.31);

\path[draw=drawColor,line width= 0.6pt,line join=round] ( 37.27, 80.81) --
	( 41.62, 80.81);

\path[draw=drawColor,line width= 0.6pt,line join=round] ( 39.45, 80.81) --
	( 39.45, 79.55);

\path[draw=drawColor,line width= 0.6pt,line join=round] ( 37.27, 79.55) --
	( 41.62, 79.55);

\path[draw=drawColor,line width= 0.6pt,line join=round] ( 40.06, 82.09) --
	( 44.41, 82.09);

\path[draw=drawColor,line width= 0.6pt,line join=round] ( 42.23, 82.09) --
	( 42.23, 81.00);

\path[draw=drawColor,line width= 0.6pt,line join=round] ( 40.06, 81.00) --
	( 44.41, 81.00);

\path[draw=drawColor,line width= 0.6pt,line join=round] ( 45.63, 88.36) --
	( 49.98, 88.36);

\path[draw=drawColor,line width= 0.6pt,line join=round] ( 47.80, 88.36) --
	( 47.80, 83.65);

\path[draw=drawColor,line width= 0.6pt,line join=round] ( 45.63, 83.65) --
	( 49.98, 83.65);

\path[draw=drawColor,line width= 0.6pt,line join=round] ( 56.77, 92.52) --
	( 61.13, 92.52);

\path[draw=drawColor,line width= 0.6pt,line join=round] ( 58.95, 92.52) --
	( 58.95, 89.01);

\path[draw=drawColor,line width= 0.6pt,line join=round] ( 56.77, 89.01) --
	( 61.13, 89.01);

\path[draw=drawColor,line width= 0.6pt,line join=round] ( 79.06,103.07) --
	( 83.41,103.07);

\path[draw=drawColor,line width= 0.6pt,line join=round] ( 81.24,103.07) --
	( 81.24, 99.52);

\path[draw=drawColor,line width= 0.6pt,line join=round] ( 79.06, 99.52) --
	( 83.41, 99.52);

\path[draw=drawColor,line width= 0.6pt,line join=round] (123.64,126.47) --
	(127.99,126.47);

\path[draw=drawColor,line width= 0.6pt,line join=round] (125.81,126.47) --
	(125.81,124.29);

\path[draw=drawColor,line width= 0.6pt,line join=round] (123.64,124.29) --
	(127.99,124.29);
\definecolor{drawColor}{gray}{0.20}

\path[draw=drawColor,line width= 0.6pt,line join=round,line cap=round] ( 30.54, 28.25) rectangle (132.63,160.72);
\end{scope}
\begin{scope}
\path[clip] (138.13, 28.25) rectangle (240.22,160.72);
\definecolor{fillColor}{RGB}{255,255,255}

\path[fill=fillColor] (138.13, 28.25) rectangle (240.22,160.72);
\definecolor{drawColor}{gray}{0.92}

\path[draw=drawColor,line width= 0.3pt,line join=round] (138.13, 49.32) --
	(240.22, 49.32);

\path[draw=drawColor,line width= 0.3pt,line join=round] (138.13, 79.43) --
	(240.22, 79.43);

\path[draw=drawColor,line width= 0.3pt,line join=round] (138.13,109.54) --
	(240.22,109.54);

\path[draw=drawColor,line width= 0.3pt,line join=round] (138.13,139.65) --
	(240.22,139.65);

\path[draw=drawColor,line width= 0.3pt,line join=round] (155.13, 28.25) --
	(155.13,160.72);

\path[draw=drawColor,line width= 0.3pt,line join=round] (176.90, 28.25) --
	(176.90,160.72);

\path[draw=drawColor,line width= 0.3pt,line join=round] (198.66, 28.25) --
	(198.66,160.72);

\path[draw=drawColor,line width= 0.3pt,line join=round] (220.43, 28.25) --
	(220.43,160.72);

\path[draw=drawColor,line width= 0.6pt,line join=round] (138.13, 34.27) --
	(240.22, 34.27);

\path[draw=drawColor,line width= 0.6pt,line join=round] (138.13, 64.37) --
	(240.22, 64.37);

\path[draw=drawColor,line width= 0.6pt,line join=round] (138.13, 94.48) --
	(240.22, 94.48);

\path[draw=drawColor,line width= 0.6pt,line join=round] (138.13,124.59) --
	(240.22,124.59);

\path[draw=drawColor,line width= 0.6pt,line join=round] (138.13,154.70) --
	(240.22,154.70);

\path[draw=drawColor,line width= 0.6pt,line join=round] (144.25, 28.25) --
	(144.25,160.72);

\path[draw=drawColor,line width= 0.6pt,line join=round] (166.02, 28.25) --
	(166.02,160.72);

\path[draw=drawColor,line width= 0.6pt,line join=round] (187.78, 28.25) --
	(187.78,160.72);

\path[draw=drawColor,line width= 0.6pt,line join=round] (209.55, 28.25) --
	(209.55,160.72);

\path[draw=drawColor,line width= 0.6pt,line join=round] (231.31, 28.25) --
	(231.31,160.72);
\definecolor{drawColor}{RGB}{228,26,28}

\path[draw=drawColor,line width= 0.6pt,line join=round] (144.95, 80.19) --
	(145.64, 81.94) --
	(147.04, 84.99) --
	(149.82, 82.35) --
	(155.39, 85.33) --
	(166.54, 92.81) --
	(188.83,102.92) --
	(233.40,125.82);
\definecolor{drawColor}{RGB}{55,126,184}

\path[draw=drawColor,line width= 0.6pt,line join=round] (144.95, 58.78) --
	(145.64, 60.10) --
	(147.04, 59.07) --
	(149.82, 59.68) --
	(155.39, 62.12) --
	(166.54, 67.52) --
	(188.83, 78.01) --
	(233.40,104.09);
\definecolor{drawColor}{RGB}{77,175,74}

\path[draw=drawColor,line width= 0.6pt,line join=round] (144.95, 90.80) --
	(145.64, 91.96) --
	(147.04, 90.34) --
	(149.82, 91.81) --
	(155.39, 95.83) --
	(166.54,102.74) --
	(188.83,110.21) --
	(233.40,139.47);
\definecolor{drawColor}{RGB}{152,78,163}

\path[draw=drawColor,line width= 0.6pt,line join=round] (144.95, 49.74) --
	(145.64, 50.86) --
	(147.04, 49.96) --
	(149.82, 51.46) --
	(155.39, 53.88) --
	(166.54, 59.31) --
	(188.83, 69.64) --
	(233.40, 95.10);

\path[draw=drawColor,line width= 0.6pt,line join=round,line cap=round] (141.11, 49.74) -- (148.78, 49.74);

\path[draw=drawColor,line width= 0.6pt,line join=round,line cap=round] (144.95, 45.90) -- (144.95, 53.57);

\path[draw=drawColor,line width= 0.6pt,line join=round,line cap=round] (141.81, 50.86) -- (149.48, 50.86);

\path[draw=drawColor,line width= 0.6pt,line join=round,line cap=round] (145.64, 47.02) -- (145.64, 54.69);

\path[draw=drawColor,line width= 0.6pt,line join=round,line cap=round] (143.20, 49.96) -- (150.87, 49.96);

\path[draw=drawColor,line width= 0.6pt,line join=round,line cap=round] (147.04, 46.13) -- (147.04, 53.79);

\path[draw=drawColor,line width= 0.6pt,line join=round,line cap=round] (145.99, 51.46) -- (153.66, 51.46);

\path[draw=drawColor,line width= 0.6pt,line join=round,line cap=round] (149.82, 47.63) -- (149.82, 55.29);

\path[draw=drawColor,line width= 0.6pt,line join=round,line cap=round] (151.56, 53.88) -- (159.23, 53.88);

\path[draw=drawColor,line width= 0.6pt,line join=round,line cap=round] (155.39, 50.05) -- (155.39, 57.72);

\path[draw=drawColor,line width= 0.6pt,line join=round,line cap=round] (162.70, 59.31) -- (170.37, 59.31);

\path[draw=drawColor,line width= 0.6pt,line join=round,line cap=round] (166.54, 55.47) -- (166.54, 63.14);

\path[draw=drawColor,line width= 0.6pt,line join=round,line cap=round] (184.99, 69.64) -- (192.66, 69.64);

\path[draw=drawColor,line width= 0.6pt,line join=round,line cap=round] (188.83, 65.80) -- (188.83, 73.47);

\path[draw=drawColor,line width= 0.6pt,line join=round,line cap=round] (229.57, 95.10) -- (237.23, 95.10);

\path[draw=drawColor,line width= 0.6pt,line join=round,line cap=round] (233.40, 91.26) -- (233.40, 98.93);
\definecolor{drawColor}{RGB}{55,126,184}

\path[draw=drawColor,line width= 0.6pt,line join=round,line cap=round] (144.95, 63.00) --
	(148.60, 56.68) --
	(141.29, 56.68) --
	(144.95, 63.00);

\path[draw=drawColor,line width= 0.6pt,line join=round,line cap=round] (145.64, 64.32) --
	(149.29, 57.99) --
	(141.99, 57.99) --
	(145.64, 64.32);

\path[draw=drawColor,line width= 0.6pt,line join=round,line cap=round] (147.04, 63.28) --
	(150.69, 56.96) --
	(143.38, 56.96) --
	(147.04, 63.28);

\path[draw=drawColor,line width= 0.6pt,line join=round,line cap=round] (149.82, 63.89) --
	(153.47, 57.57) --
	(146.17, 57.57) --
	(149.82, 63.89);

\path[draw=drawColor,line width= 0.6pt,line join=round,line cap=round] (155.39, 66.34) --
	(159.04, 60.02) --
	(151.74, 60.02) --
	(155.39, 66.34);

\path[draw=drawColor,line width= 0.6pt,line join=round,line cap=round] (166.54, 71.74) --
	(170.19, 65.41) --
	(162.89, 65.41) --
	(166.54, 71.74);

\path[draw=drawColor,line width= 0.6pt,line join=round,line cap=round] (188.83, 82.23) --
	(192.48, 75.90) --
	(185.17, 75.90) --
	(188.83, 82.23);

\path[draw=drawColor,line width= 0.6pt,line join=round,line cap=round] (233.40,108.31) --
	(237.05,101.98) --
	(229.75,101.98) --
	(233.40,108.31);
\definecolor{drawColor}{RGB}{228,26,28}

\path[draw=drawColor,line width= 0.6pt,line join=round,line cap=round] (144.95, 80.19) circle (  2.71);

\path[draw=drawColor,line width= 0.6pt,line join=round,line cap=round] (145.64, 81.94) circle (  2.71);

\path[draw=drawColor,line width= 0.6pt,line join=round,line cap=round] (147.04, 84.99) circle (  2.71);

\path[draw=drawColor,line width= 0.6pt,line join=round,line cap=round] (149.82, 82.35) circle (  2.71);

\path[draw=drawColor,line width= 0.6pt,line join=round,line cap=round] (155.39, 85.33) circle (  2.71);

\path[draw=drawColor,line width= 0.6pt,line join=round,line cap=round] (166.54, 92.81) circle (  2.71);

\path[draw=drawColor,line width= 0.6pt,line join=round,line cap=round] (188.83,102.92) circle (  2.71);

\path[draw=drawColor,line width= 0.6pt,line join=round,line cap=round] (233.40,125.82) circle (  2.71);
\definecolor{drawColor}{RGB}{77,175,74}

\path[draw=drawColor,line width= 0.6pt,line join=round,line cap=round] (142.24, 88.08) -- (147.66, 93.51);

\path[draw=drawColor,line width= 0.6pt,line join=round,line cap=round] (142.24, 93.51) -- (147.66, 88.08);

\path[draw=drawColor,line width= 0.6pt,line join=round,line cap=round] (142.93, 89.25) -- (148.35, 94.68);

\path[draw=drawColor,line width= 0.6pt,line join=round,line cap=round] (142.93, 94.68) -- (148.35, 89.25);

\path[draw=drawColor,line width= 0.6pt,line join=round,line cap=round] (144.32, 87.63) -- (149.75, 93.05);

\path[draw=drawColor,line width= 0.6pt,line join=round,line cap=round] (144.32, 93.05) -- (149.75, 87.63);

\path[draw=drawColor,line width= 0.6pt,line join=round,line cap=round] (147.11, 89.10) -- (152.53, 94.52);

\path[draw=drawColor,line width= 0.6pt,line join=round,line cap=round] (147.11, 94.52) -- (152.53, 89.10);

\path[draw=drawColor,line width= 0.6pt,line join=round,line cap=round] (152.68, 93.12) -- (158.10, 98.54);

\path[draw=drawColor,line width= 0.6pt,line join=round,line cap=round] (152.68, 98.54) -- (158.10, 93.12);

\path[draw=drawColor,line width= 0.6pt,line join=round,line cap=round] (163.83,100.03) -- (169.25,105.45);

\path[draw=drawColor,line width= 0.6pt,line join=round,line cap=round] (163.83,105.45) -- (169.25,100.03);

\path[draw=drawColor,line width= 0.6pt,line join=round,line cap=round] (186.11,107.50) -- (191.54,112.92);

\path[draw=drawColor,line width= 0.6pt,line join=round,line cap=round] (186.11,112.92) -- (191.54,107.50);

\path[draw=drawColor,line width= 0.6pt,line join=round,line cap=round] (230.69,136.75) -- (236.11,142.18);

\path[draw=drawColor,line width= 0.6pt,line join=round,line cap=round] (230.69,142.18) -- (236.11,136.75);
\definecolor{drawColor}{RGB}{152,78,163}

\path[draw=drawColor,line width= 0.6pt,line join=round] (142.77, 50.25) --
	(147.12, 50.25);

\path[draw=drawColor,line width= 0.6pt,line join=round] (144.95, 50.25) --
	(144.95, 49.23);

\path[draw=drawColor,line width= 0.6pt,line join=round] (142.77, 49.23) --
	(147.12, 49.23);

\path[draw=drawColor,line width= 0.6pt,line join=round] (143.47, 51.16) --
	(147.82, 51.16);

\path[draw=drawColor,line width= 0.6pt,line join=round] (145.64, 51.16) --
	(145.64, 50.56);

\path[draw=drawColor,line width= 0.6pt,line join=round] (143.47, 50.56) --
	(147.82, 50.56);

\path[draw=drawColor,line width= 0.6pt,line join=round] (144.86, 50.59) --
	(149.21, 50.59);

\path[draw=drawColor,line width= 0.6pt,line join=round] (147.04, 50.59) --
	(147.04, 49.33);

\path[draw=drawColor,line width= 0.6pt,line join=round] (144.86, 49.33) --
	(149.21, 49.33);

\path[draw=drawColor,line width= 0.6pt,line join=round] (147.64, 52.21) --
	(152.00, 52.21);

\path[draw=drawColor,line width= 0.6pt,line join=round] (149.82, 52.21) --
	(149.82, 50.71);

\path[draw=drawColor,line width= 0.6pt,line join=round] (147.64, 50.71) --
	(152.00, 50.71);

\path[draw=drawColor,line width= 0.6pt,line join=round] (153.22, 54.50) --
	(157.57, 54.50);

\path[draw=drawColor,line width= 0.6pt,line join=round] (155.39, 54.50) --
	(155.39, 53.26);

\path[draw=drawColor,line width= 0.6pt,line join=round] (153.22, 53.26) --
	(157.57, 53.26);

\path[draw=drawColor,line width= 0.6pt,line join=round] (164.36, 60.00) --
	(168.71, 60.00);

\path[draw=drawColor,line width= 0.6pt,line join=round] (166.54, 60.00) --
	(166.54, 58.61);

\path[draw=drawColor,line width= 0.6pt,line join=round] (164.36, 58.61) --
	(168.71, 58.61);

\path[draw=drawColor,line width= 0.6pt,line join=round] (186.65, 70.30) --
	(191.00, 70.30);

\path[draw=drawColor,line width= 0.6pt,line join=round] (188.83, 70.30) --
	(188.83, 68.98);

\path[draw=drawColor,line width= 0.6pt,line join=round] (186.65, 68.98) --
	(191.00, 68.98);

\path[draw=drawColor,line width= 0.6pt,line join=round] (231.22, 95.75) --
	(235.58, 95.75);

\path[draw=drawColor,line width= 0.6pt,line join=round] (233.40, 95.75) --
	(233.40, 94.45);

\path[draw=drawColor,line width= 0.6pt,line join=round] (231.22, 94.45) --
	(235.58, 94.45);
\definecolor{drawColor}{RGB}{55,126,184}

\path[draw=drawColor,line width= 0.6pt,line join=round] (142.77, 59.91) --
	(147.12, 59.91);

\path[draw=drawColor,line width= 0.6pt,line join=round] (144.95, 59.91) --
	(144.95, 57.66);

\path[draw=drawColor,line width= 0.6pt,line join=round] (142.77, 57.66) --
	(147.12, 57.66);

\path[draw=drawColor,line width= 0.6pt,line join=round] (143.47, 60.54) --
	(147.82, 60.54);

\path[draw=drawColor,line width= 0.6pt,line join=round] (145.64, 60.54) --
	(145.64, 59.66);

\path[draw=drawColor,line width= 0.6pt,line join=round] (143.47, 59.66) --
	(147.82, 59.66);

\path[draw=drawColor,line width= 0.6pt,line join=round] (144.86, 60.18) --
	(149.21, 60.18);

\path[draw=drawColor,line width= 0.6pt,line join=round] (147.04, 60.18) --
	(147.04, 57.96);

\path[draw=drawColor,line width= 0.6pt,line join=round] (144.86, 57.96) --
	(149.21, 57.96);

\path[draw=drawColor,line width= 0.6pt,line join=round] (147.64, 60.19) --
	(152.00, 60.19);

\path[draw=drawColor,line width= 0.6pt,line join=round] (149.82, 60.19) --
	(149.82, 59.16);

\path[draw=drawColor,line width= 0.6pt,line join=round] (147.64, 59.16) --
	(152.00, 59.16);

\path[draw=drawColor,line width= 0.6pt,line join=round] (153.22, 62.33) --
	(157.57, 62.33);

\path[draw=drawColor,line width= 0.6pt,line join=round] (155.39, 62.33) --
	(155.39, 61.92);

\path[draw=drawColor,line width= 0.6pt,line join=round] (153.22, 61.92) --
	(157.57, 61.92);

\path[draw=drawColor,line width= 0.6pt,line join=round] (164.36, 67.94) --
	(168.71, 67.94);

\path[draw=drawColor,line width= 0.6pt,line join=round] (166.54, 67.94) --
	(166.54, 67.10);

\path[draw=drawColor,line width= 0.6pt,line join=round] (164.36, 67.10) --
	(168.71, 67.10);

\path[draw=drawColor,line width= 0.6pt,line join=round] (186.65, 78.67) --
	(191.00, 78.67);

\path[draw=drawColor,line width= 0.6pt,line join=round] (188.83, 78.67) --
	(188.83, 77.35);

\path[draw=drawColor,line width= 0.6pt,line join=round] (186.65, 77.35) --
	(191.00, 77.35);

\path[draw=drawColor,line width= 0.6pt,line join=round] (231.22,105.22) --
	(235.58,105.22);

\path[draw=drawColor,line width= 0.6pt,line join=round] (233.40,105.22) --
	(233.40,102.96);

\path[draw=drawColor,line width= 0.6pt,line join=round] (231.22,102.96) --
	(235.58,102.96);
\definecolor{drawColor}{RGB}{228,26,28}

\path[draw=drawColor,line width= 0.6pt,line join=round] (142.77, 80.92) --
	(147.12, 80.92);

\path[draw=drawColor,line width= 0.6pt,line join=round] (144.95, 80.92) --
	(144.95, 79.46);

\path[draw=drawColor,line width= 0.6pt,line join=round] (142.77, 79.46) --
	(147.12, 79.46);

\path[draw=drawColor,line width= 0.6pt,line join=round] (143.47, 82.75) --
	(147.82, 82.75);

\path[draw=drawColor,line width= 0.6pt,line join=round] (145.64, 82.75) --
	(145.64, 81.13);

\path[draw=drawColor,line width= 0.6pt,line join=round] (143.47, 81.13) --
	(147.82, 81.13);

\path[draw=drawColor,line width= 0.6pt,line join=round] (144.86, 87.57) --
	(149.21, 87.57);

\path[draw=drawColor,line width= 0.6pt,line join=round] (147.04, 87.57) --
	(147.04, 82.41);

\path[draw=drawColor,line width= 0.6pt,line join=round] (144.86, 82.41) --
	(149.21, 82.41);

\path[draw=drawColor,line width= 0.6pt,line join=round] (147.64, 83.39) --
	(152.00, 83.39);

\path[draw=drawColor,line width= 0.6pt,line join=round] (149.82, 83.39) --
	(149.82, 81.30);

\path[draw=drawColor,line width= 0.6pt,line join=round] (147.64, 81.30) --
	(152.00, 81.30);

\path[draw=drawColor,line width= 0.6pt,line join=round] (153.22, 87.35) --
	(157.57, 87.35);

\path[draw=drawColor,line width= 0.6pt,line join=round] (155.39, 87.35) --
	(155.39, 83.31);

\path[draw=drawColor,line width= 0.6pt,line join=round] (153.22, 83.31) --
	(157.57, 83.31);

\path[draw=drawColor,line width= 0.6pt,line join=round] (164.36,100.17) --
	(168.71,100.17);

\path[draw=drawColor,line width= 0.6pt,line join=round] (166.54,100.17) --
	(166.54, 85.44);

\path[draw=drawColor,line width= 0.6pt,line join=round] (164.36, 85.44) --
	(168.71, 85.44);

\path[draw=drawColor,line width= 0.6pt,line join=round] (186.65,106.96) --
	(191.00,106.96);

\path[draw=drawColor,line width= 0.6pt,line join=round] (188.83,106.96) --
	(188.83, 98.88);

\path[draw=drawColor,line width= 0.6pt,line join=round] (186.65, 98.88) --
	(191.00, 98.88);

\path[draw=drawColor,line width= 0.6pt,line join=round] (231.22,127.46) --
	(235.58,127.46);

\path[draw=drawColor,line width= 0.6pt,line join=round] (233.40,127.46) --
	(233.40,124.19);

\path[draw=drawColor,line width= 0.6pt,line join=round] (231.22,124.19) --
	(235.58,124.19);
\definecolor{drawColor}{RGB}{77,175,74}

\path[draw=drawColor,line width= 0.6pt,line join=round] (142.77, 92.06) --
	(147.12, 92.06);

\path[draw=drawColor,line width= 0.6pt,line join=round] (144.95, 92.06) --
	(144.95, 89.53);

\path[draw=drawColor,line width= 0.6pt,line join=round] (142.77, 89.53) --
	(147.12, 89.53);

\path[draw=drawColor,line width= 0.6pt,line join=round] (143.47, 92.60) --
	(147.82, 92.60);

\path[draw=drawColor,line width= 0.6pt,line join=round] (145.64, 92.60) --
	(145.64, 91.33);

\path[draw=drawColor,line width= 0.6pt,line join=round] (143.47, 91.33) --
	(147.82, 91.33);

\path[draw=drawColor,line width= 0.6pt,line join=round] (144.86, 91.19) --
	(149.21, 91.19);

\path[draw=drawColor,line width= 0.6pt,line join=round] (147.04, 91.19) --
	(147.04, 89.48);

\path[draw=drawColor,line width= 0.6pt,line join=round] (144.86, 89.48) --
	(149.21, 89.48);

\path[draw=drawColor,line width= 0.6pt,line join=round] (147.64, 92.77) --
	(152.00, 92.77);

\path[draw=drawColor,line width= 0.6pt,line join=round] (149.82, 92.77) --
	(149.82, 90.85);

\path[draw=drawColor,line width= 0.6pt,line join=round] (147.64, 90.85) --
	(152.00, 90.85);

\path[draw=drawColor,line width= 0.6pt,line join=round] (153.22, 97.21) --
	(157.57, 97.21);

\path[draw=drawColor,line width= 0.6pt,line join=round] (155.39, 97.21) --
	(155.39, 94.46);

\path[draw=drawColor,line width= 0.6pt,line join=round] (153.22, 94.46) --
	(157.57, 94.46);

\path[draw=drawColor,line width= 0.6pt,line join=round] (164.36,106.34) --
	(168.71,106.34);

\path[draw=drawColor,line width= 0.6pt,line join=round] (166.54,106.34) --
	(166.54, 99.14);

\path[draw=drawColor,line width= 0.6pt,line join=round] (164.36, 99.14) --
	(168.71, 99.14);

\path[draw=drawColor,line width= 0.6pt,line join=round] (186.65,111.36) --
	(191.00,111.36);

\path[draw=drawColor,line width= 0.6pt,line join=round] (188.83,111.36) --
	(188.83,109.07);

\path[draw=drawColor,line width= 0.6pt,line join=round] (186.65,109.07) --
	(191.00,109.07);

\path[draw=drawColor,line width= 0.6pt,line join=round] (231.22,145.11) --
	(235.58,145.11);

\path[draw=drawColor,line width= 0.6pt,line join=round] (233.40,145.11) --
	(233.40,133.82);

\path[draw=drawColor,line width= 0.6pt,line join=round] (231.22,133.82) --
	(235.58,133.82);
\definecolor{drawColor}{gray}{0.20}

\path[draw=drawColor,line width= 0.6pt,line join=round,line cap=round] (138.13, 28.25) rectangle (240.22,160.72);
\end{scope}
\begin{scope}
\path[clip] (  0.00,  0.00) rectangle (245.72,166.22);
\definecolor{drawColor}{gray}{0.20}

\path[draw=drawColor,line width= 0.6pt,line join=round] ( 36.66, 25.50) --
	( 36.66, 28.25);

\path[draw=drawColor,line width= 0.6pt,line join=round] ( 58.43, 25.50) --
	( 58.43, 28.25);

\path[draw=drawColor,line width= 0.6pt,line join=round] ( 80.19, 25.50) --
	( 80.19, 28.25);

\path[draw=drawColor,line width= 0.6pt,line join=round] (101.96, 25.50) --
	(101.96, 28.25);

\path[draw=drawColor,line width= 0.6pt,line join=round] (123.72, 25.50) --
	(123.72, 28.25);
\end{scope}
\begin{scope}
\path[clip] (  0.00,  0.00) rectangle (245.72,166.22);
\definecolor{drawColor}{gray}{0.30}

\node[text=drawColor,anchor=base,inner sep=0pt, outer sep=0pt, scale=  0.72] at ( 36.66, 18.34) {0};

\node[text=drawColor,anchor=base,inner sep=0pt, outer sep=0pt, scale=  0.72] at ( 58.43, 18.34) {2000};

\node[text=drawColor,anchor=base,inner sep=0pt, outer sep=0pt, scale=  0.72] at ( 80.19, 18.34) {4000};

\node[text=drawColor,anchor=base,inner sep=0pt, outer sep=0pt, scale=  0.72] at (101.96, 18.34) {6000};

\node[text=drawColor,anchor=base,inner sep=0pt, outer sep=0pt, scale=  0.72] at (123.72, 18.34) {8000};
\end{scope}
\begin{scope}
\path[clip] (  0.00,  0.00) rectangle (245.72,166.22);
\definecolor{drawColor}{gray}{0.20}

\path[draw=drawColor,line width= 0.6pt,line join=round] (144.25, 25.50) --
	(144.25, 28.25);

\path[draw=drawColor,line width= 0.6pt,line join=round] (166.02, 25.50) --
	(166.02, 28.25);

\path[draw=drawColor,line width= 0.6pt,line join=round] (187.78, 25.50) --
	(187.78, 28.25);

\path[draw=drawColor,line width= 0.6pt,line join=round] (209.55, 25.50) --
	(209.55, 28.25);

\path[draw=drawColor,line width= 0.6pt,line join=round] (231.31, 25.50) --
	(231.31, 28.25);
\end{scope}
\begin{scope}
\path[clip] (  0.00,  0.00) rectangle (245.72,166.22);
\definecolor{drawColor}{gray}{0.30}

\node[text=drawColor,anchor=base,inner sep=0pt, outer sep=0pt, scale=  0.72] at (144.25, 18.34) {0};

\node[text=drawColor,anchor=base,inner sep=0pt, outer sep=0pt, scale=  0.72] at (166.02, 18.34) {2000};

\node[text=drawColor,anchor=base,inner sep=0pt, outer sep=0pt, scale=  0.72] at (187.78, 18.34) {4000};

\node[text=drawColor,anchor=base,inner sep=0pt, outer sep=0pt, scale=  0.72] at (209.55, 18.34) {6000};

\node[text=drawColor,anchor=base,inner sep=0pt, outer sep=0pt, scale=  0.72] at (231.31, 18.34) {8000};
\end{scope}
\begin{scope}
\path[clip] (  0.00,  0.00) rectangle (245.72,166.22);
\definecolor{drawColor}{gray}{0.30}

\node[text=drawColor,anchor=base east,inner sep=0pt, outer sep=0pt, scale=  0.72] at ( 25.59, 31.79) {0.0};

\node[text=drawColor,anchor=base east,inner sep=0pt, outer sep=0pt, scale=  0.72] at ( 25.59, 61.90) {0.1};

\node[text=drawColor,anchor=base east,inner sep=0pt, outer sep=0pt, scale=  0.72] at ( 25.59, 92.00) {0.2};

\node[text=drawColor,anchor=base east,inner sep=0pt, outer sep=0pt, scale=  0.72] at ( 25.59,122.11) {0.3};

\node[text=drawColor,anchor=base east,inner sep=0pt, outer sep=0pt, scale=  0.72] at ( 25.59,152.22) {0.4};
\end{scope}
\begin{scope}
\path[clip] (  0.00,  0.00) rectangle (245.72,166.22);
\definecolor{drawColor}{gray}{0.20}

\path[draw=drawColor,line width= 0.6pt,line join=round] ( 27.79, 34.27) --
	( 30.54, 34.27);

\path[draw=drawColor,line width= 0.6pt,line join=round] ( 27.79, 64.37) --
	( 30.54, 64.37);

\path[draw=drawColor,line width= 0.6pt,line join=round] ( 27.79, 94.48) --
	( 30.54, 94.48);

\path[draw=drawColor,line width= 0.6pt,line join=round] ( 27.79,124.59) --
	( 30.54,124.59);

\path[draw=drawColor,line width= 0.6pt,line join=round] ( 27.79,154.70) --
	( 30.54,154.70);
\end{scope}
\begin{scope}
\path[clip] (  0.00,  0.00) rectangle (245.72,166.22);
\definecolor{drawColor}{RGB}{0,0,0}

\node[text=drawColor,anchor=base,inner sep=0pt, outer sep=0pt, scale=  0.90] at (135.38,  7.44) {Payload Size [bytes]};
\end{scope}
\begin{scope}
\path[clip] (  0.00,  0.00) rectangle (245.72,166.22);
\definecolor{drawColor}{RGB}{0,0,0}

\node[text=drawColor,rotate= 90.00,anchor=base,inner sep=0pt, outer sep=0pt, scale=  0.90] at ( 11.70, 94.48) {Packet Preparation Time [us]};
\end{scope}
\begin{scope}
\path[clip] (  0.00,  0.00) rectangle (245.72,166.22);
\definecolor{drawColor}{RGB}{0,0,0}
\definecolor{fillColor}{RGB}{255,255,255}

\path[draw=drawColor,line width= 0.3pt,line join=round,line cap=round,fill=fillColor] ( 33.54,108.18) rectangle (112.00,157.72);
\end{scope}
\begin{scope}
\path[clip] (  0.00,  0.00) rectangle (245.72,166.22);
\definecolor{fillColor}{RGB}{255,255,255}

\path[fill=fillColor] ( 39.04,142.59) rectangle ( 48.68,152.22);
\end{scope}
\begin{scope}
\path[clip] (  0.00,  0.00) rectangle (245.72,166.22);
\definecolor{drawColor}{RGB}{228,26,28}

\path[draw=drawColor,line width= 0.6pt,line join=round] ( 40.00,147.40) -- ( 47.71,147.40);
\end{scope}
\begin{scope}
\path[clip] (  0.00,  0.00) rectangle (245.72,166.22);
\definecolor{drawColor}{RGB}{228,26,28}

\path[draw=drawColor,line width= 0.6pt,line join=round,line cap=round] ( 43.86,147.40) circle (  2.71);
\end{scope}
\begin{scope}
\path[clip] (  0.00,  0.00) rectangle (245.72,166.22);
\definecolor{drawColor}{RGB}{228,26,28}

\path[draw=drawColor,line width= 0.6pt,line join=round] ( 40.00,147.40) -- ( 47.71,147.40);
\end{scope}
\begin{scope}
\path[clip] (  0.00,  0.00) rectangle (245.72,166.22);
\definecolor{fillColor}{RGB}{255,255,255}

\path[fill=fillColor] ( 39.04,132.95) rectangle ( 48.68,142.59);
\end{scope}
\begin{scope}
\path[clip] (  0.00,  0.00) rectangle (245.72,166.22);
\definecolor{drawColor}{RGB}{55,126,184}

\path[draw=drawColor,line width= 0.6pt,line join=round] ( 40.00,137.77) -- ( 47.71,137.77);
\end{scope}
\begin{scope}
\path[clip] (  0.00,  0.00) rectangle (245.72,166.22);
\definecolor{drawColor}{RGB}{55,126,184}

\path[draw=drawColor,line width= 0.6pt,line join=round,line cap=round] ( 43.86,141.98) --
	( 47.51,135.66) --
	( 40.21,135.66) --
	( 43.86,141.98);
\end{scope}
\begin{scope}
\path[clip] (  0.00,  0.00) rectangle (245.72,166.22);
\definecolor{drawColor}{RGB}{55,126,184}

\path[draw=drawColor,line width= 0.6pt,line join=round] ( 40.00,137.77) -- ( 47.71,137.77);
\end{scope}
\begin{scope}
\path[clip] (  0.00,  0.00) rectangle (245.72,166.22);
\definecolor{fillColor}{RGB}{255,255,255}

\path[fill=fillColor] ( 39.04,123.31) rectangle ( 48.68,132.95);
\end{scope}
\begin{scope}
\path[clip] (  0.00,  0.00) rectangle (245.72,166.22);
\definecolor{drawColor}{RGB}{77,175,74}

\path[draw=drawColor,line width= 0.6pt,line join=round] ( 40.00,128.13) -- ( 47.71,128.13);
\end{scope}
\begin{scope}
\path[clip] (  0.00,  0.00) rectangle (245.72,166.22);
\definecolor{drawColor}{RGB}{77,175,74}

\path[draw=drawColor,line width= 0.6pt,line join=round,line cap=round] ( 41.15,125.42) -- ( 46.57,130.84);

\path[draw=drawColor,line width= 0.6pt,line join=round,line cap=round] ( 41.15,130.84) -- ( 46.57,125.42);
\end{scope}
\begin{scope}
\path[clip] (  0.00,  0.00) rectangle (245.72,166.22);
\definecolor{drawColor}{RGB}{77,175,74}

\path[draw=drawColor,line width= 0.6pt,line join=round] ( 40.00,128.13) -- ( 47.71,128.13);
\end{scope}
\begin{scope}
\path[clip] (  0.00,  0.00) rectangle (245.72,166.22);
\definecolor{fillColor}{RGB}{255,255,255}

\path[fill=fillColor] ( 39.04,113.68) rectangle ( 48.68,123.31);
\end{scope}
\begin{scope}
\path[clip] (  0.00,  0.00) rectangle (245.72,166.22);
\definecolor{drawColor}{RGB}{152,78,163}

\path[draw=drawColor,line width= 0.6pt,line join=round] ( 40.00,118.49) -- ( 47.71,118.49);
\end{scope}
\begin{scope}
\path[clip] (  0.00,  0.00) rectangle (245.72,166.22);
\definecolor{drawColor}{RGB}{152,78,163}

\path[draw=drawColor,line width= 0.6pt,line join=round,line cap=round] ( 40.02,118.49) -- ( 47.69,118.49);

\path[draw=drawColor,line width= 0.6pt,line join=round,line cap=round] ( 43.86,114.66) -- ( 43.86,122.33);
\end{scope}
\begin{scope}
\path[clip] (  0.00,  0.00) rectangle (245.72,166.22);
\definecolor{drawColor}{RGB}{152,78,163}

\path[draw=drawColor,line width= 0.6pt,line join=round] ( 40.00,118.49) -- ( 47.71,118.49);
\end{scope}
\begin{scope}
\path[clip] (  0.00,  0.00) rectangle (245.72,166.22);
\definecolor{drawColor}{RGB}{0,0,0}

\node[text=drawColor,anchor=base west,inner sep=0pt, outer sep=0pt, scale=  0.72] at ( 48.68,144.92) {BASP};
\end{scope}
\begin{scope}
\path[clip] (  0.00,  0.00) rectangle (245.72,166.22);
\definecolor{drawColor}{RGB}{0,0,0}

\node[text=drawColor,anchor=base west,inner sep=0pt, outer sep=0pt, scale=  0.72] at ( 48.68,135.29) {Ordering};
\end{scope}
\begin{scope}
\path[clip] (  0.00,  0.00) rectangle (245.72,166.22);
\definecolor{drawColor}{RGB}{0,0,0}

\node[text=drawColor,anchor=base west,inner sep=0pt, outer sep=0pt, scale=  0.72] at ( 48.68,125.65) {Ordering + BASP};
\end{scope}
\begin{scope}
\path[clip] (  0.00,  0.00) rectangle (245.72,166.22);
\definecolor{drawColor}{RGB}{0,0,0}

\node[text=drawColor,anchor=base west,inner sep=0pt, outer sep=0pt, scale=  0.72] at ( 48.68,116.02) {Raw};
\end{scope}
\end{tikzpicture}

%% file: receive_combined.tikz
\begin{tikzpicture}[x=1pt,y=1pt]
\definecolor{fillColor}{RGB}{255,255,255}
\path[use as bounding box,fill=fillColor,fill opacity=0.00] (0,0) rectangle (245.72,166.22);
\begin{scope}
\path[clip] (  0.00,  0.00) rectangle (245.72,166.22);
\definecolor{drawColor}{RGB}{255,255,255}
\definecolor{fillColor}{RGB}{255,255,255}

\path[draw=drawColor,line width= 0.6pt,line join=round,line cap=round,fill=fillColor] (  0.00,  0.00) rectangle (245.72,166.22);
\end{scope}
\begin{scope}
\path[clip] ( 30.54, 28.25) rectangle (132.63,160.72);
\definecolor{fillColor}{RGB}{255,255,255}

\path[fill=fillColor] ( 30.54, 28.25) rectangle (132.63,160.72);
\definecolor{drawColor}{gray}{0.92}

\path[draw=drawColor,line width= 0.3pt,line join=round] ( 30.54, 49.32) --
	(132.63, 49.32);

\path[draw=drawColor,line width= 0.3pt,line join=round] ( 30.54, 79.43) --
	(132.63, 79.43);

\path[draw=drawColor,line width= 0.3pt,line join=round] ( 30.54,109.54) --
	(132.63,109.54);

\path[draw=drawColor,line width= 0.3pt,line join=round] ( 30.54,139.65) --
	(132.63,139.65);

\path[draw=drawColor,line width= 0.3pt,line join=round] ( 47.54, 28.25) --
	( 47.54,160.72);

\path[draw=drawColor,line width= 0.3pt,line join=round] ( 69.31, 28.25) --
	( 69.31,160.72);

\path[draw=drawColor,line width= 0.3pt,line join=round] ( 91.07, 28.25) --
	( 91.07,160.72);

\path[draw=drawColor,line width= 0.3pt,line join=round] (112.84, 28.25) --
	(112.84,160.72);

\path[draw=drawColor,line width= 0.6pt,line join=round] ( 30.54, 34.27) --
	(132.63, 34.27);

\path[draw=drawColor,line width= 0.6pt,line join=round] ( 30.54, 64.37) --
	(132.63, 64.37);

\path[draw=drawColor,line width= 0.6pt,line join=round] ( 30.54, 94.48) --
	(132.63, 94.48);

\path[draw=drawColor,line width= 0.6pt,line join=round] ( 30.54,124.59) --
	(132.63,124.59);

\path[draw=drawColor,line width= 0.6pt,line join=round] ( 30.54,154.70) --
	(132.63,154.70);

\path[draw=drawColor,line width= 0.6pt,line join=round] ( 36.66, 28.25) --
	( 36.66,160.72);

\path[draw=drawColor,line width= 0.6pt,line join=round] ( 58.43, 28.25) --
	( 58.43,160.72);

\path[draw=drawColor,line width= 0.6pt,line join=round] ( 80.19, 28.25) --
	( 80.19,160.72);

\path[draw=drawColor,line width= 0.6pt,line join=round] (101.96, 28.25) --
	(101.96,160.72);

\path[draw=drawColor,line width= 0.6pt,line join=round] (123.72, 28.25) --
	(123.72,160.72);
\definecolor{drawColor}{RGB}{228,26,28}

\path[draw=drawColor,line width= 0.6pt,line join=round] ( 37.36,102.57) --
	( 38.05,102.83) --
	( 39.45,102.16) --
	( 42.23,102.24) --
	( 47.80,104.18) --
	( 58.95,102.84) --
	( 81.24,102.60) --
	(125.81,102.68);
\definecolor{drawColor}{RGB}{152,78,163}

\path[draw=drawColor,line width= 0.6pt,line join=round] ( 37.36, 67.30) --
	( 38.05, 67.19) --
	( 39.45, 67.15) --
	( 42.23, 66.99) --
	( 47.80, 67.18) --
	( 58.95, 67.10) --
	( 81.24, 67.08) --
	(125.81, 66.98);

\path[draw=drawColor,line width= 0.6pt,line join=round,line cap=round] ( 33.52, 67.30) -- ( 41.19, 67.30);

\path[draw=drawColor,line width= 0.6pt,line join=round,line cap=round] ( 37.36, 63.46) -- ( 37.36, 71.13);

\path[draw=drawColor,line width= 0.6pt,line join=round,line cap=round] ( 34.22, 67.19) -- ( 41.89, 67.19);

\path[draw=drawColor,line width= 0.6pt,line join=round,line cap=round] ( 38.05, 63.36) -- ( 38.05, 71.03);

\path[draw=drawColor,line width= 0.6pt,line join=round,line cap=round] ( 35.61, 67.15) -- ( 43.28, 67.15);

\path[draw=drawColor,line width= 0.6pt,line join=round,line cap=round] ( 39.45, 63.32) -- ( 39.45, 70.99);

\path[draw=drawColor,line width= 0.6pt,line join=round,line cap=round] ( 38.40, 66.99) -- ( 46.07, 66.99);

\path[draw=drawColor,line width= 0.6pt,line join=round,line cap=round] ( 42.23, 63.15) -- ( 42.23, 70.82);

\path[draw=drawColor,line width= 0.6pt,line join=round,line cap=round] ( 43.97, 67.18) -- ( 51.64, 67.18);

\path[draw=drawColor,line width= 0.6pt,line join=round,line cap=round] ( 47.80, 63.35) -- ( 47.80, 71.02);

\path[draw=drawColor,line width= 0.6pt,line join=round,line cap=round] ( 55.11, 67.10) -- ( 62.78, 67.10);

\path[draw=drawColor,line width= 0.6pt,line join=round,line cap=round] ( 58.95, 63.27) -- ( 58.95, 70.94);

\path[draw=drawColor,line width= 0.6pt,line join=round,line cap=round] ( 77.40, 67.08) -- ( 85.07, 67.08);

\path[draw=drawColor,line width= 0.6pt,line join=round,line cap=round] ( 81.24, 63.24) -- ( 81.24, 70.91);

\path[draw=drawColor,line width= 0.6pt,line join=round,line cap=round] (121.98, 66.98) -- (129.65, 66.98);

\path[draw=drawColor,line width= 0.6pt,line join=round,line cap=round] (125.81, 63.15) -- (125.81, 70.82);
\definecolor{drawColor}{RGB}{228,26,28}

\path[draw=drawColor,line width= 0.6pt,line join=round,line cap=round] ( 37.36,102.57) circle (  2.71);

\path[draw=drawColor,line width= 0.6pt,line join=round,line cap=round] ( 38.05,102.83) circle (  2.71);

\path[draw=drawColor,line width= 0.6pt,line join=round,line cap=round] ( 39.45,102.16) circle (  2.71);

\path[draw=drawColor,line width= 0.6pt,line join=round,line cap=round] ( 42.23,102.24) circle (  2.71);

\path[draw=drawColor,line width= 0.6pt,line join=round,line cap=round] ( 47.80,104.18) circle (  2.71);

\path[draw=drawColor,line width= 0.6pt,line join=round,line cap=round] ( 58.95,102.84) circle (  2.71);

\path[draw=drawColor,line width= 0.6pt,line join=round,line cap=round] ( 81.24,102.60) circle (  2.71);

\path[draw=drawColor,line width= 0.6pt,line join=round,line cap=round] (125.81,102.68) circle (  2.71);
\definecolor{drawColor}{RGB}{152,78,163}

\path[draw=drawColor,line width= 0.6pt,line join=round] ( 35.18, 68.24) --
	( 39.53, 68.24);

\path[draw=drawColor,line width= 0.6pt,line join=round] ( 37.36, 68.24) --
	( 37.36, 66.36);

\path[draw=drawColor,line width= 0.6pt,line join=round] ( 35.18, 66.36) --
	( 39.53, 66.36);

\path[draw=drawColor,line width= 0.6pt,line join=round] ( 35.88, 67.79) --
	( 40.23, 67.79);

\path[draw=drawColor,line width= 0.6pt,line join=round] ( 38.05, 67.79) --
	( 38.05, 66.60);

\path[draw=drawColor,line width= 0.6pt,line join=round] ( 35.88, 66.60) --
	( 40.23, 66.60);

\path[draw=drawColor,line width= 0.6pt,line join=round] ( 37.27, 67.97) --
	( 41.62, 67.97);

\path[draw=drawColor,line width= 0.6pt,line join=round] ( 39.45, 67.97) --
	( 39.45, 66.34);

\path[draw=drawColor,line width= 0.6pt,line join=round] ( 37.27, 66.34) --
	( 41.62, 66.34);

\path[draw=drawColor,line width= 0.6pt,line join=round] ( 40.06, 67.55) --
	( 44.41, 67.55);

\path[draw=drawColor,line width= 0.6pt,line join=round] ( 42.23, 67.55) --
	( 42.23, 66.42);

\path[draw=drawColor,line width= 0.6pt,line join=round] ( 40.06, 66.42) --
	( 44.41, 66.42);

\path[draw=drawColor,line width= 0.6pt,line join=round] ( 45.63, 67.78) --
	( 49.98, 67.78);

\path[draw=drawColor,line width= 0.6pt,line join=round] ( 47.80, 67.78) --
	( 47.80, 66.59);

\path[draw=drawColor,line width= 0.6pt,line join=round] ( 45.63, 66.59) --
	( 49.98, 66.59);

\path[draw=drawColor,line width= 0.6pt,line join=round] ( 56.77, 67.92) --
	( 61.13, 67.92);

\path[draw=drawColor,line width= 0.6pt,line join=round] ( 58.95, 67.92) --
	( 58.95, 66.29);

\path[draw=drawColor,line width= 0.6pt,line join=round] ( 56.77, 66.29) --
	( 61.13, 66.29);

\path[draw=drawColor,line width= 0.6pt,line join=round] ( 79.06, 67.63) --
	( 83.41, 67.63);

\path[draw=drawColor,line width= 0.6pt,line join=round] ( 81.24, 67.63) --
	( 81.24, 66.52);

\path[draw=drawColor,line width= 0.6pt,line join=round] ( 79.06, 66.52) --
	( 83.41, 66.52);

\path[draw=drawColor,line width= 0.6pt,line join=round] (123.64, 67.59) --
	(127.99, 67.59);

\path[draw=drawColor,line width= 0.6pt,line join=round] (125.81, 67.59) --
	(125.81, 66.38);

\path[draw=drawColor,line width= 0.6pt,line join=round] (123.64, 66.38) --
	(127.99, 66.38);
\definecolor{drawColor}{RGB}{228,26,28}

\path[draw=drawColor,line width= 0.6pt,line join=round] ( 35.18,103.86) --
	( 39.53,103.86);

\path[draw=drawColor,line width= 0.6pt,line join=round] ( 37.36,103.86) --
	( 37.36,101.28);

\path[draw=drawColor,line width= 0.6pt,line join=round] ( 35.18,101.28) --
	( 39.53,101.28);

\path[draw=drawColor,line width= 0.6pt,line join=round] ( 35.88,104.06) --
	( 40.23,104.06);

\path[draw=drawColor,line width= 0.6pt,line join=round] ( 38.05,104.06) --
	( 38.05,101.60);

\path[draw=drawColor,line width= 0.6pt,line join=round] ( 35.88,101.60) --
	( 40.23,101.60);

\path[draw=drawColor,line width= 0.6pt,line join=round] ( 37.27,103.30) --
	( 41.62,103.30);

\path[draw=drawColor,line width= 0.6pt,line join=round] ( 39.45,103.30) --
	( 39.45,101.02);

\path[draw=drawColor,line width= 0.6pt,line join=round] ( 37.27,101.02) --
	( 41.62,101.02);

\path[draw=drawColor,line width= 0.6pt,line join=round] ( 40.06,103.48) --
	( 44.41,103.48);

\path[draw=drawColor,line width= 0.6pt,line join=round] ( 42.23,103.48) --
	( 42.23,101.00);

\path[draw=drawColor,line width= 0.6pt,line join=round] ( 40.06,101.00) --
	( 44.41,101.00);

\path[draw=drawColor,line width= 0.6pt,line join=round] ( 45.63,105.83) --
	( 49.98,105.83);

\path[draw=drawColor,line width= 0.6pt,line join=round] ( 47.80,105.83) --
	( 47.80,102.53);

\path[draw=drawColor,line width= 0.6pt,line join=round] ( 45.63,102.53) --
	( 49.98,102.53);

\path[draw=drawColor,line width= 0.6pt,line join=round] ( 56.77,104.28) --
	( 61.13,104.28);

\path[draw=drawColor,line width= 0.6pt,line join=round] ( 58.95,104.28) --
	( 58.95,101.40);

\path[draw=drawColor,line width= 0.6pt,line join=round] ( 56.77,101.40) --
	( 61.13,101.40);

\path[draw=drawColor,line width= 0.6pt,line join=round] ( 79.06,103.83) --
	( 83.41,103.83);

\path[draw=drawColor,line width= 0.6pt,line join=round] ( 81.24,103.83) --
	( 81.24,101.38);

\path[draw=drawColor,line width= 0.6pt,line join=round] ( 79.06,101.38) --
	( 83.41,101.38);

\path[draw=drawColor,line width= 0.6pt,line join=round] (123.64,103.58) --
	(127.99,103.58);

\path[draw=drawColor,line width= 0.6pt,line join=round] (125.81,103.58) --
	(125.81,101.78);

\path[draw=drawColor,line width= 0.6pt,line join=round] (123.64,101.78) --
	(127.99,101.78);
\definecolor{drawColor}{gray}{0.20}

\path[draw=drawColor,line width= 0.6pt,line join=round,line cap=round] ( 30.54, 28.25) rectangle (132.63,160.72);
\end{scope}
\begin{scope}
\path[clip] (138.13, 28.25) rectangle (240.22,160.72);
\definecolor{fillColor}{RGB}{255,255,255}

\path[fill=fillColor] (138.13, 28.25) rectangle (240.22,160.72);
\definecolor{drawColor}{gray}{0.92}

\path[draw=drawColor,line width= 0.3pt,line join=round] (138.13, 49.32) --
	(240.22, 49.32);

\path[draw=drawColor,line width= 0.3pt,line join=round] (138.13, 79.43) --
	(240.22, 79.43);

\path[draw=drawColor,line width= 0.3pt,line join=round] (138.13,109.54) --
	(240.22,109.54);

\path[draw=drawColor,line width= 0.3pt,line join=round] (138.13,139.65) --
	(240.22,139.65);

\path[draw=drawColor,line width= 0.3pt,line join=round] (155.13, 28.25) --
	(155.13,160.72);

\path[draw=drawColor,line width= 0.3pt,line join=round] (176.90, 28.25) --
	(176.90,160.72);

\path[draw=drawColor,line width= 0.3pt,line join=round] (198.66, 28.25) --
	(198.66,160.72);

\path[draw=drawColor,line width= 0.3pt,line join=round] (220.43, 28.25) --
	(220.43,160.72);

\path[draw=drawColor,line width= 0.6pt,line join=round] (138.13, 34.27) --
	(240.22, 34.27);

\path[draw=drawColor,line width= 0.6pt,line join=round] (138.13, 64.37) --
	(240.22, 64.37);

\path[draw=drawColor,line width= 0.6pt,line join=round] (138.13, 94.48) --
	(240.22, 94.48);

\path[draw=drawColor,line width= 0.6pt,line join=round] (138.13,124.59) --
	(240.22,124.59);

\path[draw=drawColor,line width= 0.6pt,line join=round] (138.13,154.70) --
	(240.22,154.70);

\path[draw=drawColor,line width= 0.6pt,line join=round] (144.25, 28.25) --
	(144.25,160.72);

\path[draw=drawColor,line width= 0.6pt,line join=round] (166.02, 28.25) --
	(166.02,160.72);

\path[draw=drawColor,line width= 0.6pt,line join=round] (187.78, 28.25) --
	(187.78,160.72);

\path[draw=drawColor,line width= 0.6pt,line join=round] (209.55, 28.25) --
	(209.55,160.72);

\path[draw=drawColor,line width= 0.6pt,line join=round] (231.31, 28.25) --
	(231.31,160.72);
\definecolor{drawColor}{RGB}{228,26,28}

\path[draw=drawColor,line width= 0.6pt,line join=round] (144.95, 88.65) --
	(145.64, 89.10) --
	(147.04, 88.86) --
	(149.82, 88.60) --
	(155.39, 88.68) --
	(166.54, 88.86) --
	(188.83, 89.49) --
	(233.40, 89.08);
\definecolor{drawColor}{RGB}{55,126,184}

\path[draw=drawColor,line width= 0.6pt,line join=round] (144.95, 84.09) --
	(145.64, 84.51) --
	(147.04, 84.25) --
	(149.82, 84.20) --
	(155.39, 84.60) --
	(166.54, 84.27) --
	(188.83, 84.72) --
	(233.40, 84.28);
\definecolor{drawColor}{RGB}{77,175,74}

\path[draw=drawColor,line width= 0.6pt,line join=round] (144.95,106.40) --
	(145.64,107.01) --
	(147.04,106.26) --
	(149.82,106.35) --
	(155.39,107.38) --
	(166.54,108.58) --
	(188.83,107.40) --
	(233.40,106.47);
\definecolor{drawColor}{RGB}{152,78,163}

\path[draw=drawColor,line width= 0.6pt,line join=round] (144.95, 67.68) --
	(145.64, 67.25) --
	(147.04, 67.46) --
	(149.82, 67.15) --
	(155.39, 68.30) --
	(166.54, 67.76) --
	(188.83, 67.26) --
	(233.40, 67.19);

\path[draw=drawColor,line width= 0.6pt,line join=round,line cap=round] (141.11, 67.68) -- (148.78, 67.68);

\path[draw=drawColor,line width= 0.6pt,line join=round,line cap=round] (144.95, 63.85) -- (144.95, 71.52);

\path[draw=drawColor,line width= 0.6pt,line join=round,line cap=round] (141.81, 67.25) -- (149.48, 67.25);

\path[draw=drawColor,line width= 0.6pt,line join=round,line cap=round] (145.64, 63.41) -- (145.64, 71.08);

\path[draw=drawColor,line width= 0.6pt,line join=round,line cap=round] (143.20, 67.46) -- (150.87, 67.46);

\path[draw=drawColor,line width= 0.6pt,line join=round,line cap=round] (147.04, 63.62) -- (147.04, 71.29);

\path[draw=drawColor,line width= 0.6pt,line join=round,line cap=round] (145.99, 67.15) -- (153.66, 67.15);

\path[draw=drawColor,line width= 0.6pt,line join=round,line cap=round] (149.82, 63.32) -- (149.82, 70.98);

\path[draw=drawColor,line width= 0.6pt,line join=round,line cap=round] (151.56, 68.30) -- (159.23, 68.30);

\path[draw=drawColor,line width= 0.6pt,line join=round,line cap=round] (155.39, 64.46) -- (155.39, 72.13);

\path[draw=drawColor,line width= 0.6pt,line join=round,line cap=round] (162.70, 67.76) -- (170.37, 67.76);

\path[draw=drawColor,line width= 0.6pt,line join=round,line cap=round] (166.54, 63.93) -- (166.54, 71.60);

\path[draw=drawColor,line width= 0.6pt,line join=round,line cap=round] (184.99, 67.26) -- (192.66, 67.26);

\path[draw=drawColor,line width= 0.6pt,line join=round,line cap=round] (188.83, 63.42) -- (188.83, 71.09);

\path[draw=drawColor,line width= 0.6pt,line join=round,line cap=round] (229.57, 67.19) -- (237.23, 67.19);

\path[draw=drawColor,line width= 0.6pt,line join=round,line cap=round] (233.40, 63.36) -- (233.40, 71.03);
\definecolor{drawColor}{RGB}{55,126,184}

\path[draw=drawColor,line width= 0.6pt,line join=round,line cap=round] (144.95, 88.31) --
	(148.60, 81.99) --
	(141.29, 81.99) --
	(144.95, 88.31);

\path[draw=drawColor,line width= 0.6pt,line join=round,line cap=round] (145.64, 88.72) --
	(149.29, 82.40) --
	(141.99, 82.40) --
	(145.64, 88.72);

\path[draw=drawColor,line width= 0.6pt,line join=round,line cap=round] (147.04, 88.47) --
	(150.69, 82.14) --
	(143.38, 82.14) --
	(147.04, 88.47);

\path[draw=drawColor,line width= 0.6pt,line join=round,line cap=round] (149.82, 88.41) --
	(153.47, 82.09) --
	(146.17, 82.09) --
	(149.82, 88.41);

\path[draw=drawColor,line width= 0.6pt,line join=round,line cap=round] (155.39, 88.82) --
	(159.04, 82.50) --
	(151.74, 82.50) --
	(155.39, 88.82);

\path[draw=drawColor,line width= 0.6pt,line join=round,line cap=round] (166.54, 88.49) --
	(170.19, 82.16) --
	(162.89, 82.16) --
	(166.54, 88.49);

\path[draw=drawColor,line width= 0.6pt,line join=round,line cap=round] (188.83, 88.94) --
	(192.48, 82.61) --
	(185.17, 82.61) --
	(188.83, 88.94);

\path[draw=drawColor,line width= 0.6pt,line join=round,line cap=round] (233.40, 88.49) --
	(237.05, 82.17) --
	(229.75, 82.17) --
	(233.40, 88.49);
\definecolor{drawColor}{RGB}{228,26,28}

\path[draw=drawColor,line width= 0.6pt,line join=round,line cap=round] (144.95, 88.65) circle (  2.71);

\path[draw=drawColor,line width= 0.6pt,line join=round,line cap=round] (145.64, 89.10) circle (  2.71);

\path[draw=drawColor,line width= 0.6pt,line join=round,line cap=round] (147.04, 88.86) circle (  2.71);

\path[draw=drawColor,line width= 0.6pt,line join=round,line cap=round] (149.82, 88.60) circle (  2.71);

\path[draw=drawColor,line width= 0.6pt,line join=round,line cap=round] (155.39, 88.68) circle (  2.71);

\path[draw=drawColor,line width= 0.6pt,line join=round,line cap=round] (166.54, 88.86) circle (  2.71);

\path[draw=drawColor,line width= 0.6pt,line join=round,line cap=round] (188.83, 89.49) circle (  2.71);

\path[draw=drawColor,line width= 0.6pt,line join=round,line cap=round] (233.40, 89.08) circle (  2.71);
\definecolor{drawColor}{RGB}{77,175,74}

\path[draw=drawColor,line width= 0.6pt,line join=round,line cap=round] (142.24,103.69) -- (147.66,109.11);

\path[draw=drawColor,line width= 0.6pt,line join=round,line cap=round] (142.24,109.11) -- (147.66,103.69);

\path[draw=drawColor,line width= 0.6pt,line join=round,line cap=round] (142.93,104.30) -- (148.35,109.72);

\path[draw=drawColor,line width= 0.6pt,line join=round,line cap=round] (142.93,109.72) -- (148.35,104.30);

\path[draw=drawColor,line width= 0.6pt,line join=round,line cap=round] (144.32,103.54) -- (149.75,108.97);

\path[draw=drawColor,line width= 0.6pt,line join=round,line cap=round] (144.32,108.97) -- (149.75,103.54);

\path[draw=drawColor,line width= 0.6pt,line join=round,line cap=round] (147.11,103.64) -- (152.53,109.06);

\path[draw=drawColor,line width= 0.6pt,line join=round,line cap=round] (147.11,109.06) -- (152.53,103.64);

\path[draw=drawColor,line width= 0.6pt,line join=round,line cap=round] (152.68,104.67) -- (158.10,110.09);

\path[draw=drawColor,line width= 0.6pt,line join=round,line cap=round] (152.68,110.09) -- (158.10,104.67);

\path[draw=drawColor,line width= 0.6pt,line join=round,line cap=round] (163.83,105.87) -- (169.25,111.29);

\path[draw=drawColor,line width= 0.6pt,line join=round,line cap=round] (163.83,111.29) -- (169.25,105.87);

\path[draw=drawColor,line width= 0.6pt,line join=round,line cap=round] (186.11,104.69) -- (191.54,110.11);

\path[draw=drawColor,line width= 0.6pt,line join=round,line cap=round] (186.11,110.11) -- (191.54,104.69);

\path[draw=drawColor,line width= 0.6pt,line join=round,line cap=round] (230.69,103.75) -- (236.11,109.18);

\path[draw=drawColor,line width= 0.6pt,line join=round,line cap=round] (230.69,109.18) -- (236.11,103.75);
\definecolor{drawColor}{RGB}{152,78,163}

\path[draw=drawColor,line width= 0.6pt,line join=round] (142.77, 68.86) --
	(147.12, 68.86);

\path[draw=drawColor,line width= 0.6pt,line join=round] (144.95, 68.86) --
	(144.95, 66.51);

\path[draw=drawColor,line width= 0.6pt,line join=round] (142.77, 66.51) --
	(147.12, 66.51);

\path[draw=drawColor,line width= 0.6pt,line join=round] (143.47, 68.03) --
	(147.82, 68.03);

\path[draw=drawColor,line width= 0.6pt,line join=round] (145.64, 68.03) --
	(145.64, 66.47);

\path[draw=drawColor,line width= 0.6pt,line join=round] (143.47, 66.47) --
	(147.82, 66.47);

\path[draw=drawColor,line width= 0.6pt,line join=round] (144.86, 68.26) --
	(149.21, 68.26);

\path[draw=drawColor,line width= 0.6pt,line join=round] (147.04, 68.26) --
	(147.04, 66.65);

\path[draw=drawColor,line width= 0.6pt,line join=round] (144.86, 66.65) --
	(149.21, 66.65);

\path[draw=drawColor,line width= 0.6pt,line join=round] (147.64, 67.83) --
	(152.00, 67.83);

\path[draw=drawColor,line width= 0.6pt,line join=round] (149.82, 67.83) --
	(149.82, 66.47);

\path[draw=drawColor,line width= 0.6pt,line join=round] (147.64, 66.47) --
	(152.00, 66.47);

\path[draw=drawColor,line width= 0.6pt,line join=round] (153.22, 69.68) --
	(157.57, 69.68);

\path[draw=drawColor,line width= 0.6pt,line join=round] (155.39, 69.68) --
	(155.39, 66.91);

\path[draw=drawColor,line width= 0.6pt,line join=round] (153.22, 66.91) --
	(157.57, 66.91);

\path[draw=drawColor,line width= 0.6pt,line join=round] (164.36, 69.18) --
	(168.71, 69.18);

\path[draw=drawColor,line width= 0.6pt,line join=round] (166.54, 69.18) --
	(166.54, 66.35);

\path[draw=drawColor,line width= 0.6pt,line join=round] (164.36, 66.35) --
	(168.71, 66.35);

\path[draw=drawColor,line width= 0.6pt,line join=round] (186.65, 67.66) --
	(191.00, 67.66);

\path[draw=drawColor,line width= 0.6pt,line join=round] (188.83, 67.66) --
	(188.83, 66.85);

\path[draw=drawColor,line width= 0.6pt,line join=round] (186.65, 66.85) --
	(191.00, 66.85);

\path[draw=drawColor,line width= 0.6pt,line join=round] (231.22, 68.04) --
	(235.58, 68.04);

\path[draw=drawColor,line width= 0.6pt,line join=round] (233.40, 68.04) --
	(233.40, 66.34);

\path[draw=drawColor,line width= 0.6pt,line join=round] (231.22, 66.34) --
	(235.58, 66.34);
\definecolor{drawColor}{RGB}{55,126,184}

\path[draw=drawColor,line width= 0.6pt,line join=round] (142.77, 84.76) --
	(147.12, 84.76);

\path[draw=drawColor,line width= 0.6pt,line join=round] (144.95, 84.76) --
	(144.95, 83.42);

\path[draw=drawColor,line width= 0.6pt,line join=round] (142.77, 83.42) --
	(147.12, 83.42);

\path[draw=drawColor,line width= 0.6pt,line join=round] (143.47, 85.25) --
	(147.82, 85.25);

\path[draw=drawColor,line width= 0.6pt,line join=round] (145.64, 85.25) --
	(145.64, 83.76);

\path[draw=drawColor,line width= 0.6pt,line join=round] (143.47, 83.76) --
	(147.82, 83.76);

\path[draw=drawColor,line width= 0.6pt,line join=round] (144.86, 84.94) --
	(149.21, 84.94);

\path[draw=drawColor,line width= 0.6pt,line join=round] (147.04, 84.94) --
	(147.04, 83.56);

\path[draw=drawColor,line width= 0.6pt,line join=round] (144.86, 83.56) --
	(149.21, 83.56);

\path[draw=drawColor,line width= 0.6pt,line join=round] (147.64, 84.86) --
	(152.00, 84.86);

\path[draw=drawColor,line width= 0.6pt,line join=round] (149.82, 84.86) --
	(149.82, 83.53);

\path[draw=drawColor,line width= 0.6pt,line join=round] (147.64, 83.53) --
	(152.00, 83.53);

\path[draw=drawColor,line width= 0.6pt,line join=round] (153.22, 85.69) --
	(157.57, 85.69);

\path[draw=drawColor,line width= 0.6pt,line join=round] (155.39, 85.69) --
	(155.39, 83.52);

\path[draw=drawColor,line width= 0.6pt,line join=round] (153.22, 83.52) --
	(157.57, 83.52);

\path[draw=drawColor,line width= 0.6pt,line join=round] (164.36, 85.14) --
	(168.71, 85.14);

\path[draw=drawColor,line width= 0.6pt,line join=round] (166.54, 85.14) --
	(166.54, 83.41);

\path[draw=drawColor,line width= 0.6pt,line join=round] (164.36, 83.41) --
	(168.71, 83.41);

\path[draw=drawColor,line width= 0.6pt,line join=round] (186.65, 85.83) --
	(191.00, 85.83);

\path[draw=drawColor,line width= 0.6pt,line join=round] (188.83, 85.83) --
	(188.83, 83.61);

\path[draw=drawColor,line width= 0.6pt,line join=round] (186.65, 83.61) --
	(191.00, 83.61);

\path[draw=drawColor,line width= 0.6pt,line join=round] (231.22, 85.38) --
	(235.58, 85.38);

\path[draw=drawColor,line width= 0.6pt,line join=round] (233.40, 85.38) --
	(233.40, 83.17);

\path[draw=drawColor,line width= 0.6pt,line join=round] (231.22, 83.17) --
	(235.58, 83.17);
\definecolor{drawColor}{RGB}{228,26,28}

\path[draw=drawColor,line width= 0.6pt,line join=round] (142.77, 89.43) --
	(147.12, 89.43);

\path[draw=drawColor,line width= 0.6pt,line join=round] (144.95, 89.43) --
	(144.95, 87.87);

\path[draw=drawColor,line width= 0.6pt,line join=round] (142.77, 87.87) --
	(147.12, 87.87);

\path[draw=drawColor,line width= 0.6pt,line join=round] (143.47, 90.38) --
	(147.82, 90.38);

\path[draw=drawColor,line width= 0.6pt,line join=round] (145.64, 90.38) --
	(145.64, 87.81);

\path[draw=drawColor,line width= 0.6pt,line join=round] (143.47, 87.81) --
	(147.82, 87.81);

\path[draw=drawColor,line width= 0.6pt,line join=round] (144.86, 89.85) --
	(149.21, 89.85);

\path[draw=drawColor,line width= 0.6pt,line join=round] (147.04, 89.85) --
	(147.04, 87.87);

\path[draw=drawColor,line width= 0.6pt,line join=round] (144.86, 87.87) --
	(149.21, 87.87);

\path[draw=drawColor,line width= 0.6pt,line join=round] (147.64, 89.44) --
	(152.00, 89.44);

\path[draw=drawColor,line width= 0.6pt,line join=round] (149.82, 89.44) --
	(149.82, 87.77);

\path[draw=drawColor,line width= 0.6pt,line join=round] (147.64, 87.77) --
	(152.00, 87.77);

\path[draw=drawColor,line width= 0.6pt,line join=round] (153.22, 89.56) --
	(157.57, 89.56);

\path[draw=drawColor,line width= 0.6pt,line join=round] (155.39, 89.56) --
	(155.39, 87.80);

\path[draw=drawColor,line width= 0.6pt,line join=round] (153.22, 87.80) --
	(157.57, 87.80);

\path[draw=drawColor,line width= 0.6pt,line join=round] (164.36, 89.71) --
	(168.71, 89.71);

\path[draw=drawColor,line width= 0.6pt,line join=round] (166.54, 89.71) --
	(166.54, 88.00);

\path[draw=drawColor,line width= 0.6pt,line join=round] (164.36, 88.00) --
	(168.71, 88.00);

\path[draw=drawColor,line width= 0.6pt,line join=round] (186.65, 90.80) --
	(191.00, 90.80);

\path[draw=drawColor,line width= 0.6pt,line join=round] (188.83, 90.80) --
	(188.83, 88.18);

\path[draw=drawColor,line width= 0.6pt,line join=round] (186.65, 88.18) --
	(191.00, 88.18);

\path[draw=drawColor,line width= 0.6pt,line join=round] (231.22, 90.05) --
	(235.58, 90.05);

\path[draw=drawColor,line width= 0.6pt,line join=round] (233.40, 90.05) --
	(233.40, 88.10);

\path[draw=drawColor,line width= 0.6pt,line join=round] (231.22, 88.10) --
	(235.58, 88.10);
\definecolor{drawColor}{RGB}{77,175,74}

\path[draw=drawColor,line width= 0.6pt,line join=round] (142.77,107.52) --
	(147.12,107.52);

\path[draw=drawColor,line width= 0.6pt,line join=round] (144.95,107.52) --
	(144.95,105.27);

\path[draw=drawColor,line width= 0.6pt,line join=round] (142.77,105.27) --
	(147.12,105.27);

\path[draw=drawColor,line width= 0.6pt,line join=round] (143.47,108.72) --
	(147.82,108.72);

\path[draw=drawColor,line width= 0.6pt,line join=round] (145.64,108.72) --
	(145.64,105.29);

\path[draw=drawColor,line width= 0.6pt,line join=round] (143.47,105.29) --
	(147.82,105.29);

\path[draw=drawColor,line width= 0.6pt,line join=round] (144.86,107.37) --
	(149.21,107.37);

\path[draw=drawColor,line width= 0.6pt,line join=round] (147.04,107.37) --
	(147.04,105.14);

\path[draw=drawColor,line width= 0.6pt,line join=round] (144.86,105.14) --
	(149.21,105.14);

\path[draw=drawColor,line width= 0.6pt,line join=round] (147.64,107.54) --
	(152.00,107.54);

\path[draw=drawColor,line width= 0.6pt,line join=round] (149.82,107.54) --
	(149.82,105.15);

\path[draw=drawColor,line width= 0.6pt,line join=round] (147.64,105.15) --
	(152.00,105.15);

\path[draw=drawColor,line width= 0.6pt,line join=round] (153.22,108.51) --
	(157.57,108.51);

\path[draw=drawColor,line width= 0.6pt,line join=round] (155.39,108.51) --
	(155.39,106.26);

\path[draw=drawColor,line width= 0.6pt,line join=round] (153.22,106.26) --
	(157.57,106.26);

\path[draw=drawColor,line width= 0.6pt,line join=round] (164.36,109.91) --
	(168.71,109.91);

\path[draw=drawColor,line width= 0.6pt,line join=round] (166.54,109.91) --
	(166.54,107.24);

\path[draw=drawColor,line width= 0.6pt,line join=round] (164.36,107.24) --
	(168.71,107.24);

\path[draw=drawColor,line width= 0.6pt,line join=round] (186.65,109.48) --
	(191.00,109.48);

\path[draw=drawColor,line width= 0.6pt,line join=round] (188.83,109.48) --
	(188.83,105.33);

\path[draw=drawColor,line width= 0.6pt,line join=round] (186.65,105.33) --
	(191.00,105.33);

\path[draw=drawColor,line width= 0.6pt,line join=round] (231.22,107.48) --
	(235.58,107.48);

\path[draw=drawColor,line width= 0.6pt,line join=round] (233.40,107.48) --
	(233.40,105.45);

\path[draw=drawColor,line width= 0.6pt,line join=round] (231.22,105.45) --
	(235.58,105.45);
\definecolor{drawColor}{gray}{0.20}

\path[draw=drawColor,line width= 0.6pt,line join=round,line cap=round] (138.13, 28.25) rectangle (240.22,160.72);
\end{scope}
\begin{scope}
\path[clip] (  0.00,  0.00) rectangle (245.72,166.22);
\definecolor{drawColor}{gray}{0.20}

\path[draw=drawColor,line width= 0.6pt,line join=round] ( 36.66, 25.50) --
	( 36.66, 28.25);

\path[draw=drawColor,line width= 0.6pt,line join=round] ( 58.43, 25.50) --
	( 58.43, 28.25);

\path[draw=drawColor,line width= 0.6pt,line join=round] ( 80.19, 25.50) --
	( 80.19, 28.25);

\path[draw=drawColor,line width= 0.6pt,line join=round] (101.96, 25.50) --
	(101.96, 28.25);

\path[draw=drawColor,line width= 0.6pt,line join=round] (123.72, 25.50) --
	(123.72, 28.25);
\end{scope}
\begin{scope}
\path[clip] (  0.00,  0.00) rectangle (245.72,166.22);
\definecolor{drawColor}{gray}{0.30}

\node[text=drawColor,anchor=base,inner sep=0pt, outer sep=0pt, scale=  0.72] at ( 36.66, 18.34) {0};

\node[text=drawColor,anchor=base,inner sep=0pt, outer sep=0pt, scale=  0.72] at ( 58.43, 18.34) {2000};

\node[text=drawColor,anchor=base,inner sep=0pt, outer sep=0pt, scale=  0.72] at ( 80.19, 18.34) {4000};

\node[text=drawColor,anchor=base,inner sep=0pt, outer sep=0pt, scale=  0.72] at (101.96, 18.34) {6000};

\node[text=drawColor,anchor=base,inner sep=0pt, outer sep=0pt, scale=  0.72] at (123.72, 18.34) {8000};
\end{scope}
\begin{scope}
\path[clip] (  0.00,  0.00) rectangle (245.72,166.22);
\definecolor{drawColor}{gray}{0.20}

\path[draw=drawColor,line width= 0.6pt,line join=round] (144.25, 25.50) --
	(144.25, 28.25);

\path[draw=drawColor,line width= 0.6pt,line join=round] (166.02, 25.50) --
	(166.02, 28.25);

\path[draw=drawColor,line width= 0.6pt,line join=round] (187.78, 25.50) --
	(187.78, 28.25);

\path[draw=drawColor,line width= 0.6pt,line join=round] (209.55, 25.50) --
	(209.55, 28.25);

\path[draw=drawColor,line width= 0.6pt,line join=round] (231.31, 25.50) --
	(231.31, 28.25);
\end{scope}
\begin{scope}
\path[clip] (  0.00,  0.00) rectangle (245.72,166.22);
\definecolor{drawColor}{gray}{0.30}

\node[text=drawColor,anchor=base,inner sep=0pt, outer sep=0pt, scale=  0.72] at (144.25, 18.34) {0};

\node[text=drawColor,anchor=base,inner sep=0pt, outer sep=0pt, scale=  0.72] at (166.02, 18.34) {2000};

\node[text=drawColor,anchor=base,inner sep=0pt, outer sep=0pt, scale=  0.72] at (187.78, 18.34) {4000};

\node[text=drawColor,anchor=base,inner sep=0pt, outer sep=0pt, scale=  0.72] at (209.55, 18.34) {6000};

\node[text=drawColor,anchor=base,inner sep=0pt, outer sep=0pt, scale=  0.72] at (231.31, 18.34) {8000};
\end{scope}
\begin{scope}
\path[clip] (  0.00,  0.00) rectangle (245.72,166.22);
\definecolor{drawColor}{gray}{0.30}

\node[text=drawColor,anchor=base east,inner sep=0pt, outer sep=0pt, scale=  0.72] at ( 25.59, 31.79) {0.0};

\node[text=drawColor,anchor=base east,inner sep=0pt, outer sep=0pt, scale=  0.72] at ( 25.59, 61.90) {0.1};

\node[text=drawColor,anchor=base east,inner sep=0pt, outer sep=0pt, scale=  0.72] at ( 25.59, 92.00) {0.2};

\node[text=drawColor,anchor=base east,inner sep=0pt, outer sep=0pt, scale=  0.72] at ( 25.59,122.11) {0.3};

\node[text=drawColor,anchor=base east,inner sep=0pt, outer sep=0pt, scale=  0.72] at ( 25.59,152.22) {0.4};
\end{scope}
\begin{scope}
\path[clip] (  0.00,  0.00) rectangle (245.72,166.22);
\definecolor{drawColor}{gray}{0.20}

\path[draw=drawColor,line width= 0.6pt,line join=round] ( 27.79, 34.27) --
	( 30.54, 34.27);

\path[draw=drawColor,line width= 0.6pt,line join=round] ( 27.79, 64.37) --
	( 30.54, 64.37);

\path[draw=drawColor,line width= 0.6pt,line join=round] ( 27.79, 94.48) --
	( 30.54, 94.48);

\path[draw=drawColor,line width= 0.6pt,line join=round] ( 27.79,124.59) --
	( 30.54,124.59);

\path[draw=drawColor,line width= 0.6pt,line join=round] ( 27.79,154.70) --
	( 30.54,154.70);
\end{scope}
\begin{scope}
\path[clip] (  0.00,  0.00) rectangle (245.72,166.22);
\definecolor{drawColor}{RGB}{0,0,0}

\node[text=drawColor,anchor=base,inner sep=0pt, outer sep=0pt, scale=  0.90] at (135.38,  7.44) {Payload Size [bytes]};
\end{scope}
\begin{scope}
\path[clip] (  0.00,  0.00) rectangle (245.72,166.22);
\definecolor{drawColor}{RGB}{0,0,0}

\node[text=drawColor,rotate= 90.00,anchor=base,inner sep=0pt, outer sep=0pt, scale=  0.90] at ( 11.70, 94.48) {Packet Preparation Time [us]};
\end{scope}
\begin{scope}
\path[clip] (  0.00,  0.00) rectangle (245.72,166.22);
\definecolor{drawColor}{RGB}{0,0,0}
\definecolor{fillColor}{RGB}{255,255,255}

\path[draw=drawColor,line width= 0.3pt,line join=round,line cap=round,fill=fillColor] ( 33.54,108.18) rectangle (112.00,157.72);
\end{scope}
\begin{scope}
\path[clip] (  0.00,  0.00) rectangle (245.72,166.22);
\definecolor{fillColor}{RGB}{255,255,255}

\path[fill=fillColor] ( 39.04,142.59) rectangle ( 48.68,152.22);
\end{scope}
\begin{scope}
\path[clip] (  0.00,  0.00) rectangle (245.72,166.22);
\definecolor{drawColor}{RGB}{228,26,28}

\path[draw=drawColor,line width= 0.6pt,line join=round] ( 40.00,147.40) -- ( 47.71,147.40);
\end{scope}
\begin{scope}
\path[clip] (  0.00,  0.00) rectangle (245.72,166.22);
\definecolor{drawColor}{RGB}{228,26,28}

\path[draw=drawColor,line width= 0.6pt,line join=round,line cap=round] ( 43.86,147.40) circle (  2.71);
\end{scope}
\begin{scope}
\path[clip] (  0.00,  0.00) rectangle (245.72,166.22);
\definecolor{drawColor}{RGB}{228,26,28}

\path[draw=drawColor,line width= 0.6pt,line join=round] ( 40.00,147.40) -- ( 47.71,147.40);
\end{scope}
\begin{scope}
\path[clip] (  0.00,  0.00) rectangle (245.72,166.22);
\definecolor{fillColor}{RGB}{255,255,255}

\path[fill=fillColor] ( 39.04,132.95) rectangle ( 48.68,142.59);
\end{scope}
\begin{scope}
\path[clip] (  0.00,  0.00) rectangle (245.72,166.22);
\definecolor{drawColor}{RGB}{55,126,184}

\path[draw=drawColor,line width= 0.6pt,line join=round] ( 40.00,137.77) -- ( 47.71,137.77);
\end{scope}
\begin{scope}
\path[clip] (  0.00,  0.00) rectangle (245.72,166.22);
\definecolor{drawColor}{RGB}{55,126,184}

\path[draw=drawColor,line width= 0.6pt,line join=round,line cap=round] ( 43.86,141.98) --
	( 47.51,135.66) --
	( 40.21,135.66) --
	( 43.86,141.98);
\end{scope}
\begin{scope}
\path[clip] (  0.00,  0.00) rectangle (245.72,166.22);
\definecolor{drawColor}{RGB}{55,126,184}

\path[draw=drawColor,line width= 0.6pt,line join=round] ( 40.00,137.77) -- ( 47.71,137.77);
\end{scope}
\begin{scope}
\path[clip] (  0.00,  0.00) rectangle (245.72,166.22);
\definecolor{fillColor}{RGB}{255,255,255}

\path[fill=fillColor] ( 39.04,123.31) rectangle ( 48.68,132.95);
\end{scope}
\begin{scope}
\path[clip] (  0.00,  0.00) rectangle (245.72,166.22);
\definecolor{drawColor}{RGB}{77,175,74}

\path[draw=drawColor,line width= 0.6pt,line join=round] ( 40.00,128.13) -- ( 47.71,128.13);
\end{scope}
\begin{scope}
\path[clip] (  0.00,  0.00) rectangle (245.72,166.22);
\definecolor{drawColor}{RGB}{77,175,74}

\path[draw=drawColor,line width= 0.6pt,line join=round,line cap=round] ( 41.15,125.42) -- ( 46.57,130.84);

\path[draw=drawColor,line width= 0.6pt,line join=round,line cap=round] ( 41.15,130.84) -- ( 46.57,125.42);
\end{scope}
\begin{scope}
\path[clip] (  0.00,  0.00) rectangle (245.72,166.22);
\definecolor{drawColor}{RGB}{77,175,74}

\path[draw=drawColor,line width= 0.6pt,line join=round] ( 40.00,128.13) -- ( 47.71,128.13);
\end{scope}
\begin{scope}
\path[clip] (  0.00,  0.00) rectangle (245.72,166.22);
\definecolor{fillColor}{RGB}{255,255,255}

\path[fill=fillColor] ( 39.04,113.68) rectangle ( 48.68,123.31);
\end{scope}
\begin{scope}
\path[clip] (  0.00,  0.00) rectangle (245.72,166.22);
\definecolor{drawColor}{RGB}{152,78,163}

\path[draw=drawColor,line width= 0.6pt,line join=round] ( 40.00,118.49) -- ( 47.71,118.49);
\end{scope}
\begin{scope}
\path[clip] (  0.00,  0.00) rectangle (245.72,166.22);
\definecolor{drawColor}{RGB}{152,78,163}

\path[draw=drawColor,line width= 0.6pt,line join=round,line cap=round] ( 40.02,118.49) -- ( 47.69,118.49);

\path[draw=drawColor,line width= 0.6pt,line join=round,line cap=round] ( 43.86,114.66) -- ( 43.86,122.33);
\end{scope}
\begin{scope}
\path[clip] (  0.00,  0.00) rectangle (245.72,166.22);
\definecolor{drawColor}{RGB}{152,78,163}

\path[draw=drawColor,line width= 0.6pt,line join=round] ( 40.00,118.49) -- ( 47.71,118.49);
\end{scope}
\begin{scope}
\path[clip] (  0.00,  0.00) rectangle (245.72,166.22);
\definecolor{drawColor}{RGB}{0,0,0}

\node[text=drawColor,anchor=base west,inner sep=0pt, outer sep=0pt, scale=  0.72] at ( 48.68,144.92) {BASP};
\end{scope}
\begin{scope}
\path[clip] (  0.00,  0.00) rectangle (245.72,166.22);
\definecolor{drawColor}{RGB}{0,0,0}

\node[text=drawColor,anchor=base west,inner sep=0pt, outer sep=0pt, scale=  0.72] at ( 48.68,135.29) {Ordering};
\end{scope}
\begin{scope}
\path[clip] (  0.00,  0.00) rectangle (245.72,166.22);
\definecolor{drawColor}{RGB}{0,0,0}

\node[text=drawColor,anchor=base west,inner sep=0pt, outer sep=0pt, scale=  0.72] at ( 48.68,125.65) {Ordering + BASP};
\end{scope}
\begin{scope}
\path[clip] (  0.00,  0.00) rectangle (245.72,166.22);
\definecolor{drawColor}{RGB}{0,0,0}

\node[text=drawColor,anchor=base west,inner sep=0pt, outer sep=0pt, scale=  0.72] at ( 48.68,116.02) {Raw};
\end{scope}
\end{tikzpicture}

%% file: udp_receive_sequence.tikz
\begin{tikzpicture}[x=1pt,y=1pt]
\definecolor{fillColor}{RGB}{255,255,255}
\path[use as bounding box,fill=fillColor,fill opacity=0.00] (0,0) rectangle (245.72,166.22);
\begin{scope}
\path[clip] (  0.00,  0.00) rectangle (245.72,166.22);
\definecolor{drawColor}{RGB}{255,255,255}
\definecolor{fillColor}{RGB}{255,255,255}

\path[draw=drawColor,line width= 0.6pt,line join=round,line cap=round,fill=fillColor] (  0.00,  0.00) rectangle (245.72,166.22);
\end{scope}
\begin{scope}
\path[clip] ( 24.94, 28.25) rectangle (240.22,160.72);
\definecolor{fillColor}{RGB}{255,255,255}

\path[fill=fillColor] ( 24.94, 28.25) rectangle (240.22,160.72);
\definecolor{drawColor}{gray}{0.92}

\path[draw=drawColor,line width= 0.3pt,line join=round] ( 24.94, 46.31) --
	(240.22, 46.31);

\path[draw=drawColor,line width= 0.3pt,line join=round] ( 24.94, 70.40) --
	(240.22, 70.40);

\path[draw=drawColor,line width= 0.3pt,line join=round] ( 24.94, 94.48) --
	(240.22, 94.48);

\path[draw=drawColor,line width= 0.3pt,line join=round] ( 24.94,118.57) --
	(240.22,118.57);

\path[draw=drawColor,line width= 0.3pt,line join=round] ( 24.94,142.66) --
	(240.22,142.66);

\path[draw=drawColor,line width= 0.3pt,line join=round] ( 59.07, 28.25) --
	( 59.07,160.72);

\path[draw=drawColor,line width= 0.3pt,line join=round] (106.07, 28.25) --
	(106.07,160.72);

\path[draw=drawColor,line width= 0.3pt,line join=round] (153.07, 28.25) --
	(153.07,160.72);

\path[draw=drawColor,line width= 0.3pt,line join=round] (200.07, 28.25) --
	(200.07,160.72);

\path[draw=drawColor,line width= 0.6pt,line join=round] ( 24.94, 34.27) --
	(240.22, 34.27);

\path[draw=drawColor,line width= 0.6pt,line join=round] ( 24.94, 58.35) --
	(240.22, 58.35);

\path[draw=drawColor,line width= 0.6pt,line join=round] ( 24.94, 82.44) --
	(240.22, 82.44);

\path[draw=drawColor,line width= 0.6pt,line join=round] ( 24.94,106.53) --
	(240.22,106.53);

\path[draw=drawColor,line width= 0.6pt,line join=round] ( 24.94,130.61) --
	(240.22,130.61);

\path[draw=drawColor,line width= 0.6pt,line join=round] ( 24.94,154.70) --
	(240.22,154.70);

\path[draw=drawColor,line width= 0.6pt,line join=round] ( 35.57, 28.25) --
	( 35.57,160.72);

\path[draw=drawColor,line width= 0.6pt,line join=round] ( 82.57, 28.25) --
	( 82.57,160.72);

\path[draw=drawColor,line width= 0.6pt,line join=round] (129.57, 28.25) --
	(129.57,160.72);

\path[draw=drawColor,line width= 0.6pt,line join=round] (176.57, 28.25) --
	(176.57,160.72);

\path[draw=drawColor,line width= 0.6pt,line join=round] (223.57, 28.25) --
	(223.57,160.72);
\definecolor{drawColor}{RGB}{228,26,28}

\path[draw=drawColor,line width= 0.6pt,line join=round] ( 37.08,104.62) --
	( 38.58,106.42) --
	( 41.59,109.44) --
	( 47.60,110.11) --
	( 59.64,120.36) --
	( 83.70,120.64) --
	(131.83,125.60) --
	(228.08,137.05);
\definecolor{drawColor}{RGB}{55,126,184}

\path[draw=drawColor,line width= 0.6pt,line join=round] ( 37.08, 78.89) --
	( 38.58, 79.34) --
	( 41.59, 80.75) --
	( 47.60, 80.81) --
	( 59.64, 82.86) --
	( 83.70, 82.62) --
	(131.83, 83.33) --
	(228.08, 85.09);
\definecolor{drawColor}{RGB}{77,175,74}

\path[draw=drawColor,line width= 0.6pt,line join=round] ( 37.08, 74.00) --
	( 38.58, 74.65) --
	( 41.59, 73.98) --
	( 47.60, 74.07) --
	( 59.64, 73.89) --
	( 83.70, 74.01) --
	(131.83, 74.09) --
	(228.08, 73.96);

\path[draw=drawColor,line width= 0.6pt,line join=round,line cap=round] ( 37.08, 69.79) --
	( 40.73, 76.11) --
	( 33.43, 76.11) --
	( 37.08, 69.79);

\path[draw=drawColor,line width= 0.6pt,line join=round,line cap=round] ( 38.58, 70.43) --
	( 42.23, 76.75) --
	( 34.93, 76.75) --
	( 38.58, 70.43);

\path[draw=drawColor,line width= 0.6pt,line join=round,line cap=round] ( 41.59, 69.77) --
	( 45.24, 76.09) --
	( 37.94, 76.09) --
	( 41.59, 69.77);

\path[draw=drawColor,line width= 0.6pt,line join=round,line cap=round] ( 47.60, 69.85) --
	( 51.26, 76.18) --
	( 43.95, 76.18) --
	( 47.60, 69.85);

\path[draw=drawColor,line width= 0.6pt,line join=round,line cap=round] ( 59.64, 69.67) --
	( 63.29, 76.00) --
	( 55.99, 76.00) --
	( 59.64, 69.67);

\path[draw=drawColor,line width= 0.6pt,line join=round,line cap=round] ( 83.70, 69.80) --
	( 87.35, 76.12) --
	( 80.05, 76.12) --
	( 83.70, 69.80);

\path[draw=drawColor,line width= 0.6pt,line join=round,line cap=round] (131.83, 69.88) --
	(135.48, 76.20) --
	(128.18, 76.20) --
	(131.83, 69.88);

\path[draw=drawColor,line width= 0.6pt,line join=round,line cap=round] (228.08, 69.75) --
	(231.73, 76.07) --
	(224.43, 76.07) --
	(228.08, 69.75);
\definecolor{drawColor}{RGB}{228,26,28}

\path[draw=drawColor,line width= 0.6pt,line join=round,line cap=round] ( 34.37,101.91) -- ( 39.79,107.34);

\path[draw=drawColor,line width= 0.6pt,line join=round,line cap=round] ( 34.37,107.34) -- ( 39.79,101.91);

\path[draw=drawColor,line width= 0.6pt,line join=round,line cap=round] ( 35.87,103.71) -- ( 41.29,109.14);

\path[draw=drawColor,line width= 0.6pt,line join=round,line cap=round] ( 35.87,109.14) -- ( 41.29,103.71);

\path[draw=drawColor,line width= 0.6pt,line join=round,line cap=round] ( 38.88,106.73) -- ( 44.30,112.15);

\path[draw=drawColor,line width= 0.6pt,line join=round,line cap=round] ( 38.88,112.15) -- ( 44.30,106.73);

\path[draw=drawColor,line width= 0.6pt,line join=round,line cap=round] ( 44.89,107.40) -- ( 50.32,112.82);

\path[draw=drawColor,line width= 0.6pt,line join=round,line cap=round] ( 44.89,112.82) -- ( 50.32,107.40);

\path[draw=drawColor,line width= 0.6pt,line join=round,line cap=round] ( 56.93,117.65) -- ( 62.35,123.07);

\path[draw=drawColor,line width= 0.6pt,line join=round,line cap=round] ( 56.93,123.07) -- ( 62.35,117.65);

\path[draw=drawColor,line width= 0.6pt,line join=round,line cap=round] ( 80.99,117.93) -- ( 86.41,123.35);

\path[draw=drawColor,line width= 0.6pt,line join=round,line cap=round] ( 80.99,123.35) -- ( 86.41,117.93);

\path[draw=drawColor,line width= 0.6pt,line join=round,line cap=round] (129.12,122.89) -- (134.54,128.31);

\path[draw=drawColor,line width= 0.6pt,line join=round,line cap=round] (129.12,128.31) -- (134.54,122.89);

\path[draw=drawColor,line width= 0.6pt,line join=round,line cap=round] (225.37,134.34) -- (230.79,139.76);

\path[draw=drawColor,line width= 0.6pt,line join=round,line cap=round] (225.37,139.76) -- (230.79,134.34);
\definecolor{drawColor}{RGB}{55,126,184}

\path[draw=drawColor,line width= 0.6pt,line join=round,line cap=round] ( 33.24, 78.89) --
	( 37.08, 82.73) --
	( 40.91, 78.89) --
	( 37.08, 75.06) --
	( 33.24, 78.89);

\path[draw=drawColor,line width= 0.6pt,line join=round,line cap=round] ( 34.75, 79.34) --
	( 38.58, 83.17) --
	( 42.41, 79.34) --
	( 38.58, 75.51) --
	( 34.75, 79.34);

\path[draw=drawColor,line width= 0.6pt,line join=round,line cap=round] ( 37.75, 80.75) --
	( 41.59, 84.58) --
	( 45.42, 80.75) --
	( 41.59, 76.91) --
	( 37.75, 80.75);

\path[draw=drawColor,line width= 0.6pt,line join=round,line cap=round] ( 43.77, 80.81) --
	( 47.60, 84.64) --
	( 51.44, 80.81) --
	( 47.60, 76.97) --
	( 43.77, 80.81);

\path[draw=drawColor,line width= 0.6pt,line join=round,line cap=round] ( 55.80, 82.86) --
	( 59.64, 86.69) --
	( 63.47, 82.86) --
	( 59.64, 79.03) --
	( 55.80, 82.86);

\path[draw=drawColor,line width= 0.6pt,line join=round,line cap=round] ( 79.87, 82.62) --
	( 83.70, 86.45) --
	( 87.53, 82.62) --
	( 83.70, 78.79) --
	( 79.87, 82.62);

\path[draw=drawColor,line width= 0.6pt,line join=round,line cap=round] (127.99, 83.33) --
	(131.83, 87.16) --
	(135.66, 83.33) --
	(131.83, 79.50) --
	(127.99, 83.33);

\path[draw=drawColor,line width= 0.6pt,line join=round,line cap=round] (224.25, 85.09) --
	(228.08, 88.93) --
	(231.92, 85.09) --
	(228.08, 81.26) --
	(224.25, 85.09);
\definecolor{drawColor}{RGB}{77,175,74}

\path[draw=drawColor,line width= 0.6pt,line join=round] ( 34.73, 74.76) --
	( 39.43, 74.76);

\path[draw=drawColor,line width= 0.6pt,line join=round] ( 37.08, 74.76) --
	( 37.08, 73.24);

\path[draw=drawColor,line width= 0.6pt,line join=round] ( 34.73, 73.24) --
	( 39.43, 73.24);

\path[draw=drawColor,line width= 0.6pt,line join=round] ( 36.23, 75.99) --
	( 40.93, 75.99);

\path[draw=drawColor,line width= 0.6pt,line join=round] ( 38.58, 75.99) --
	( 38.58, 73.30);

\path[draw=drawColor,line width= 0.6pt,line join=round] ( 36.23, 73.30) --
	( 40.93, 73.30);

\path[draw=drawColor,line width= 0.6pt,line join=round] ( 39.24, 74.63) --
	( 43.94, 74.63);

\path[draw=drawColor,line width= 0.6pt,line join=round] ( 41.59, 74.63) --
	( 41.59, 73.33);

\path[draw=drawColor,line width= 0.6pt,line join=round] ( 39.24, 73.33) --
	( 43.94, 73.33);

\path[draw=drawColor,line width= 0.6pt,line join=round] ( 45.25, 74.97) --
	( 49.95, 74.97);

\path[draw=drawColor,line width= 0.6pt,line join=round] ( 47.60, 74.97) --
	( 47.60, 73.16);

\path[draw=drawColor,line width= 0.6pt,line join=round] ( 45.25, 73.16) --
	( 49.95, 73.16);

\path[draw=drawColor,line width= 0.6pt,line join=round] ( 57.29, 74.71) --
	( 61.99, 74.71);

\path[draw=drawColor,line width= 0.6pt,line join=round] ( 59.64, 74.71) --
	( 59.64, 73.06);

\path[draw=drawColor,line width= 0.6pt,line join=round] ( 57.29, 73.06) --
	( 61.99, 73.06);

\path[draw=drawColor,line width= 0.6pt,line join=round] ( 81.35, 74.64) --
	( 86.05, 74.64);

\path[draw=drawColor,line width= 0.6pt,line join=round] ( 83.70, 74.64) --
	( 83.70, 73.38);

\path[draw=drawColor,line width= 0.6pt,line join=round] ( 81.35, 73.38) --
	( 86.05, 73.38);

\path[draw=drawColor,line width= 0.6pt,line join=round] (129.48, 74.83) --
	(134.18, 74.83);

\path[draw=drawColor,line width= 0.6pt,line join=round] (131.83, 74.83) --
	(131.83, 73.36);

\path[draw=drawColor,line width= 0.6pt,line join=round] (129.48, 73.36) --
	(134.18, 73.36);

\path[draw=drawColor,line width= 0.6pt,line join=round] (225.73, 74.44) --
	(230.43, 74.44);

\path[draw=drawColor,line width= 0.6pt,line join=round] (228.08, 74.44) --
	(228.08, 73.48);

\path[draw=drawColor,line width= 0.6pt,line join=round] (225.73, 73.48) --
	(230.43, 73.48);
\definecolor{drawColor}{RGB}{228,26,28}

\path[draw=drawColor,line width= 0.6pt,line join=round] ( 34.73,105.91) --
	( 39.43,105.91);

\path[draw=drawColor,line width= 0.6pt,line join=round] ( 37.08,105.91) --
	( 37.08,103.34);

\path[draw=drawColor,line width= 0.6pt,line join=round] ( 34.73,103.34) --
	( 39.43,103.34);

\path[draw=drawColor,line width= 0.6pt,line join=round] ( 36.23,107.95) --
	( 40.93,107.95);

\path[draw=drawColor,line width= 0.6pt,line join=round] ( 38.58,107.95) --
	( 38.58,104.90);

\path[draw=drawColor,line width= 0.6pt,line join=round] ( 36.23,104.90) --
	( 40.93,104.90);

\path[draw=drawColor,line width= 0.6pt,line join=round] ( 39.24,111.76) --
	( 43.94,111.76);

\path[draw=drawColor,line width= 0.6pt,line join=round] ( 41.59,111.76) --
	( 41.59,107.12);

\path[draw=drawColor,line width= 0.6pt,line join=round] ( 39.24,107.12) --
	( 43.94,107.12);

\path[draw=drawColor,line width= 0.6pt,line join=round] ( 45.25,111.32) --
	( 49.95,111.32);

\path[draw=drawColor,line width= 0.6pt,line join=round] ( 47.60,111.32) --
	( 47.60,108.90);

\path[draw=drawColor,line width= 0.6pt,line join=round] ( 45.25,108.90) --
	( 49.95,108.90);

\path[draw=drawColor,line width= 0.6pt,line join=round] ( 57.29,122.11) --
	( 61.99,122.11);

\path[draw=drawColor,line width= 0.6pt,line join=round] ( 59.64,122.11) --
	( 59.64,118.61);

\path[draw=drawColor,line width= 0.6pt,line join=round] ( 57.29,118.61) --
	( 61.99,118.61);

\path[draw=drawColor,line width= 0.6pt,line join=round] ( 81.35,121.53) --
	( 86.05,121.53);

\path[draw=drawColor,line width= 0.6pt,line join=round] ( 83.70,121.53) --
	( 83.70,119.76);

\path[draw=drawColor,line width= 0.6pt,line join=round] ( 81.35,119.76) --
	( 86.05,119.76);

\path[draw=drawColor,line width= 0.6pt,line join=round] (129.48,127.62) --
	(134.18,127.62);

\path[draw=drawColor,line width= 0.6pt,line join=round] (131.83,127.62) --
	(131.83,123.57);

\path[draw=drawColor,line width= 0.6pt,line join=round] (129.48,123.57) --
	(134.18,123.57);

\path[draw=drawColor,line width= 0.6pt,line join=round] (225.73,138.72) --
	(230.43,138.72);

\path[draw=drawColor,line width= 0.6pt,line join=round] (228.08,138.72) --
	(228.08,135.37);

\path[draw=drawColor,line width= 0.6pt,line join=round] (225.73,135.37) --
	(230.43,135.37);
\definecolor{drawColor}{RGB}{55,126,184}

\path[draw=drawColor,line width= 0.6pt,line join=round] ( 34.73, 79.37) --
	( 39.43, 79.37);

\path[draw=drawColor,line width= 0.6pt,line join=round] ( 37.08, 79.37) --
	( 37.08, 78.42);

\path[draw=drawColor,line width= 0.6pt,line join=round] ( 34.73, 78.42) --
	( 39.43, 78.42);

\path[draw=drawColor,line width= 0.6pt,line join=round] ( 36.23, 79.58) --
	( 40.93, 79.58);

\path[draw=drawColor,line width= 0.6pt,line join=round] ( 38.58, 79.58) --
	( 38.58, 79.09);

\path[draw=drawColor,line width= 0.6pt,line join=round] ( 36.23, 79.09) --
	( 40.93, 79.09);

\path[draw=drawColor,line width= 0.6pt,line join=round] ( 39.24, 81.29) --
	( 43.94, 81.29);

\path[draw=drawColor,line width= 0.6pt,line join=round] ( 41.59, 81.29) --
	( 41.59, 80.20);

\path[draw=drawColor,line width= 0.6pt,line join=round] ( 39.24, 80.20) --
	( 43.94, 80.20);

\path[draw=drawColor,line width= 0.6pt,line join=round] ( 45.25, 81.67) --
	( 49.95, 81.67);

\path[draw=drawColor,line width= 0.6pt,line join=round] ( 47.60, 81.67) --
	( 47.60, 79.94);

\path[draw=drawColor,line width= 0.6pt,line join=round] ( 45.25, 79.94) --
	( 49.95, 79.94);

\path[draw=drawColor,line width= 0.6pt,line join=round] ( 57.29, 83.25) --
	( 61.99, 83.25);

\path[draw=drawColor,line width= 0.6pt,line join=round] ( 59.64, 83.25) --
	( 59.64, 82.47);

\path[draw=drawColor,line width= 0.6pt,line join=round] ( 57.29, 82.47) --
	( 61.99, 82.47);

\path[draw=drawColor,line width= 0.6pt,line join=round] ( 81.35, 83.08) --
	( 86.05, 83.08);

\path[draw=drawColor,line width= 0.6pt,line join=round] ( 83.70, 83.08) --
	( 83.70, 82.16);

\path[draw=drawColor,line width= 0.6pt,line join=round] ( 81.35, 82.16) --
	( 86.05, 82.16);

\path[draw=drawColor,line width= 0.6pt,line join=round] (129.48, 83.76) --
	(134.18, 83.76);

\path[draw=drawColor,line width= 0.6pt,line join=round] (131.83, 83.76) --
	(131.83, 82.90);

\path[draw=drawColor,line width= 0.6pt,line join=round] (129.48, 82.90) --
	(134.18, 82.90);

\path[draw=drawColor,line width= 0.6pt,line join=round] (225.73, 85.72) --
	(230.43, 85.72);

\path[draw=drawColor,line width= 0.6pt,line join=round] (228.08, 85.72) --
	(228.08, 84.46);

\path[draw=drawColor,line width= 0.6pt,line join=round] (225.73, 84.46) --
	(230.43, 84.46);
\definecolor{drawColor}{gray}{0.20}

\path[draw=drawColor,line width= 0.6pt,line join=round,line cap=round] ( 24.94, 28.25) rectangle (240.22,160.72);
\end{scope}
\begin{scope}
\path[clip] (  0.00,  0.00) rectangle (245.72,166.22);
\definecolor{drawColor}{gray}{0.30}

\node[text=drawColor,anchor=base east,inner sep=0pt, outer sep=0pt, scale=  0.72] at ( 19.99, 31.79) {0};

\node[text=drawColor,anchor=base east,inner sep=0pt, outer sep=0pt, scale=  0.72] at ( 19.99, 55.87) {1};

\node[text=drawColor,anchor=base east,inner sep=0pt, outer sep=0pt, scale=  0.72] at ( 19.99, 79.96) {2};

\node[text=drawColor,anchor=base east,inner sep=0pt, outer sep=0pt, scale=  0.72] at ( 19.99,104.05) {3};

\node[text=drawColor,anchor=base east,inner sep=0pt, outer sep=0pt, scale=  0.72] at ( 19.99,128.13) {4};

\node[text=drawColor,anchor=base east,inner sep=0pt, outer sep=0pt, scale=  0.72] at ( 19.99,152.22) {5};
\end{scope}
\begin{scope}
\path[clip] (  0.00,  0.00) rectangle (245.72,166.22);
\definecolor{drawColor}{gray}{0.20}

\path[draw=drawColor,line width= 0.6pt,line join=round] ( 22.19, 34.27) --
	( 24.94, 34.27);

\path[draw=drawColor,line width= 0.6pt,line join=round] ( 22.19, 58.35) --
	( 24.94, 58.35);

\path[draw=drawColor,line width= 0.6pt,line join=round] ( 22.19, 82.44) --
	( 24.94, 82.44);

\path[draw=drawColor,line width= 0.6pt,line join=round] ( 22.19,106.53) --
	( 24.94,106.53);

\path[draw=drawColor,line width= 0.6pt,line join=round] ( 22.19,130.61) --
	( 24.94,130.61);

\path[draw=drawColor,line width= 0.6pt,line join=round] ( 22.19,154.70) --
	( 24.94,154.70);
\end{scope}
\begin{scope}
\path[clip] (  0.00,  0.00) rectangle (245.72,166.22);
\definecolor{drawColor}{gray}{0.20}

\path[draw=drawColor,line width= 0.6pt,line join=round] ( 35.57, 25.50) --
	( 35.57, 28.25);

\path[draw=drawColor,line width= 0.6pt,line join=round] ( 82.57, 25.50) --
	( 82.57, 28.25);

\path[draw=drawColor,line width= 0.6pt,line join=round] (129.57, 25.50) --
	(129.57, 28.25);

\path[draw=drawColor,line width= 0.6pt,line join=round] (176.57, 25.50) --
	(176.57, 28.25);

\path[draw=drawColor,line width= 0.6pt,line join=round] (223.57, 25.50) --
	(223.57, 28.25);
\end{scope}
\begin{scope}
\path[clip] (  0.00,  0.00) rectangle (245.72,166.22);
\definecolor{drawColor}{gray}{0.30}

\node[text=drawColor,anchor=base,inner sep=0pt, outer sep=0pt, scale=  0.72] at ( 35.57, 18.34) {0};

\node[text=drawColor,anchor=base,inner sep=0pt, outer sep=0pt, scale=  0.72] at ( 82.57, 18.34) {2000};

\node[text=drawColor,anchor=base,inner sep=0pt, outer sep=0pt, scale=  0.72] at (129.57, 18.34) {4000};

\node[text=drawColor,anchor=base,inner sep=0pt, outer sep=0pt, scale=  0.72] at (176.57, 18.34) {6000};

\node[text=drawColor,anchor=base,inner sep=0pt, outer sep=0pt, scale=  0.72] at (223.57, 18.34) {8000};
\end{scope}
\begin{scope}
\path[clip] (  0.00,  0.00) rectangle (245.72,166.22);
\definecolor{drawColor}{RGB}{0,0,0}

\node[text=drawColor,anchor=base,inner sep=0pt, outer sep=0pt, scale=  0.90] at (132.58,  7.44) {Payload Size [bytes]};
\end{scope}
\begin{scope}
\path[clip] (  0.00,  0.00) rectangle (245.72,166.22);
\definecolor{drawColor}{RGB}{0,0,0}

\node[text=drawColor,rotate= 90.00,anchor=base,inner sep=0pt, outer sep=0pt, scale=  0.90] at ( 11.70, 94.48) {Sequence Handling Time [us]};
\end{scope}
\begin{scope}
\path[clip] (  0.00,  0.00) rectangle (245.72,166.22);
\definecolor{drawColor}{RGB}{0,0,0}
\definecolor{fillColor}{RGB}{255,255,255}

\path[draw=drawColor,line width= 0.3pt,line join=round,line cap=round,fill=fillColor] ( 27.94,137.08) rectangle (134.89,157.72);
\end{scope}
\begin{scope}
\path[clip] (  0.00,  0.00) rectangle (245.72,166.22);
\definecolor{fillColor}{RGB}{255,255,255}

\path[fill=fillColor] ( 33.44,142.59) rectangle ( 43.08,152.22);
\end{scope}
\begin{scope}
\path[clip] (  0.00,  0.00) rectangle (245.72,166.22);
\definecolor{drawColor}{RGB}{228,26,28}

\path[draw=drawColor,line width= 0.6pt,line join=round] ( 34.41,147.40) -- ( 42.11,147.40);
\end{scope}
\begin{scope}
\path[clip] (  0.00,  0.00) rectangle (245.72,166.22);
\definecolor{drawColor}{RGB}{228,26,28}

\path[draw=drawColor,line width= 0.6pt,line join=round,line cap=round] ( 35.55,144.69) -- ( 40.97,150.11);

\path[draw=drawColor,line width= 0.6pt,line join=round,line cap=round] ( 35.55,150.11) -- ( 40.97,144.69);
\end{scope}
\begin{scope}
\path[clip] (  0.00,  0.00) rectangle (245.72,166.22);
\definecolor{drawColor}{RGB}{228,26,28}

\path[draw=drawColor,line width= 0.6pt,line join=round] ( 34.41,147.40) -- ( 42.11,147.40);
\end{scope}
\begin{scope}
\path[clip] (  0.00,  0.00) rectangle (245.72,166.22);
\definecolor{fillColor}{RGB}{255,255,255}

\path[fill=fillColor] ( 70.39,142.59) rectangle ( 80.03,152.22);
\end{scope}
\begin{scope}
\path[clip] (  0.00,  0.00) rectangle (245.72,166.22);
\definecolor{drawColor}{RGB}{55,126,184}

\path[draw=drawColor,line width= 0.6pt,line join=round] ( 71.35,147.40) -- ( 79.06,147.40);
\end{scope}
\begin{scope}
\path[clip] (  0.00,  0.00) rectangle (245.72,166.22);
\definecolor{drawColor}{RGB}{55,126,184}

\path[draw=drawColor,line width= 0.6pt,line join=round,line cap=round] ( 71.37,147.40) --
	( 75.21,151.24) --
	( 79.04,147.40) --
	( 75.21,143.57) --
	( 71.37,147.40);
\end{scope}
\begin{scope}
\path[clip] (  0.00,  0.00) rectangle (245.72,166.22);
\definecolor{drawColor}{RGB}{55,126,184}

\path[draw=drawColor,line width= 0.6pt,line join=round] ( 71.35,147.40) -- ( 79.06,147.40);
\end{scope}
\begin{scope}
\path[clip] (  0.00,  0.00) rectangle (245.72,166.22);
\definecolor{fillColor}{RGB}{255,255,255}

\path[fill=fillColor] ( 94.12,142.59) rectangle (103.76,152.22);
\end{scope}
\begin{scope}
\path[clip] (  0.00,  0.00) rectangle (245.72,166.22);
\definecolor{drawColor}{RGB}{77,175,74}

\path[draw=drawColor,line width= 0.6pt,line join=round] ( 95.09,147.40) -- (102.80,147.40);
\end{scope}
\begin{scope}
\path[clip] (  0.00,  0.00) rectangle (245.72,166.22);
\definecolor{drawColor}{RGB}{77,175,74}

\path[draw=drawColor,line width= 0.6pt,line join=round,line cap=round] ( 98.94,143.19) --
	(102.59,149.51) --
	( 95.29,149.51) --
	( 98.94,143.19);
\end{scope}
\begin{scope}
\path[clip] (  0.00,  0.00) rectangle (245.72,166.22);
\definecolor{drawColor}{RGB}{77,175,74}

\path[draw=drawColor,line width= 0.6pt,line join=round] ( 95.09,147.40) -- (102.80,147.40);
\end{scope}
\begin{scope}
\path[clip] (  0.00,  0.00) rectangle (245.72,166.22);
\definecolor{drawColor}{RGB}{0,0,0}

\node[text=drawColor,anchor=base west,inner sep=0pt, outer sep=0pt, scale=  0.72] at ( 43.08,144.92) {Dropped};
\end{scope}
\begin{scope}
\path[clip] (  0.00,  0.00) rectangle (245.72,166.22);
\definecolor{drawColor}{RGB}{0,0,0}

\node[text=drawColor,anchor=base west,inner sep=0pt, outer sep=0pt, scale=  0.72] at ( 80.03,144.92) {Late};
\end{scope}
\begin{scope}
\path[clip] (  0.00,  0.00) rectangle (245.72,166.22);
\definecolor{drawColor}{RGB}{0,0,0}

\node[text=drawColor,anchor=base west,inner sep=0pt, outer sep=0pt, scale=  0.72] at (103.76,144.92) {Ordered};
\end{scope}
\end{tikzpicture}

%% file: pingpong.tikz
\begin{tikzpicture}[x=1pt,y=1pt]
\definecolor{fillColor}{RGB}{255,255,255}
\path[use as bounding box,fill=fillColor,fill opacity=0.00] (0,0) rectangle (245.72,166.22);
\begin{scope}
\path[clip] (  0.00,  0.00) rectangle (245.72,166.22);
\definecolor{drawColor}{RGB}{255,255,255}
\definecolor{fillColor}{RGB}{255,255,255}

\path[draw=drawColor,line width= 0.6pt,line join=round,line cap=round,fill=fillColor] ( -0.00,  0.00) rectangle (245.72,166.22);
\end{scope}
\begin{scope}
\path[clip] ( 28.54, 28.25) rectangle (240.22,160.72);
\definecolor{fillColor}{RGB}{255,255,255}

\path[fill=fillColor] ( 28.54, 28.25) rectangle (240.22,160.72);
\definecolor{drawColor}{gray}{0.92}

\path[draw=drawColor,line width= 0.3pt,line join=round] ( 28.54, 49.32) --
	(240.22, 49.32);

\path[draw=drawColor,line width= 0.3pt,line join=round] ( 28.54, 79.43) --
	(240.22, 79.43);

\path[draw=drawColor,line width= 0.3pt,line join=round] ( 28.54,109.54) --
	(240.22,109.54);

\path[draw=drawColor,line width= 0.3pt,line join=round] ( 28.54,139.65) --
	(240.22,139.65);

\path[draw=drawColor,line width= 0.3pt,line join=round] ( 30.62, 28.25) --
	( 30.62,160.72);

\path[draw=drawColor,line width= 0.3pt,line join=round] ( 49.48, 28.25) --
	( 49.48,160.72);

\path[draw=drawColor,line width= 0.3pt,line join=round] ( 68.35, 28.25) --
	( 68.35,160.72);

\path[draw=drawColor,line width= 0.3pt,line join=round] ( 87.21, 28.25) --
	( 87.21,160.72);

\path[draw=drawColor,line width= 0.3pt,line join=round] (106.08, 28.25) --
	(106.08,160.72);

\path[draw=drawColor,line width= 0.3pt,line join=round] (124.95, 28.25) --
	(124.95,160.72);

\path[draw=drawColor,line width= 0.3pt,line join=round] (143.81, 28.25) --
	(143.81,160.72);

\path[draw=drawColor,line width= 0.3pt,line join=round] (162.68, 28.25) --
	(162.68,160.72);

\path[draw=drawColor,line width= 0.3pt,line join=round] (181.54, 28.25) --
	(181.54,160.72);

\path[draw=drawColor,line width= 0.3pt,line join=round] (200.41, 28.25) --
	(200.41,160.72);

\path[draw=drawColor,line width= 0.3pt,line join=round] (219.28, 28.25) --
	(219.28,160.72);

\path[draw=drawColor,line width= 0.3pt,line join=round] (238.14, 28.25) --
	(238.14,160.72);

\path[draw=drawColor,line width= 0.6pt,line join=round] ( 28.54, 34.27) --
	(240.22, 34.27);

\path[draw=drawColor,line width= 0.6pt,line join=round] ( 28.54, 64.37) --
	(240.22, 64.37);

\path[draw=drawColor,line width= 0.6pt,line join=round] ( 28.54, 94.48) --
	(240.22, 94.48);

\path[draw=drawColor,line width= 0.6pt,line join=round] ( 28.54,124.59) --
	(240.22,124.59);

\path[draw=drawColor,line width= 0.6pt,line join=round] ( 28.54,154.70) --
	(240.22,154.70);

\path[draw=drawColor,line width= 0.6pt,line join=round] ( 40.05, 28.25) --
	( 40.05,160.72);

\path[draw=drawColor,line width= 0.6pt,line join=round] ( 58.92, 28.25) --
	( 58.92,160.72);

\path[draw=drawColor,line width= 0.6pt,line join=round] ( 77.78, 28.25) --
	( 77.78,160.72);

\path[draw=drawColor,line width= 0.6pt,line join=round] ( 96.65, 28.25) --
	( 96.65,160.72);

\path[draw=drawColor,line width= 0.6pt,line join=round] (115.51, 28.25) --
	(115.51,160.72);

\path[draw=drawColor,line width= 0.6pt,line join=round] (134.38, 28.25) --
	(134.38,160.72);

\path[draw=drawColor,line width= 0.6pt,line join=round] (153.25, 28.25) --
	(153.25,160.72);

\path[draw=drawColor,line width= 0.6pt,line join=round] (172.11, 28.25) --
	(172.11,160.72);

\path[draw=drawColor,line width= 0.6pt,line join=round] (190.98, 28.25) --
	(190.98,160.72);

\path[draw=drawColor,line width= 0.6pt,line join=round] (209.84, 28.25) --
	(209.84,160.72);

\path[draw=drawColor,line width= 0.6pt,line join=round] (228.71, 28.25) --
	(228.71,160.72);
\definecolor{drawColor}{RGB}{228,26,28}

\path[draw=drawColor,line width= 0.6pt,line join=round] ( 40.05, 35.19) --
	( 58.92, 41.13) --
	( 77.78, 46.39) --
	( 96.65, 54.67) --
	(115.51, 57.03) --
	(134.38, 60.60) --
	(153.25, 69.04) --
	(172.11, 71.86) --
	(190.98, 80.60) --
	(209.84, 86.02) --
	(228.71, 91.88);
\definecolor{drawColor}{RGB}{55,126,184}

\path[draw=drawColor,line width= 0.6pt,line join=round] ( 40.05, 35.22) --
	( 58.92, 40.18) --
	( 77.78, 46.85) --
	( 96.65, 54.04) --
	(115.51, 56.06) --
	(134.38, 60.34) --
	(153.25, 67.88) --
	(172.11, 75.08) --
	(190.98, 80.05) --
	(209.84, 86.33) --
	(228.71, 89.01);
\definecolor{drawColor}{RGB}{77,175,74}

\path[draw=drawColor,line width= 0.6pt,line join=round] ( 40.05, 34.89) --
	( 58.92, 41.67) --
	( 77.78, 46.40) --
	( 96.65, 54.89) --
	(115.51, 59.43) --
	(134.38, 70.37) --
	(153.25, 79.70) --
	(172.11, 92.31) --
	(190.98, 91.61) --
	(209.84,107.99) --
	(228.71,123.53);

\path[draw=drawColor,line width= 0.6pt,line join=round,line cap=round] ( 36.22, 34.89) -- ( 43.88, 34.89);

\path[draw=drawColor,line width= 0.6pt,line join=round,line cap=round] ( 40.05, 31.06) -- ( 40.05, 38.73);

\path[draw=drawColor,line width= 0.6pt,line join=round,line cap=round] ( 55.08, 41.67) -- ( 62.75, 41.67);

\path[draw=drawColor,line width= 0.6pt,line join=round,line cap=round] ( 58.92, 37.84) -- ( 58.92, 45.51);

\path[draw=drawColor,line width= 0.6pt,line join=round,line cap=round] ( 73.95, 46.40) -- ( 81.62, 46.40);

\path[draw=drawColor,line width= 0.6pt,line join=round,line cap=round] ( 77.78, 42.57) -- ( 77.78, 50.23);

\path[draw=drawColor,line width= 0.6pt,line join=round,line cap=round] ( 92.81, 54.89) -- (100.48, 54.89);

\path[draw=drawColor,line width= 0.6pt,line join=round,line cap=round] ( 96.65, 51.06) -- ( 96.65, 58.73);

\path[draw=drawColor,line width= 0.6pt,line join=round,line cap=round] (111.68, 59.43) -- (119.35, 59.43);

\path[draw=drawColor,line width= 0.6pt,line join=round,line cap=round] (115.51, 55.59) -- (115.51, 63.26);

\path[draw=drawColor,line width= 0.6pt,line join=round,line cap=round] (130.55, 70.37) -- (138.21, 70.37);

\path[draw=drawColor,line width= 0.6pt,line join=round,line cap=round] (134.38, 66.53) -- (134.38, 74.20);

\path[draw=drawColor,line width= 0.6pt,line join=round,line cap=round] (149.41, 79.70) -- (157.08, 79.70);

\path[draw=drawColor,line width= 0.6pt,line join=round,line cap=round] (153.25, 75.86) -- (153.25, 83.53);

\path[draw=drawColor,line width= 0.6pt,line join=round,line cap=round] (168.28, 92.31) -- (175.95, 92.31);

\path[draw=drawColor,line width= 0.6pt,line join=round,line cap=round] (172.11, 88.48) -- (172.11, 96.15);

\path[draw=drawColor,line width= 0.6pt,line join=round,line cap=round] (187.14, 91.61) -- (194.81, 91.61);

\path[draw=drawColor,line width= 0.6pt,line join=round,line cap=round] (190.98, 87.77) -- (190.98, 95.44);

\path[draw=drawColor,line width= 0.6pt,line join=round,line cap=round] (206.01,107.99) -- (213.68,107.99);

\path[draw=drawColor,line width= 0.6pt,line join=round,line cap=round] (209.84,104.15) -- (209.84,111.82);

\path[draw=drawColor,line width= 0.6pt,line join=round,line cap=round] (224.88,123.53) -- (232.54,123.53);

\path[draw=drawColor,line width= 0.6pt,line join=round,line cap=round] (228.71,119.69) -- (228.71,127.36);
\definecolor{drawColor}{RGB}{55,126,184}

\path[draw=drawColor,line width= 0.6pt,line join=round,line cap=round] ( 37.34, 32.51) -- ( 42.76, 37.93);

\path[draw=drawColor,line width= 0.6pt,line join=round,line cap=round] ( 37.34, 37.93) -- ( 42.76, 32.51);

\path[draw=drawColor,line width= 0.6pt,line join=round,line cap=round] ( 56.20, 37.47) -- ( 61.63, 42.89);

\path[draw=drawColor,line width= 0.6pt,line join=round,line cap=round] ( 56.20, 42.89) -- ( 61.63, 37.47);

\path[draw=drawColor,line width= 0.6pt,line join=round,line cap=round] ( 75.07, 44.14) -- ( 80.49, 49.56);

\path[draw=drawColor,line width= 0.6pt,line join=round,line cap=round] ( 75.07, 49.56) -- ( 80.49, 44.14);

\path[draw=drawColor,line width= 0.6pt,line join=round,line cap=round] ( 93.94, 51.33) -- ( 99.36, 56.75);

\path[draw=drawColor,line width= 0.6pt,line join=round,line cap=round] ( 93.94, 56.75) -- ( 99.36, 51.33);

\path[draw=drawColor,line width= 0.6pt,line join=round,line cap=round] (112.80, 53.35) -- (118.22, 58.77);

\path[draw=drawColor,line width= 0.6pt,line join=round,line cap=round] (112.80, 58.77) -- (118.22, 53.35);

\path[draw=drawColor,line width= 0.6pt,line join=round,line cap=round] (131.67, 57.63) -- (137.09, 63.05);

\path[draw=drawColor,line width= 0.6pt,line join=round,line cap=round] (131.67, 63.05) -- (137.09, 57.63);

\path[draw=drawColor,line width= 0.6pt,line join=round,line cap=round] (150.53, 65.17) -- (155.96, 70.59);

\path[draw=drawColor,line width= 0.6pt,line join=round,line cap=round] (150.53, 70.59) -- (155.96, 65.17);

\path[draw=drawColor,line width= 0.6pt,line join=round,line cap=round] (169.40, 72.37) -- (174.82, 77.79);

\path[draw=drawColor,line width= 0.6pt,line join=round,line cap=round] (169.40, 77.79) -- (174.82, 72.37);

\path[draw=drawColor,line width= 0.6pt,line join=round,line cap=round] (188.27, 77.34) -- (193.69, 82.76);

\path[draw=drawColor,line width= 0.6pt,line join=round,line cap=round] (188.27, 82.76) -- (193.69, 77.34);

\path[draw=drawColor,line width= 0.6pt,line join=round,line cap=round] (207.13, 83.62) -- (212.55, 89.04);

\path[draw=drawColor,line width= 0.6pt,line join=round,line cap=round] (207.13, 89.04) -- (212.55, 83.62);

\path[draw=drawColor,line width= 0.6pt,line join=round,line cap=round] (226.00, 86.30) -- (231.42, 91.72);

\path[draw=drawColor,line width= 0.6pt,line join=round,line cap=round] (226.00, 91.72) -- (231.42, 86.30);
\definecolor{drawColor}{RGB}{228,26,28}

\path[draw=drawColor,line width= 0.6pt,line join=round,line cap=round] ( 40.05, 35.19) circle (  2.71);

\path[draw=drawColor,line width= 0.6pt,line join=round,line cap=round] ( 58.92, 41.13) circle (  2.71);

\path[draw=drawColor,line width= 0.6pt,line join=round,line cap=round] ( 77.78, 46.39) circle (  2.71);

\path[draw=drawColor,line width= 0.6pt,line join=round,line cap=round] ( 96.65, 54.67) circle (  2.71);

\path[draw=drawColor,line width= 0.6pt,line join=round,line cap=round] (115.51, 57.03) circle (  2.71);

\path[draw=drawColor,line width= 0.6pt,line join=round,line cap=round] (134.38, 60.60) circle (  2.71);

\path[draw=drawColor,line width= 0.6pt,line join=round,line cap=round] (153.25, 69.04) circle (  2.71);

\path[draw=drawColor,line width= 0.6pt,line join=round,line cap=round] (172.11, 71.86) circle (  2.71);

\path[draw=drawColor,line width= 0.6pt,line join=round,line cap=round] (190.98, 80.60) circle (  2.71);

\path[draw=drawColor,line width= 0.6pt,line join=round,line cap=round] (209.84, 86.02) circle (  2.71);

\path[draw=drawColor,line width= 0.6pt,line join=round,line cap=round] (228.71, 91.88) circle (  2.71);
\definecolor{drawColor}{RGB}{77,175,74}

\path[draw=drawColor,line width= 0.6pt,line join=round] ( 38.16, 35.11) --
	( 41.94, 35.11);

\path[draw=drawColor,line width= 0.6pt,line join=round] ( 40.05, 35.11) --
	( 40.05, 34.68);

\path[draw=drawColor,line width= 0.6pt,line join=round] ( 38.16, 34.68) --
	( 41.94, 34.68);

\path[draw=drawColor,line width= 0.6pt,line join=round] ( 57.03, 43.88) --
	( 60.80, 43.88);

\path[draw=drawColor,line width= 0.6pt,line join=round] ( 58.92, 43.88) --
	( 58.92, 39.46);

\path[draw=drawColor,line width= 0.6pt,line join=round] ( 57.03, 39.46) --
	( 60.80, 39.46);

\path[draw=drawColor,line width= 0.6pt,line join=round] ( 75.89, 48.29) --
	( 79.67, 48.29);

\path[draw=drawColor,line width= 0.6pt,line join=round] ( 77.78, 48.29) --
	( 77.78, 44.51);

\path[draw=drawColor,line width= 0.6pt,line join=round] ( 75.89, 44.51) --
	( 79.67, 44.51);

\path[draw=drawColor,line width= 0.6pt,line join=round] ( 94.76, 61.07) --
	( 98.53, 61.07);

\path[draw=drawColor,line width= 0.6pt,line join=round] ( 96.65, 61.07) --
	( 96.65, 48.72);

\path[draw=drawColor,line width= 0.6pt,line join=round] ( 94.76, 48.72) --
	( 98.53, 48.72);

\path[draw=drawColor,line width= 0.6pt,line join=round] (113.63, 63.41) --
	(117.40, 63.41);

\path[draw=drawColor,line width= 0.6pt,line join=round] (115.51, 63.41) --
	(115.51, 55.44);

\path[draw=drawColor,line width= 0.6pt,line join=round] (113.63, 55.44) --
	(117.40, 55.44);

\path[draw=drawColor,line width= 0.6pt,line join=round] (132.49, 75.28) --
	(136.27, 75.28);

\path[draw=drawColor,line width= 0.6pt,line join=round] (134.38, 75.28) --
	(134.38, 65.45);

\path[draw=drawColor,line width= 0.6pt,line join=round] (132.49, 65.45) --
	(136.27, 65.45);

\path[draw=drawColor,line width= 0.6pt,line join=round] (151.36, 94.87) --
	(155.13, 94.87);

\path[draw=drawColor,line width= 0.6pt,line join=round] (153.25, 94.87) --
	(153.25, 64.52);

\path[draw=drawColor,line width= 0.6pt,line join=round] (151.36, 64.52) --
	(155.13, 64.52);

\path[draw=drawColor,line width= 0.6pt,line join=round] (170.22,107.17) --
	(174.00,107.17);

\path[draw=drawColor,line width= 0.6pt,line join=round] (172.11,107.17) --
	(172.11, 77.46);

\path[draw=drawColor,line width= 0.6pt,line join=round] (170.22, 77.46) --
	(174.00, 77.46);

\path[draw=drawColor,line width= 0.6pt,line join=round] (189.09, 97.83) --
	(192.86, 97.83);

\path[draw=drawColor,line width= 0.6pt,line join=round] (190.98, 97.83) --
	(190.98, 85.39);

\path[draw=drawColor,line width= 0.6pt,line join=round] (189.09, 85.39) --
	(192.86, 85.39);

\path[draw=drawColor,line width= 0.6pt,line join=round] (207.96,122.58) --
	(211.73,122.58);

\path[draw=drawColor,line width= 0.6pt,line join=round] (209.84,122.58) --
	(209.84, 93.39);

\path[draw=drawColor,line width= 0.6pt,line join=round] (207.96, 93.39) --
	(211.73, 93.39);

\path[draw=drawColor,line width= 0.6pt,line join=round] (226.82,147.49) --
	(230.60,147.49);

\path[draw=drawColor,line width= 0.6pt,line join=round] (228.71,147.49) --
	(228.71, 99.57);

\path[draw=drawColor,line width= 0.6pt,line join=round] (226.82, 99.57) --
	(230.60, 99.57);
\definecolor{drawColor}{RGB}{55,126,184}

\path[draw=drawColor,line width= 0.6pt,line join=round] ( 38.16, 35.39) --
	( 41.94, 35.39);

\path[draw=drawColor,line width= 0.6pt,line join=round] ( 40.05, 35.39) --
	( 40.05, 35.05);

\path[draw=drawColor,line width= 0.6pt,line join=round] ( 38.16, 35.05) --
	( 41.94, 35.05);

\path[draw=drawColor,line width= 0.6pt,line join=round] ( 57.03, 41.89) --
	( 60.80, 41.89);

\path[draw=drawColor,line width= 0.6pt,line join=round] ( 58.92, 41.89) --
	( 58.92, 38.48);

\path[draw=drawColor,line width= 0.6pt,line join=round] ( 57.03, 38.48) --
	( 60.80, 38.48);

\path[draw=drawColor,line width= 0.6pt,line join=round] ( 75.89, 51.37) --
	( 79.67, 51.37);

\path[draw=drawColor,line width= 0.6pt,line join=round] ( 77.78, 51.37) --
	( 77.78, 42.34);

\path[draw=drawColor,line width= 0.6pt,line join=round] ( 75.89, 42.34) --
	( 79.67, 42.34);

\path[draw=drawColor,line width= 0.6pt,line join=round] ( 94.76, 59.24) --
	( 98.53, 59.24);

\path[draw=drawColor,line width= 0.6pt,line join=round] ( 96.65, 59.24) --
	( 96.65, 48.83);

\path[draw=drawColor,line width= 0.6pt,line join=round] ( 94.76, 48.83) --
	( 98.53, 48.83);

\path[draw=drawColor,line width= 0.6pt,line join=round] (113.63, 59.45) --
	(117.40, 59.45);

\path[draw=drawColor,line width= 0.6pt,line join=round] (115.51, 59.45) --
	(115.51, 52.68);

\path[draw=drawColor,line width= 0.6pt,line join=round] (113.63, 52.68) --
	(117.40, 52.68);

\path[draw=drawColor,line width= 0.6pt,line join=round] (132.49, 61.82) --
	(136.27, 61.82);

\path[draw=drawColor,line width= 0.6pt,line join=round] (134.38, 61.82) --
	(134.38, 58.85);

\path[draw=drawColor,line width= 0.6pt,line join=round] (132.49, 58.85) --
	(136.27, 58.85);

\path[draw=drawColor,line width= 0.6pt,line join=round] (151.36, 72.99) --
	(155.13, 72.99);

\path[draw=drawColor,line width= 0.6pt,line join=round] (153.25, 72.99) --
	(153.25, 62.76);

\path[draw=drawColor,line width= 0.6pt,line join=round] (151.36, 62.76) --
	(155.13, 62.76);

\path[draw=drawColor,line width= 0.6pt,line join=round] (170.22, 80.70) --
	(174.00, 80.70);

\path[draw=drawColor,line width= 0.6pt,line join=round] (172.11, 80.70) --
	(172.11, 69.46);

\path[draw=drawColor,line width= 0.6pt,line join=round] (170.22, 69.46) --
	(174.00, 69.46);

\path[draw=drawColor,line width= 0.6pt,line join=round] (189.09, 87.16) --
	(192.86, 87.16);

\path[draw=drawColor,line width= 0.6pt,line join=round] (190.98, 87.16) --
	(190.98, 72.94);

\path[draw=drawColor,line width= 0.6pt,line join=round] (189.09, 72.94) --
	(192.86, 72.94);

\path[draw=drawColor,line width= 0.6pt,line join=round] (207.96, 91.18) --
	(211.73, 91.18);

\path[draw=drawColor,line width= 0.6pt,line join=round] (209.84, 91.18) --
	(209.84, 81.47);

\path[draw=drawColor,line width= 0.6pt,line join=round] (207.96, 81.47) --
	(211.73, 81.47);

\path[draw=drawColor,line width= 0.6pt,line join=round] (226.82, 91.80) --
	(230.60, 91.80);

\path[draw=drawColor,line width= 0.6pt,line join=round] (228.71, 91.80) --
	(228.71, 86.22);

\path[draw=drawColor,line width= 0.6pt,line join=round] (226.82, 86.22) --
	(230.60, 86.22);
\definecolor{drawColor}{RGB}{228,26,28}

\path[draw=drawColor,line width= 0.6pt,line join=round] ( 38.16, 35.47) --
	( 41.94, 35.47);

\path[draw=drawColor,line width= 0.6pt,line join=round] ( 40.05, 35.47) --
	( 40.05, 34.91);

\path[draw=drawColor,line width= 0.6pt,line join=round] ( 38.16, 34.91) --
	( 41.94, 34.91);

\path[draw=drawColor,line width= 0.6pt,line join=round] ( 57.03, 42.89) --
	( 60.80, 42.89);

\path[draw=drawColor,line width= 0.6pt,line join=round] ( 58.92, 42.89) --
	( 58.92, 39.38);

\path[draw=drawColor,line width= 0.6pt,line join=round] ( 57.03, 39.38) --
	( 60.80, 39.38);

\path[draw=drawColor,line width= 0.6pt,line join=round] ( 75.89, 49.15) --
	( 79.67, 49.15);

\path[draw=drawColor,line width= 0.6pt,line join=round] ( 77.78, 49.15) --
	( 77.78, 43.63);

\path[draw=drawColor,line width= 0.6pt,line join=round] ( 75.89, 43.63) --
	( 79.67, 43.63);

\path[draw=drawColor,line width= 0.6pt,line join=round] ( 94.76, 64.64) --
	( 98.53, 64.64);

\path[draw=drawColor,line width= 0.6pt,line join=round] ( 96.65, 64.64) --
	( 96.65, 44.69);

\path[draw=drawColor,line width= 0.6pt,line join=round] ( 94.76, 44.69) --
	( 98.53, 44.69);

\path[draw=drawColor,line width= 0.6pt,line join=round] (113.63, 59.52) --
	(117.40, 59.52);

\path[draw=drawColor,line width= 0.6pt,line join=round] (115.51, 59.52) --
	(115.51, 54.55);

\path[draw=drawColor,line width= 0.6pt,line join=round] (113.63, 54.55) --
	(117.40, 54.55);

\path[draw=drawColor,line width= 0.6pt,line join=round] (132.49, 62.78) --
	(136.27, 62.78);

\path[draw=drawColor,line width= 0.6pt,line join=round] (134.38, 62.78) --
	(134.38, 58.43);

\path[draw=drawColor,line width= 0.6pt,line join=round] (132.49, 58.43) --
	(136.27, 58.43);

\path[draw=drawColor,line width= 0.6pt,line join=round] (151.36, 73.78) --
	(155.13, 73.78);

\path[draw=drawColor,line width= 0.6pt,line join=round] (153.25, 73.78) --
	(153.25, 64.31);

\path[draw=drawColor,line width= 0.6pt,line join=round] (151.36, 64.31) --
	(155.13, 64.31);

\path[draw=drawColor,line width= 0.6pt,line join=round] (170.22, 73.78) --
	(174.00, 73.78);

\path[draw=drawColor,line width= 0.6pt,line join=round] (172.11, 73.78) --
	(172.11, 69.94);

\path[draw=drawColor,line width= 0.6pt,line join=round] (170.22, 69.94) --
	(174.00, 69.94);

\path[draw=drawColor,line width= 0.6pt,line join=round] (189.09, 87.15) --
	(192.86, 87.15);

\path[draw=drawColor,line width= 0.6pt,line join=round] (190.98, 87.15) --
	(190.98, 74.05);

\path[draw=drawColor,line width= 0.6pt,line join=round] (189.09, 74.05) --
	(192.86, 74.05);

\path[draw=drawColor,line width= 0.6pt,line join=round] (207.96, 92.81) --
	(211.73, 92.81);

\path[draw=drawColor,line width= 0.6pt,line join=round] (209.84, 92.81) --
	(209.84, 79.22);

\path[draw=drawColor,line width= 0.6pt,line join=round] (207.96, 79.22) --
	(211.73, 79.22);

\path[draw=drawColor,line width= 0.6pt,line join=round] (226.82,100.33) --
	(230.60,100.33);

\path[draw=drawColor,line width= 0.6pt,line join=round] (228.71,100.33) --
	(228.71, 83.44);

\path[draw=drawColor,line width= 0.6pt,line join=round] (226.82, 83.44) --
	(230.60, 83.44);
\definecolor{drawColor}{gray}{0.20}

\path[draw=drawColor,line width= 0.6pt,line join=round,line cap=round] ( 28.54, 28.25) rectangle (240.22,160.72);
\end{scope}
\begin{scope}
\path[clip] (  0.00,  0.00) rectangle (245.72,166.22);
\definecolor{drawColor}{gray}{0.30}

\node[text=drawColor,anchor=base east,inner sep=0pt, outer sep=0pt, scale=  0.72] at ( 23.59, 31.79) {0};

\node[text=drawColor,anchor=base east,inner sep=0pt, outer sep=0pt, scale=  0.72] at ( 23.59, 61.90) {10};

\node[text=drawColor,anchor=base east,inner sep=0pt, outer sep=0pt, scale=  0.72] at ( 23.59, 92.00) {20};

\node[text=drawColor,anchor=base east,inner sep=0pt, outer sep=0pt, scale=  0.72] at ( 23.59,122.11) {30};

\node[text=drawColor,anchor=base east,inner sep=0pt, outer sep=0pt, scale=  0.72] at ( 23.59,152.22) {40};
\end{scope}
\begin{scope}
\path[clip] (  0.00,  0.00) rectangle (245.72,166.22);
\definecolor{drawColor}{gray}{0.20}

\path[draw=drawColor,line width= 0.6pt,line join=round] ( 25.79, 34.27) --
	( 28.54, 34.27);

\path[draw=drawColor,line width= 0.6pt,line join=round] ( 25.79, 64.37) --
	( 28.54, 64.37);

\path[draw=drawColor,line width= 0.6pt,line join=round] ( 25.79, 94.48) --
	( 28.54, 94.48);

\path[draw=drawColor,line width= 0.6pt,line join=round] ( 25.79,124.59) --
	( 28.54,124.59);

\path[draw=drawColor,line width= 0.6pt,line join=round] ( 25.79,154.70) --
	( 28.54,154.70);
\end{scope}
\begin{scope}
\path[clip] (  0.00,  0.00) rectangle (245.72,166.22);
\definecolor{drawColor}{gray}{0.20}

\path[draw=drawColor,line width= 0.6pt,line join=round] ( 40.05, 25.50) --
	( 40.05, 28.25);

\path[draw=drawColor,line width= 0.6pt,line join=round] ( 58.92, 25.50) --
	( 58.92, 28.25);

\path[draw=drawColor,line width= 0.6pt,line join=round] ( 77.78, 25.50) --
	( 77.78, 28.25);

\path[draw=drawColor,line width= 0.6pt,line join=round] ( 96.65, 25.50) --
	( 96.65, 28.25);

\path[draw=drawColor,line width= 0.6pt,line join=round] (115.51, 25.50) --
	(115.51, 28.25);

\path[draw=drawColor,line width= 0.6pt,line join=round] (134.38, 25.50) --
	(134.38, 28.25);

\path[draw=drawColor,line width= 0.6pt,line join=round] (153.25, 25.50) --
	(153.25, 28.25);

\path[draw=drawColor,line width= 0.6pt,line join=round] (172.11, 25.50) --
	(172.11, 28.25);

\path[draw=drawColor,line width= 0.6pt,line join=round] (190.98, 25.50) --
	(190.98, 28.25);

\path[draw=drawColor,line width= 0.6pt,line join=round] (209.84, 25.50) --
	(209.84, 28.25);

\path[draw=drawColor,line width= 0.6pt,line join=round] (228.71, 25.50) --
	(228.71, 28.25);
\end{scope}
\begin{scope}
\path[clip] (  0.00,  0.00) rectangle (245.72,166.22);
\definecolor{drawColor}{gray}{0.30}

\node[text=drawColor,anchor=base,inner sep=0pt, outer sep=0pt, scale=  0.72] at ( 40.05, 18.34) {0};

\node[text=drawColor,anchor=base,inner sep=0pt, outer sep=0pt, scale=  0.72] at ( 58.92, 18.34) {1};

\node[text=drawColor,anchor=base,inner sep=0pt, outer sep=0pt, scale=  0.72] at ( 77.78, 18.34) {2};

\node[text=drawColor,anchor=base,inner sep=0pt, outer sep=0pt, scale=  0.72] at ( 96.65, 18.34) {3};

\node[text=drawColor,anchor=base,inner sep=0pt, outer sep=0pt, scale=  0.72] at (115.51, 18.34) {4};

\node[text=drawColor,anchor=base,inner sep=0pt, outer sep=0pt, scale=  0.72] at (134.38, 18.34) {5};

\node[text=drawColor,anchor=base,inner sep=0pt, outer sep=0pt, scale=  0.72] at (153.25, 18.34) {6};

\node[text=drawColor,anchor=base,inner sep=0pt, outer sep=0pt, scale=  0.72] at (172.11, 18.34) {7};

\node[text=drawColor,anchor=base,inner sep=0pt, outer sep=0pt, scale=  0.72] at (190.98, 18.34) {8};

\node[text=drawColor,anchor=base,inner sep=0pt, outer sep=0pt, scale=  0.72] at (209.84, 18.34) {9};

\node[text=drawColor,anchor=base,inner sep=0pt, outer sep=0pt, scale=  0.72] at (228.71, 18.34) {10};
\end{scope}
\begin{scope}
\path[clip] (  0.00,  0.00) rectangle (245.72,166.22);
\definecolor{drawColor}{RGB}{0,0,0}

\node[text=drawColor,anchor=base,inner sep=0pt, outer sep=0pt, scale=  0.90] at (134.38,  7.44) {Relative Loss on Link [{\%}]};
\end{scope}
\begin{scope}
\path[clip] (  0.00,  0.00) rectangle (245.72,166.22);
\definecolor{drawColor}{RGB}{0,0,0}

\node[text=drawColor,rotate= 90.00,anchor=base,inner sep=0pt, outer sep=0pt, scale=  0.90] at ( 11.70, 94.48) {Runtime [s]};
\end{scope}
\begin{scope}
\path[clip] (  0.00,  0.00) rectangle (245.72,166.22);
\definecolor{drawColor}{RGB}{0,0,0}
\definecolor{fillColor}{RGB}{255,255,255}

\path[draw=drawColor,line width= 0.3pt,line join=round,line cap=round,fill=fillColor] ( 31.54,117.81) rectangle (123.70,157.72);
\end{scope}
\begin{scope}
\path[clip] (  0.00,  0.00) rectangle (245.72,166.22);
\definecolor{fillColor}{RGB}{255,255,255}

\path[fill=fillColor] ( 37.04,142.59) rectangle ( 46.68,152.22);
\end{scope}
\begin{scope}
\path[clip] (  0.00,  0.00) rectangle (245.72,166.22);
\definecolor{drawColor}{RGB}{228,26,28}

\path[draw=drawColor,line width= 0.6pt,line join=round] ( 38.00,147.40) -- ( 45.71,147.40);
\end{scope}
\begin{scope}
\path[clip] (  0.00,  0.00) rectangle (245.72,166.22);
\definecolor{drawColor}{RGB}{228,26,28}

\path[draw=drawColor,line width= 0.6pt,line join=round,line cap=round] ( 41.86,147.40) circle (  2.71);
\end{scope}
\begin{scope}
\path[clip] (  0.00,  0.00) rectangle (245.72,166.22);
\definecolor{drawColor}{RGB}{228,26,28}

\path[draw=drawColor,line width= 0.6pt,line join=round] ( 38.00,147.40) -- ( 45.71,147.40);
\end{scope}
\begin{scope}
\path[clip] (  0.00,  0.00) rectangle (245.72,166.22);
\definecolor{fillColor}{RGB}{255,255,255}

\path[fill=fillColor] ( 37.04,132.95) rectangle ( 46.68,142.58);
\end{scope}
\begin{scope}
\path[clip] (  0.00,  0.00) rectangle (245.72,166.22);
\definecolor{drawColor}{RGB}{55,126,184}

\path[draw=drawColor,line width= 0.6pt,line join=round] ( 38.00,137.77) -- ( 45.71,137.77);
\end{scope}
\begin{scope}
\path[clip] (  0.00,  0.00) rectangle (245.72,166.22);
\definecolor{drawColor}{RGB}{55,126,184}

\path[draw=drawColor,line width= 0.6pt,line join=round,line cap=round] ( 39.15,135.06) -- ( 44.57,140.48);

\path[draw=drawColor,line width= 0.6pt,line join=round,line cap=round] ( 39.15,140.48) -- ( 44.57,135.06);
\end{scope}
\begin{scope}
\path[clip] (  0.00,  0.00) rectangle (245.72,166.22);
\definecolor{drawColor}{RGB}{55,126,184}

\path[draw=drawColor,line width= 0.6pt,line join=round] ( 38.00,137.77) -- ( 45.71,137.77);
\end{scope}
\begin{scope}
\path[clip] (  0.00,  0.00) rectangle (245.72,166.22);
\definecolor{fillColor}{RGB}{255,255,255}

\path[fill=fillColor] ( 37.04,123.31) rectangle ( 46.68,132.95);
\end{scope}
\begin{scope}
\path[clip] (  0.00,  0.00) rectangle (245.72,166.22);
\definecolor{drawColor}{RGB}{77,175,74}

\path[draw=drawColor,line width= 0.6pt,line join=round] ( 38.00,128.13) -- ( 45.71,128.13);
\end{scope}
\begin{scope}
\path[clip] (  0.00,  0.00) rectangle (245.72,166.22);
\definecolor{drawColor}{RGB}{77,175,74}

\path[draw=drawColor,line width= 0.6pt,line join=round,line cap=round] ( 38.02,128.13) -- ( 45.69,128.13);

\path[draw=drawColor,line width= 0.6pt,line join=round,line cap=round] ( 41.86,124.30) -- ( 41.86,131.96);
\end{scope}
\begin{scope}
\path[clip] (  0.00,  0.00) rectangle (245.72,166.22);
\definecolor{drawColor}{RGB}{77,175,74}

\path[draw=drawColor,line width= 0.6pt,line join=round] ( 38.00,128.13) -- ( 45.71,128.13);
\end{scope}
\begin{scope}
\path[clip] (  0.00,  0.00) rectangle (245.72,166.22);
\definecolor{drawColor}{RGB}{0,0,0}

\node[text=drawColor,anchor=base west,inner sep=0pt, outer sep=0pt, scale=  0.72] at ( 46.68,144.92) {Reliable Ordered UDP};
\end{scope}
\begin{scope}
\path[clip] (  0.00,  0.00) rectangle (245.72,166.22);
\definecolor{drawColor}{RGB}{0,0,0}

\node[text=drawColor,anchor=base west,inner sep=0pt, outer sep=0pt, scale=  0.72] at ( 46.68,135.29) {Reliable UDP};
\end{scope}
\begin{scope}
\path[clip] (  0.00,  0.00) rectangle (245.72,166.22);
\definecolor{drawColor}{RGB}{0,0,0}

\node[text=drawColor,anchor=base west,inner sep=0pt, outer sep=0pt, scale=  0.72] at ( 46.68,125.65) {TCP};
\end{scope}
\end{tikzpicture}

%% file: pingpong-10.tikz
\begin{tikzpicture}[x=1pt,y=1pt]
\definecolor{fillColor}{RGB}{255,255,255}
\path[use as bounding box,fill=fillColor,fill opacity=0.00] (0,0) rectangle (245.72,166.22);
\begin{scope}
\path[clip] (  0.00,  0.00) rectangle (245.72,166.22);
\definecolor{drawColor}{RGB}{255,255,255}
\definecolor{fillColor}{RGB}{255,255,255}

\path[draw=drawColor,line width= 0.6pt,line join=round,line cap=round,fill=fillColor] (  0.00,  0.00) rectangle (245.72,166.22);
\end{scope}
\begin{scope}
\path[clip] ( 32.14, 28.25) rectangle (240.22,160.72);
\definecolor{fillColor}{RGB}{255,255,255}

\path[fill=fillColor] ( 32.14, 28.25) rectangle (240.22,160.72);
\definecolor{drawColor}{gray}{0.92}

\path[draw=drawColor,line width= 0.3pt,line join=round] ( 32.14, 28.79) --
	(240.22, 28.79);

\path[draw=drawColor,line width= 0.3pt,line join=round] ( 32.14, 39.74) --
	(240.22, 39.74);

\path[draw=drawColor,line width= 0.3pt,line join=round] ( 32.14, 50.69) --
	(240.22, 50.69);

\path[draw=drawColor,line width= 0.3pt,line join=round] ( 32.14, 61.64) --
	(240.22, 61.64);

\path[draw=drawColor,line width= 0.3pt,line join=round] ( 32.14, 72.59) --
	(240.22, 72.59);

\path[draw=drawColor,line width= 0.3pt,line join=round] ( 32.14, 83.53) --
	(240.22, 83.53);

\path[draw=drawColor,line width= 0.3pt,line join=round] ( 32.14, 94.48) --
	(240.22, 94.48);

\path[draw=drawColor,line width= 0.3pt,line join=round] ( 32.14,105.43) --
	(240.22,105.43);

\path[draw=drawColor,line width= 0.3pt,line join=round] ( 32.14,116.38) --
	(240.22,116.38);

\path[draw=drawColor,line width= 0.3pt,line join=round] ( 32.14,127.33) --
	(240.22,127.33);

\path[draw=drawColor,line width= 0.3pt,line join=round] ( 32.14,138.28) --
	(240.22,138.28);

\path[draw=drawColor,line width= 0.3pt,line join=round] ( 32.14,149.23) --
	(240.22,149.23);

\path[draw=drawColor,line width= 0.3pt,line join=round] ( 32.14,160.17) --
	(240.22,160.17);

\path[draw=drawColor,line width= 0.3pt,line join=round] ( 34.18, 28.25) --
	( 34.18,160.72);

\path[draw=drawColor,line width= 0.3pt,line join=round] ( 52.73, 28.25) --
	( 52.73,160.72);

\path[draw=drawColor,line width= 0.3pt,line join=round] ( 71.27, 28.25) --
	( 71.27,160.72);

\path[draw=drawColor,line width= 0.3pt,line join=round] ( 89.82, 28.25) --
	( 89.82,160.72);

\path[draw=drawColor,line width= 0.3pt,line join=round] (108.36, 28.25) --
	(108.36,160.72);

\path[draw=drawColor,line width= 0.3pt,line join=round] (126.91, 28.25) --
	(126.91,160.72);

\path[draw=drawColor,line width= 0.3pt,line join=round] (145.45, 28.25) --
	(145.45,160.72);

\path[draw=drawColor,line width= 0.3pt,line join=round] (164.00, 28.25) --
	(164.00,160.72);

\path[draw=drawColor,line width= 0.3pt,line join=round] (182.54, 28.25) --
	(182.54,160.72);

\path[draw=drawColor,line width= 0.3pt,line join=round] (201.09, 28.25) --
	(201.09,160.72);

\path[draw=drawColor,line width= 0.3pt,line join=round] (219.63, 28.25) --
	(219.63,160.72);

\path[draw=drawColor,line width= 0.3pt,line join=round] (238.18, 28.25) --
	(238.18,160.72);

\path[draw=drawColor,line width= 0.6pt,line join=round] ( 32.14, 34.27) --
	(240.22, 34.27);

\path[draw=drawColor,line width= 0.6pt,line join=round] ( 32.14, 45.22) --
	(240.22, 45.22);

\path[draw=drawColor,line width= 0.6pt,line join=round] ( 32.14, 56.16) --
	(240.22, 56.16);

\path[draw=drawColor,line width= 0.6pt,line join=round] ( 32.14, 67.11) --
	(240.22, 67.11);

\path[draw=drawColor,line width= 0.6pt,line join=round] ( 32.14, 78.06) --
	(240.22, 78.06);

\path[draw=drawColor,line width= 0.6pt,line join=round] ( 32.14, 89.01) --
	(240.22, 89.01);

\path[draw=drawColor,line width= 0.6pt,line join=round] ( 32.14, 99.96) --
	(240.22, 99.96);

\path[draw=drawColor,line width= 0.6pt,line join=round] ( 32.14,110.91) --
	(240.22,110.91);

\path[draw=drawColor,line width= 0.6pt,line join=round] ( 32.14,121.85) --
	(240.22,121.85);

\path[draw=drawColor,line width= 0.6pt,line join=round] ( 32.14,132.80) --
	(240.22,132.80);

\path[draw=drawColor,line width= 0.6pt,line join=round] ( 32.14,143.75) --
	(240.22,143.75);

\path[draw=drawColor,line width= 0.6pt,line join=round] ( 32.14,154.70) --
	(240.22,154.70);

\path[draw=drawColor,line width= 0.6pt,line join=round] ( 43.45, 28.25) --
	( 43.45,160.72);

\path[draw=drawColor,line width= 0.6pt,line join=round] ( 62.00, 28.25) --
	( 62.00,160.72);

\path[draw=drawColor,line width= 0.6pt,line join=round] ( 80.54, 28.25) --
	( 80.54,160.72);

\path[draw=drawColor,line width= 0.6pt,line join=round] ( 99.09, 28.25) --
	( 99.09,160.72);

\path[draw=drawColor,line width= 0.6pt,line join=round] (117.63, 28.25) --
	(117.63,160.72);

\path[draw=drawColor,line width= 0.6pt,line join=round] (136.18, 28.25) --
	(136.18,160.72);

\path[draw=drawColor,line width= 0.6pt,line join=round] (154.72, 28.25) --
	(154.72,160.72);

\path[draw=drawColor,line width= 0.6pt,line join=round] (173.27, 28.25) --
	(173.27,160.72);

\path[draw=drawColor,line width= 0.6pt,line join=round] (191.81, 28.25) --
	(191.81,160.72);

\path[draw=drawColor,line width= 0.6pt,line join=round] (210.36, 28.25) --
	(210.36,160.72);

\path[draw=drawColor,line width= 0.6pt,line join=round] (228.91, 28.25) --
	(228.91,160.72);
\definecolor{drawColor}{RGB}{228,26,28}

\path[draw=drawColor,line width= 0.6pt,line join=round] ( 43.45, 80.41) --
	( 62.00, 80.69) --
	( 80.54, 84.35) --
	( 99.09, 84.77) --
	(117.63, 86.81) --
	(136.18, 89.59) --
	(154.72, 90.14) --
	(173.27, 93.06) --
	(191.81, 94.02) --
	(210.36, 97.05) --
	(228.91, 98.91);
\definecolor{drawColor}{RGB}{55,126,184}

\path[draw=drawColor,line width= 0.6pt,line join=round] ( 43.45, 79.44) --
	( 62.00, 81.21) --
	( 80.54, 82.92) --
	( 99.09, 84.09) --
	(117.63, 86.74) --
	(136.18, 88.76) --
	(154.72, 90.11) --
	(173.27, 93.50) --
	(191.81, 94.44) --
	(210.36,100.38) --
	(228.91,100.29);
\definecolor{drawColor}{RGB}{77,175,74}

\path[draw=drawColor,line width= 0.6pt,line join=round] ( 43.45, 79.36) --
	( 62.00, 82.57) --
	( 80.54, 89.74) --
	( 99.09, 93.65) --
	(117.63, 99.29) --
	(136.18,103.40) --
	(154.72,114.05) --
	(173.27,104.69) --
	(191.81,120.55) --
	(210.36,113.61) --
	(228.91,126.61);

\path[draw=drawColor,line width= 0.6pt,line join=round,line cap=round] ( 39.62, 79.36) -- ( 47.29, 79.36);

\path[draw=drawColor,line width= 0.6pt,line join=round,line cap=round] ( 43.45, 75.52) -- ( 43.45, 83.19);

\path[draw=drawColor,line width= 0.6pt,line join=round,line cap=round] ( 58.16, 82.57) -- ( 65.83, 82.57);

\path[draw=drawColor,line width= 0.6pt,line join=round,line cap=round] ( 62.00, 78.74) -- ( 62.00, 86.40);

\path[draw=drawColor,line width= 0.6pt,line join=round,line cap=round] ( 76.71, 89.74) -- ( 84.38, 89.74);

\path[draw=drawColor,line width= 0.6pt,line join=round,line cap=round] ( 80.54, 85.90) -- ( 80.54, 93.57);

\path[draw=drawColor,line width= 0.6pt,line join=round,line cap=round] ( 95.25, 93.65) -- (102.92, 93.65);

\path[draw=drawColor,line width= 0.6pt,line join=round,line cap=round] ( 99.09, 89.82) -- ( 99.09, 97.49);

\path[draw=drawColor,line width= 0.6pt,line join=round,line cap=round] (113.80, 99.29) -- (121.47, 99.29);

\path[draw=drawColor,line width= 0.6pt,line join=round,line cap=round] (117.63, 95.45) -- (117.63,103.12);

\path[draw=drawColor,line width= 0.6pt,line join=round,line cap=round] (132.34,103.40) -- (140.01,103.40);

\path[draw=drawColor,line width= 0.6pt,line join=round,line cap=round] (136.18, 99.56) -- (136.18,107.23);

\path[draw=drawColor,line width= 0.6pt,line join=round,line cap=round] (150.89,114.05) -- (158.56,114.05);

\path[draw=drawColor,line width= 0.6pt,line join=round,line cap=round] (154.72,110.22) -- (154.72,117.89);

\path[draw=drawColor,line width= 0.6pt,line join=round,line cap=round] (169.44,104.69) -- (177.10,104.69);

\path[draw=drawColor,line width= 0.6pt,line join=round,line cap=round] (173.27,100.86) -- (173.27,108.53);

\path[draw=drawColor,line width= 0.6pt,line join=round,line cap=round] (187.98,120.55) -- (195.65,120.55);

\path[draw=drawColor,line width= 0.6pt,line join=round,line cap=round] (191.81,116.72) -- (191.81,124.39);

\path[draw=drawColor,line width= 0.6pt,line join=round,line cap=round] (206.53,113.61) -- (214.19,113.61);

\path[draw=drawColor,line width= 0.6pt,line join=round,line cap=round] (210.36,109.78) -- (210.36,117.44);

\path[draw=drawColor,line width= 0.6pt,line join=round,line cap=round] (225.07,126.61) -- (232.74,126.61);

\path[draw=drawColor,line width= 0.6pt,line join=round,line cap=round] (228.91,122.78) -- (228.91,130.45);
\definecolor{drawColor}{RGB}{55,126,184}

\path[draw=drawColor,line width= 0.6pt,line join=round,line cap=round] ( 40.74, 76.73) -- ( 46.16, 82.16);

\path[draw=drawColor,line width= 0.6pt,line join=round,line cap=round] ( 40.74, 82.16) -- ( 46.16, 76.73);

\path[draw=drawColor,line width= 0.6pt,line join=round,line cap=round] ( 59.29, 78.50) -- ( 64.71, 83.92);

\path[draw=drawColor,line width= 0.6pt,line join=round,line cap=round] ( 59.29, 83.92) -- ( 64.71, 78.50);

\path[draw=drawColor,line width= 0.6pt,line join=round,line cap=round] ( 77.83, 80.21) -- ( 83.25, 85.63);

\path[draw=drawColor,line width= 0.6pt,line join=round,line cap=round] ( 77.83, 85.63) -- ( 83.25, 80.21);

\path[draw=drawColor,line width= 0.6pt,line join=round,line cap=round] ( 96.38, 81.38) -- (101.80, 86.80);

\path[draw=drawColor,line width= 0.6pt,line join=round,line cap=round] ( 96.38, 86.80) -- (101.80, 81.38);

\path[draw=drawColor,line width= 0.6pt,line join=round,line cap=round] (114.92, 84.03) -- (120.34, 89.45);

\path[draw=drawColor,line width= 0.6pt,line join=round,line cap=round] (114.92, 89.45) -- (120.34, 84.03);

\path[draw=drawColor,line width= 0.6pt,line join=round,line cap=round] (133.47, 86.04) -- (138.89, 91.47);

\path[draw=drawColor,line width= 0.6pt,line join=round,line cap=round] (133.47, 91.47) -- (138.89, 86.04);

\path[draw=drawColor,line width= 0.6pt,line join=round,line cap=round] (152.01, 87.40) -- (157.44, 92.82);

\path[draw=drawColor,line width= 0.6pt,line join=round,line cap=round] (152.01, 92.82) -- (157.44, 87.40);

\path[draw=drawColor,line width= 0.6pt,line join=round,line cap=round] (170.56, 90.79) -- (175.98, 96.21);

\path[draw=drawColor,line width= 0.6pt,line join=round,line cap=round] (170.56, 96.21) -- (175.98, 90.79);

\path[draw=drawColor,line width= 0.6pt,line join=round,line cap=round] (189.10, 91.73) -- (194.53, 97.15);

\path[draw=drawColor,line width= 0.6pt,line join=round,line cap=round] (189.10, 97.15) -- (194.53, 91.73);

\path[draw=drawColor,line width= 0.6pt,line join=round,line cap=round] (207.65, 97.67) -- (213.07,103.09);

\path[draw=drawColor,line width= 0.6pt,line join=round,line cap=round] (207.65,103.09) -- (213.07, 97.67);

\path[draw=drawColor,line width= 0.6pt,line join=round,line cap=round] (226.19, 97.58) -- (231.62,103.00);

\path[draw=drawColor,line width= 0.6pt,line join=round,line cap=round] (226.19,103.00) -- (231.62, 97.58);
\definecolor{drawColor}{RGB}{228,26,28}

\path[draw=drawColor,line width= 0.6pt,line join=round,line cap=round] ( 43.45, 80.41) circle (  2.71);

\path[draw=drawColor,line width= 0.6pt,line join=round,line cap=round] ( 62.00, 80.69) circle (  2.71);

\path[draw=drawColor,line width= 0.6pt,line join=round,line cap=round] ( 80.54, 84.35) circle (  2.71);

\path[draw=drawColor,line width= 0.6pt,line join=round,line cap=round] ( 99.09, 84.77) circle (  2.71);

\path[draw=drawColor,line width= 0.6pt,line join=round,line cap=round] (117.63, 86.81) circle (  2.71);

\path[draw=drawColor,line width= 0.6pt,line join=round,line cap=round] (136.18, 89.59) circle (  2.71);

\path[draw=drawColor,line width= 0.6pt,line join=round,line cap=round] (154.72, 90.14) circle (  2.71);

\path[draw=drawColor,line width= 0.6pt,line join=round,line cap=round] (173.27, 93.06) circle (  2.71);

\path[draw=drawColor,line width= 0.6pt,line join=round,line cap=round] (191.81, 94.02) circle (  2.71);

\path[draw=drawColor,line width= 0.6pt,line join=round,line cap=round] (210.36, 97.05) circle (  2.71);

\path[draw=drawColor,line width= 0.6pt,line join=round,line cap=round] (228.91, 98.91) circle (  2.71);
\definecolor{drawColor}{RGB}{77,175,74}

\path[draw=drawColor,line width= 0.6pt,line join=round] ( 41.60, 80.72) --
	( 45.31, 80.72);

\path[draw=drawColor,line width= 0.6pt,line join=round] ( 43.45, 80.72) --
	( 43.45, 77.99);

\path[draw=drawColor,line width= 0.6pt,line join=round] ( 41.60, 77.99) --
	( 45.31, 77.99);

\path[draw=drawColor,line width= 0.6pt,line join=round] ( 60.14, 84.11) --
	( 63.85, 84.11);

\path[draw=drawColor,line width= 0.6pt,line join=round] ( 62.00, 84.11) --
	( 62.00, 81.03);

\path[draw=drawColor,line width= 0.6pt,line join=round] ( 60.14, 81.03) --
	( 63.85, 81.03);

\path[draw=drawColor,line width= 0.6pt,line join=round] ( 78.69, 97.32) --
	( 82.40, 97.32);

\path[draw=drawColor,line width= 0.6pt,line join=round] ( 80.54, 97.32) --
	( 80.54, 82.15);

\path[draw=drawColor,line width= 0.6pt,line join=round] ( 78.69, 82.15) --
	( 82.40, 82.15);

\path[draw=drawColor,line width= 0.6pt,line join=round] ( 97.23,104.99) --
	(100.94,104.99);

\path[draw=drawColor,line width= 0.6pt,line join=round] ( 99.09,104.99) --
	( 99.09, 82.32);

\path[draw=drawColor,line width= 0.6pt,line join=round] ( 97.23, 82.32) --
	(100.94, 82.32);

\path[draw=drawColor,line width= 0.6pt,line join=round] (115.78,108.88) --
	(119.49,108.88);

\path[draw=drawColor,line width= 0.6pt,line join=round] (117.63,108.88) --
	(117.63, 89.69);

\path[draw=drawColor,line width= 0.6pt,line join=round] (115.78, 89.69) --
	(119.49, 89.69);

\path[draw=drawColor,line width= 0.6pt,line join=round] (134.32,116.40) --
	(138.03,116.40);

\path[draw=drawColor,line width= 0.6pt,line join=round] (136.18,116.40) --
	(136.18, 90.40);

\path[draw=drawColor,line width= 0.6pt,line join=round] (134.32, 90.40) --
	(138.03, 90.40);

\path[draw=drawColor,line width= 0.6pt,line join=round] (152.87,143.12) --
	(156.58,143.12);

\path[draw=drawColor,line width= 0.6pt,line join=round] (154.72,143.12) --
	(154.72, 84.98);

\path[draw=drawColor,line width= 0.6pt,line join=round] (152.87, 84.98) --
	(156.58, 84.98);

\path[draw=drawColor,line width= 0.6pt,line join=round] (171.41,108.02) --
	(175.12,108.02);

\path[draw=drawColor,line width= 0.6pt,line join=round] (173.27,108.02) --
	(173.27,101.36);

\path[draw=drawColor,line width= 0.6pt,line join=round] (171.41,101.36) --
	(175.12,101.36);

\path[draw=drawColor,line width= 0.6pt,line join=round] (189.96,142.30) --
	(193.67,142.30);

\path[draw=drawColor,line width= 0.6pt,line join=round] (191.81,142.30) --
	(191.81, 98.80);

\path[draw=drawColor,line width= 0.6pt,line join=round] (189.96, 98.80) --
	(193.67, 98.80);

\path[draw=drawColor,line width= 0.6pt,line join=round] (208.51,116.39) --
	(212.21,116.39);

\path[draw=drawColor,line width= 0.6pt,line join=round] (210.36,116.39) --
	(210.36,110.83);

\path[draw=drawColor,line width= 0.6pt,line join=round] (208.51,110.83) --
	(212.21,110.83);

\path[draw=drawColor,line width= 0.6pt,line join=round] (227.05,138.68) --
	(230.76,138.68);

\path[draw=drawColor,line width= 0.6pt,line join=round] (228.91,138.68) --
	(228.91,114.54);

\path[draw=drawColor,line width= 0.6pt,line join=round] (227.05,114.54) --
	(230.76,114.54);
\definecolor{drawColor}{RGB}{55,126,184}

\path[draw=drawColor,line width= 0.6pt,line join=round] ( 41.60, 81.24) --
	( 45.31, 81.24);

\path[draw=drawColor,line width= 0.6pt,line join=round] ( 43.45, 81.24) --
	( 43.45, 77.65);

\path[draw=drawColor,line width= 0.6pt,line join=round] ( 41.60, 77.65) --
	( 45.31, 77.65);

\path[draw=drawColor,line width= 0.6pt,line join=round] ( 60.14, 83.50) --
	( 63.85, 83.50);

\path[draw=drawColor,line width= 0.6pt,line join=round] ( 62.00, 83.50) --
	( 62.00, 78.91);

\path[draw=drawColor,line width= 0.6pt,line join=round] ( 60.14, 78.91) --
	( 63.85, 78.91);

\path[draw=drawColor,line width= 0.6pt,line join=round] ( 78.69, 85.18) --
	( 82.40, 85.18);

\path[draw=drawColor,line width= 0.6pt,line join=round] ( 80.54, 85.18) --
	( 80.54, 80.66);

\path[draw=drawColor,line width= 0.6pt,line join=round] ( 78.69, 80.66) --
	( 82.40, 80.66);

\path[draw=drawColor,line width= 0.6pt,line join=round] ( 97.23, 84.59) --
	(100.94, 84.59);

\path[draw=drawColor,line width= 0.6pt,line join=round] ( 99.09, 84.59) --
	( 99.09, 83.59);

\path[draw=drawColor,line width= 0.6pt,line join=round] ( 97.23, 83.59) --
	(100.94, 83.59);

\path[draw=drawColor,line width= 0.6pt,line join=round] (115.78, 88.78) --
	(119.49, 88.78);

\path[draw=drawColor,line width= 0.6pt,line join=round] (117.63, 88.78) --
	(117.63, 84.70);

\path[draw=drawColor,line width= 0.6pt,line join=round] (115.78, 84.70) --
	(119.49, 84.70);

\path[draw=drawColor,line width= 0.6pt,line join=round] (134.32, 89.69) --
	(138.03, 89.69);

\path[draw=drawColor,line width= 0.6pt,line join=round] (136.18, 89.69) --
	(136.18, 87.82);

\path[draw=drawColor,line width= 0.6pt,line join=round] (134.32, 87.82) --
	(138.03, 87.82);

\path[draw=drawColor,line width= 0.6pt,line join=round] (152.87, 90.78) --
	(156.58, 90.78);

\path[draw=drawColor,line width= 0.6pt,line join=round] (154.72, 90.78) --
	(154.72, 89.43);

\path[draw=drawColor,line width= 0.6pt,line join=round] (152.87, 89.43) --
	(156.58, 89.43);

\path[draw=drawColor,line width= 0.6pt,line join=round] (171.41, 96.17) --
	(175.12, 96.17);

\path[draw=drawColor,line width= 0.6pt,line join=round] (173.27, 96.17) --
	(173.27, 90.84);

\path[draw=drawColor,line width= 0.6pt,line join=round] (171.41, 90.84) --
	(175.12, 90.84);

\path[draw=drawColor,line width= 0.6pt,line join=round] (189.96, 95.99) --
	(193.67, 95.99);

\path[draw=drawColor,line width= 0.6pt,line join=round] (191.81, 95.99) --
	(191.81, 92.89);

\path[draw=drawColor,line width= 0.6pt,line join=round] (189.96, 92.89) --
	(193.67, 92.89);

\path[draw=drawColor,line width= 0.6pt,line join=round] (208.51,112.19) --
	(212.21,112.19);

\path[draw=drawColor,line width= 0.6pt,line join=round] (210.36,112.19) --
	(210.36, 88.57);

\path[draw=drawColor,line width= 0.6pt,line join=round] (208.51, 88.57) --
	(212.21, 88.57);

\path[draw=drawColor,line width= 0.6pt,line join=round] (227.05,104.66) --
	(230.76,104.66);

\path[draw=drawColor,line width= 0.6pt,line join=round] (228.91,104.66) --
	(228.91, 95.91);

\path[draw=drawColor,line width= 0.6pt,line join=round] (227.05, 95.91) --
	(230.76, 95.91);
\definecolor{drawColor}{RGB}{228,26,28}

\path[draw=drawColor,line width= 0.6pt,line join=round] ( 41.60, 85.64) --
	( 45.31, 85.64);

\path[draw=drawColor,line width= 0.6pt,line join=round] ( 43.45, 85.64) --
	( 43.45, 75.17);

\path[draw=drawColor,line width= 0.6pt,line join=round] ( 41.60, 75.17) --
	( 45.31, 75.17);

\path[draw=drawColor,line width= 0.6pt,line join=round] ( 60.14, 81.95) --
	( 63.85, 81.95);

\path[draw=drawColor,line width= 0.6pt,line join=round] ( 62.00, 81.95) --
	( 62.00, 79.42);

\path[draw=drawColor,line width= 0.6pt,line join=round] ( 60.14, 79.42) --
	( 63.85, 79.42);

\path[draw=drawColor,line width= 0.6pt,line join=round] ( 78.69, 88.02) --
	( 82.40, 88.02);

\path[draw=drawColor,line width= 0.6pt,line join=round] ( 80.54, 88.02) --
	( 80.54, 80.67);

\path[draw=drawColor,line width= 0.6pt,line join=round] ( 78.69, 80.67) --
	( 82.40, 80.67);

\path[draw=drawColor,line width= 0.6pt,line join=round] ( 97.23, 85.97) --
	(100.94, 85.97);

\path[draw=drawColor,line width= 0.6pt,line join=round] ( 99.09, 85.97) --
	( 99.09, 83.57);

\path[draw=drawColor,line width= 0.6pt,line join=round] ( 97.23, 83.57) --
	(100.94, 83.57);

\path[draw=drawColor,line width= 0.6pt,line join=round] (115.78, 88.28) --
	(119.49, 88.28);

\path[draw=drawColor,line width= 0.6pt,line join=round] (117.63, 88.28) --
	(117.63, 85.34);

\path[draw=drawColor,line width= 0.6pt,line join=round] (115.78, 85.34) --
	(119.49, 85.34);

\path[draw=drawColor,line width= 0.6pt,line join=round] (134.32, 92.47) --
	(138.03, 92.47);

\path[draw=drawColor,line width= 0.6pt,line join=round] (136.18, 92.47) --
	(136.18, 86.70);

\path[draw=drawColor,line width= 0.6pt,line join=round] (134.32, 86.70) --
	(138.03, 86.70);

\path[draw=drawColor,line width= 0.6pt,line join=round] (152.87, 90.88) --
	(156.58, 90.88);

\path[draw=drawColor,line width= 0.6pt,line join=round] (154.72, 90.88) --
	(154.72, 89.41);

\path[draw=drawColor,line width= 0.6pt,line join=round] (152.87, 89.41) --
	(156.58, 89.41);

\path[draw=drawColor,line width= 0.6pt,line join=round] (171.41, 95.39) --
	(175.12, 95.39);

\path[draw=drawColor,line width= 0.6pt,line join=round] (173.27, 95.39) --
	(173.27, 90.74);

\path[draw=drawColor,line width= 0.6pt,line join=round] (171.41, 90.74) --
	(175.12, 90.74);

\path[draw=drawColor,line width= 0.6pt,line join=round] (189.96, 95.11) --
	(193.67, 95.11);

\path[draw=drawColor,line width= 0.6pt,line join=round] (191.81, 95.11) --
	(191.81, 92.93);

\path[draw=drawColor,line width= 0.6pt,line join=round] (189.96, 92.93) --
	(193.67, 92.93);

\path[draw=drawColor,line width= 0.6pt,line join=round] (208.51, 99.72) --
	(212.21, 99.72);

\path[draw=drawColor,line width= 0.6pt,line join=round] (210.36, 99.72) --
	(210.36, 94.38);

\path[draw=drawColor,line width= 0.6pt,line join=round] (208.51, 94.38) --
	(212.21, 94.38);

\path[draw=drawColor,line width= 0.6pt,line join=round] (227.05,100.13) --
	(230.76,100.13);

\path[draw=drawColor,line width= 0.6pt,line join=round] (228.91,100.13) --
	(228.91, 97.69);

\path[draw=drawColor,line width= 0.6pt,line join=round] (227.05, 97.69) --
	(230.76, 97.69);
\definecolor{drawColor}{gray}{0.20}

\path[draw=drawColor,line width= 0.6pt,line join=round,line cap=round] ( 32.14, 28.25) rectangle (240.22,160.72);
\end{scope}
\begin{scope}
\path[clip] (  0.00,  0.00) rectangle (245.72,166.22);
\definecolor{drawColor}{gray}{0.30}

\node[text=drawColor,anchor=base east,inner sep=0pt, outer sep=0pt, scale=  0.72] at ( 27.19, 31.79) {0};

\node[text=drawColor,anchor=base east,inner sep=0pt, outer sep=0pt, scale=  0.72] at ( 27.19, 42.74) {10};

\node[text=drawColor,anchor=base east,inner sep=0pt, outer sep=0pt, scale=  0.72] at ( 27.19, 53.68) {20};

\node[text=drawColor,anchor=base east,inner sep=0pt, outer sep=0pt, scale=  0.72] at ( 27.19, 64.63) {30};

\node[text=drawColor,anchor=base east,inner sep=0pt, outer sep=0pt, scale=  0.72] at ( 27.19, 75.58) {40};

\node[text=drawColor,anchor=base east,inner sep=0pt, outer sep=0pt, scale=  0.72] at ( 27.19, 86.53) {50};

\node[text=drawColor,anchor=base east,inner sep=0pt, outer sep=0pt, scale=  0.72] at ( 27.19, 97.48) {60};

\node[text=drawColor,anchor=base east,inner sep=0pt, outer sep=0pt, scale=  0.72] at ( 27.19,108.43) {70};

\node[text=drawColor,anchor=base east,inner sep=0pt, outer sep=0pt, scale=  0.72] at ( 27.19,119.37) {80};

\node[text=drawColor,anchor=base east,inner sep=0pt, outer sep=0pt, scale=  0.72] at ( 27.19,130.32) {90};

\node[text=drawColor,anchor=base east,inner sep=0pt, outer sep=0pt, scale=  0.72] at ( 27.19,141.27) {100};

\node[text=drawColor,anchor=base east,inner sep=0pt, outer sep=0pt, scale=  0.72] at ( 27.19,152.22) {110};
\end{scope}
\begin{scope}
\path[clip] (  0.00,  0.00) rectangle (245.72,166.22);
\definecolor{drawColor}{gray}{0.20}

\path[draw=drawColor,line width= 0.6pt,line join=round] ( 29.39, 34.27) --
	( 32.14, 34.27);

\path[draw=drawColor,line width= 0.6pt,line join=round] ( 29.39, 45.22) --
	( 32.14, 45.22);

\path[draw=drawColor,line width= 0.6pt,line join=round] ( 29.39, 56.16) --
	( 32.14, 56.16);

\path[draw=drawColor,line width= 0.6pt,line join=round] ( 29.39, 67.11) --
	( 32.14, 67.11);

\path[draw=drawColor,line width= 0.6pt,line join=round] ( 29.39, 78.06) --
	( 32.14, 78.06);

\path[draw=drawColor,line width= 0.6pt,line join=round] ( 29.39, 89.01) --
	( 32.14, 89.01);

\path[draw=drawColor,line width= 0.6pt,line join=round] ( 29.39, 99.96) --
	( 32.14, 99.96);

\path[draw=drawColor,line width= 0.6pt,line join=round] ( 29.39,110.91) --
	( 32.14,110.91);

\path[draw=drawColor,line width= 0.6pt,line join=round] ( 29.39,121.85) --
	( 32.14,121.85);

\path[draw=drawColor,line width= 0.6pt,line join=round] ( 29.39,132.80) --
	( 32.14,132.80);

\path[draw=drawColor,line width= 0.6pt,line join=round] ( 29.39,143.75) --
	( 32.14,143.75);

\path[draw=drawColor,line width= 0.6pt,line join=round] ( 29.39,154.70) --
	( 32.14,154.70);
\end{scope}
\begin{scope}
\path[clip] (  0.00,  0.00) rectangle (245.72,166.22);
\definecolor{drawColor}{gray}{0.20}

\path[draw=drawColor,line width= 0.6pt,line join=round] ( 43.45, 25.50) --
	( 43.45, 28.25);

\path[draw=drawColor,line width= 0.6pt,line join=round] ( 62.00, 25.50) --
	( 62.00, 28.25);

\path[draw=drawColor,line width= 0.6pt,line join=round] ( 80.54, 25.50) --
	( 80.54, 28.25);

\path[draw=drawColor,line width= 0.6pt,line join=round] ( 99.09, 25.50) --
	( 99.09, 28.25);

\path[draw=drawColor,line width= 0.6pt,line join=round] (117.63, 25.50) --
	(117.63, 28.25);

\path[draw=drawColor,line width= 0.6pt,line join=round] (136.18, 25.50) --
	(136.18, 28.25);

\path[draw=drawColor,line width= 0.6pt,line join=round] (154.72, 25.50) --
	(154.72, 28.25);

\path[draw=drawColor,line width= 0.6pt,line join=round] (173.27, 25.50) --
	(173.27, 28.25);

\path[draw=drawColor,line width= 0.6pt,line join=round] (191.81, 25.50) --
	(191.81, 28.25);

\path[draw=drawColor,line width= 0.6pt,line join=round] (210.36, 25.50) --
	(210.36, 28.25);

\path[draw=drawColor,line width= 0.6pt,line join=round] (228.91, 25.50) --
	(228.91, 28.25);
\end{scope}
\begin{scope}
\path[clip] (  0.00,  0.00) rectangle (245.72,166.22);
\definecolor{drawColor}{gray}{0.30}

\node[text=drawColor,anchor=base,inner sep=0pt, outer sep=0pt, scale=  0.72] at ( 43.45, 18.34) {0};

\node[text=drawColor,anchor=base,inner sep=0pt, outer sep=0pt, scale=  0.72] at ( 62.00, 18.34) {1};

\node[text=drawColor,anchor=base,inner sep=0pt, outer sep=0pt, scale=  0.72] at ( 80.54, 18.34) {2};

\node[text=drawColor,anchor=base,inner sep=0pt, outer sep=0pt, scale=  0.72] at ( 99.09, 18.34) {3};

\node[text=drawColor,anchor=base,inner sep=0pt, outer sep=0pt, scale=  0.72] at (117.63, 18.34) {4};

\node[text=drawColor,anchor=base,inner sep=0pt, outer sep=0pt, scale=  0.72] at (136.18, 18.34) {5};

\node[text=drawColor,anchor=base,inner sep=0pt, outer sep=0pt, scale=  0.72] at (154.72, 18.34) {6};

\node[text=drawColor,anchor=base,inner sep=0pt, outer sep=0pt, scale=  0.72] at (173.27, 18.34) {7};

\node[text=drawColor,anchor=base,inner sep=0pt, outer sep=0pt, scale=  0.72] at (191.81, 18.34) {8};

\node[text=drawColor,anchor=base,inner sep=0pt, outer sep=0pt, scale=  0.72] at (210.36, 18.34) {9};

\node[text=drawColor,anchor=base,inner sep=0pt, outer sep=0pt, scale=  0.72] at (228.91, 18.34) {10};
\end{scope}
\begin{scope}
\path[clip] (  0.00,  0.00) rectangle (245.72,166.22);
\definecolor{drawColor}{RGB}{0,0,0}

\node[text=drawColor,anchor=base,inner sep=0pt, outer sep=0pt, scale=  0.90] at (136.18,  7.44) {Relative Loss on Link [{\%}]};
\end{scope}
\begin{scope}
\path[clip] (  0.00,  0.00) rectangle (245.72,166.22);
\definecolor{drawColor}{RGB}{0,0,0}

\node[text=drawColor,rotate= 90.00,anchor=base,inner sep=0pt, outer sep=0pt, scale=  0.90] at ( 11.70, 94.48) {Runtime [s]};
\end{scope}
\begin{scope}
\path[clip] (  0.00,  0.00) rectangle (245.72,166.22);
\definecolor{drawColor}{RGB}{0,0,0}
\definecolor{fillColor}{RGB}{255,255,255}

\path[draw=drawColor,line width= 0.3pt,line join=round,line cap=round,fill=fillColor] ( 35.14,117.81) rectangle (127.30,157.72);
\end{scope}
\begin{scope}
\path[clip] (  0.00,  0.00) rectangle (245.72,166.22);
\definecolor{fillColor}{RGB}{255,255,255}

\path[fill=fillColor] ( 40.64,142.59) rectangle ( 50.28,152.22);
\end{scope}
\begin{scope}
\path[clip] (  0.00,  0.00) rectangle (245.72,166.22);
\definecolor{drawColor}{RGB}{228,26,28}

\path[draw=drawColor,line width= 0.6pt,line join=round] ( 41.60,147.40) -- ( 49.31,147.40);
\end{scope}
\begin{scope}
\path[clip] (  0.00,  0.00) rectangle (245.72,166.22);
\definecolor{drawColor}{RGB}{228,26,28}

\path[draw=drawColor,line width= 0.6pt,line join=round,line cap=round] ( 45.46,147.40) circle (  2.71);
\end{scope}
\begin{scope}
\path[clip] (  0.00,  0.00) rectangle (245.72,166.22);
\definecolor{drawColor}{RGB}{228,26,28}

\path[draw=drawColor,line width= 0.6pt,line join=round] ( 41.60,147.40) -- ( 49.31,147.40);
\end{scope}
\begin{scope}
\path[clip] (  0.00,  0.00) rectangle (245.72,166.22);
\definecolor{fillColor}{RGB}{255,255,255}

\path[fill=fillColor] ( 40.64,132.95) rectangle ( 50.28,142.58);
\end{scope}
\begin{scope}
\path[clip] (  0.00,  0.00) rectangle (245.72,166.22);
\definecolor{drawColor}{RGB}{55,126,184}

\path[draw=drawColor,line width= 0.6pt,line join=round] ( 41.60,137.77) -- ( 49.31,137.77);
\end{scope}
\begin{scope}
\path[clip] (  0.00,  0.00) rectangle (245.72,166.22);
\definecolor{drawColor}{RGB}{55,126,184}

\path[draw=drawColor,line width= 0.6pt,line join=round,line cap=round] ( 42.75,135.06) -- ( 48.17,140.48);

\path[draw=drawColor,line width= 0.6pt,line join=round,line cap=round] ( 42.75,140.48) -- ( 48.17,135.06);
\end{scope}
\begin{scope}
\path[clip] (  0.00,  0.00) rectangle (245.72,166.22);
\definecolor{drawColor}{RGB}{55,126,184}

\path[draw=drawColor,line width= 0.6pt,line join=round] ( 41.60,137.77) -- ( 49.31,137.77);
\end{scope}
\begin{scope}
\path[clip] (  0.00,  0.00) rectangle (245.72,166.22);
\definecolor{fillColor}{RGB}{255,255,255}

\path[fill=fillColor] ( 40.64,123.31) rectangle ( 50.28,132.95);
\end{scope}
\begin{scope}
\path[clip] (  0.00,  0.00) rectangle (245.72,166.22);
\definecolor{drawColor}{RGB}{77,175,74}

\path[draw=drawColor,line width= 0.6pt,line join=round] ( 41.60,128.13) -- ( 49.31,128.13);
\end{scope}
\begin{scope}
\path[clip] (  0.00,  0.00) rectangle (245.72,166.22);
\definecolor{drawColor}{RGB}{77,175,74}

\path[draw=drawColor,line width= 0.6pt,line join=round,line cap=round] ( 41.62,128.13) -- ( 49.29,128.13);

\path[draw=drawColor,line width= 0.6pt,line join=round,line cap=round] ( 45.46,124.30) -- ( 45.46,131.96);
\end{scope}
\begin{scope}
\path[clip] (  0.00,  0.00) rectangle (245.72,166.22);
\definecolor{drawColor}{RGB}{77,175,74}

\path[draw=drawColor,line width= 0.6pt,line join=round] ( 41.60,128.13) -- ( 49.31,128.13);
\end{scope}
\begin{scope}
\path[clip] (  0.00,  0.00) rectangle (245.72,166.22);
\definecolor{drawColor}{RGB}{0,0,0}

\node[text=drawColor,anchor=base west,inner sep=0pt, outer sep=0pt, scale=  0.72] at ( 50.28,144.92) {Reliable Ordered UDP};
\end{scope}
\begin{scope}
\path[clip] (  0.00,  0.00) rectangle (245.72,166.22);
\definecolor{drawColor}{RGB}{0,0,0}

\node[text=drawColor,anchor=base west,inner sep=0pt, outer sep=0pt, scale=  0.72] at ( 50.28,135.29) {Reliable UDP};
\end{scope}
\begin{scope}
\path[clip] (  0.00,  0.00) rectangle (245.72,166.22);
\definecolor{drawColor}{RGB}{0,0,0}

\node[text=drawColor,anchor=base west,inner sep=0pt, outer sep=0pt, scale=  0.72] at ( 50.28,125.65) {TCP};
\end{scope}
\end{tikzpicture}

%% file: conclusion.tex
\section{Conclusion and Outlook}
\label{sec:conclusion}

The characteristics of actor communication lack a common design and often change with context and implementation. Most notably guarantees often change when moving from local to remote contexts.

This work examined reliable delivery and ordering in the context of actor communication and found three notable delivery guarantees. First, a ``fire and forget'' approach that bares little overhead. It allows developers to build more complex systems on top but requires explicit error handling as part of the application. Second, guaranteed delivery to the mailbox of the receiving actor. This aligns the guarantees between local and remote contexts thus increasing the transparency of distribution. Third and last, guaranteed processing feedback bares great value when considering end-to-end communication. However, it induces overhead and might not be required in all cases.

With regard to ordering, the discrepancy between local and remote contexts weakens guarantees from \textit{causal} ordering to \textit{FIFO} or none. While algorithms exist to establish a \textit{causal} order in distributed systems, these come at significant cost. Ensuring \textit{FIFO} ordering already provides valuable information, helps developers to reason about their code, and comes at comparably little cost.

Many implementations inherit their guarantees for remote messaging from TCP. This is problematic as transport protocols offer more than guarantees and can adjust applications to specific environments. To enhance transport bindings in CAF we implemented a composable network stack that allows bundling a transport protocol with additional layers to add new functionality. An evaluation shows that our layer design introduces minimal overhead. Additionally a reliability layer was implemented and tested with a varying degree of packet loss to showcase a more complex layer.

Examining existing actor systems and laying out the implementation space for reliable delivery and ordering is a first step towards a more detailed discussion on the message passing guarantees for actors. While our implementation shows that a lightweight implementation is possible, generalization is required to make our results translatable to a wide range of frameworks.

There are several directions for future work. As a first step, the system broker and SSL module should be ported to the new design and thoroughly benchmarked against their previous implementations. Next, we want to examine the possibility to integrate the streaming capabilities of CAF into the new brokers. Streaming adds a backchannel to actor communication to avoid overburdening actors and with this addition could take the network behavior into account. Splitting the monolithic system broker into smaller light-weight brokers is a first step towards a multi-threaded network back-end. Finally, there are aspects that were disregarded in this work and are left for future work such as reachability and security.